\documentclass[preprint, 3p]{elsarticle}

\usepackage[utf8]{inputenc}
\usepackage{textcomp}
\usepackage{graphicx}
\usepackage{amsmath}
\usepackage[version=4]{mhchem}
\usepackage{siunitx}
\usepackage{longtable,tabularx}
\usepackage{subcaption}
\usepackage{soul}
\usepackage{multicol}
\usepackage{multirow}
\usepackage{url}
\newcommand{\dist}[1]{\left\lVert #1 \right\rVert}

\usepackage{amssymb} 
\usepackage{amsmath}  
\allowdisplaybreaks

\journal{Computers and Fluids}

\begin{document}

\begin{frontmatter}


\title{Large airfoil models}

\author{Howon Lee\corref{cor1}}
\cortext[cor1]{Corresponding author.}
\ead{hlee981@gatech.edu}
\author{Aanchal Save}
\author{Pranay Seshadri}
\author{Juergen Rauleder}
\address{Georgia Institute of Technology, North Ave NW, Atlanta, 30332, GA, USA}

\begin{abstract}
The development of a Large Airfoil Model (LAM), a transformative approach for answering technical questions on airfoil aerodynamics, requires a vast dataset and a model to leverage it. To build this foundation, a novel probabilistic machine learning approach, A Deep Airfoil Prediction Tool (ADAPT), has been developed. ADAPT makes uncertainty-aware predictions of airfoil pressure coefficient ($C_p$) distributions by harnessing experimental data and incorporating measurement uncertainties. By employing deep kernel learning, performing Gaussian Process Regression in a ten-dimensional latent space learned by a neural network, ADAPT effectively handles unstructured experimental datasets. In tandem, Airfoil Surface Pressure Information Repository of Experiments (ASPIRE), the first large-scale, open-source repository of airfoil experimental data, has been developed. ASPIRE integrates century-old historical data with modern reports, forming an unparalleled resource of real-world pressure measurements. This addresses a critical gap left by prior repositories, which relied primarily on numerical simulations. Demonstrative results for three airfoils show that ADAPT accurately predicts $C_p$ distributions and aerodynamic coefficients across varied flow conditions, achieving a mean absolute error in enclosed area ($\text{MAE}_\text{enclosed}$) of $0.029$. ASPIRE and ADAPT lay the foundation for an interactive airfoil analysis tool driven by a large language model, enabling users to perform design tasks based on natural language questions rather than explicit technical input.
\end{abstract}

\begin{keyword}
Deep kernel learning \sep Gaussian processes \sep Bayesian inference \sep Airfoils \sep Aerodynamics
\end{keyword}

\end{frontmatter}

\section{Introduction}\label{1_intro}
Large language models (LLMs) such as ChatGPT~\cite{brown}, Claude~\cite{claude}, and Gemini~\cite{gemini}, are now at the forefront of artificial intelligence (AI), rapidly gaining popularity as they make learning and understanding complex topics more accessible. Beyond general-purpose LLMs, it is also possible to create specialized models, designed to answer questions and provide insights on specific topics or datasets \cite{LLM_med, AviationGPT, aeroBERT-NER}.

In the context of aerodynamics, there are several key questions that aerodynamicists have during the wing (fixed-wing, rotary-wing, or wind turbine) design process: What is the maximum lift coefficient? Does stall occur at the leading or trailing edge? How do drag and stall behavior change with Mach number? Is there a significant pitching moment? These questions inherently involve operations on sectional pressure coefficients, $C_p$. This motivates the idea that a LLM for airfoil aerodynamics, or a \emph{large airfoil model} (LAM),  could be used to answer these queries. To accurately respond to user inquiries, the LAM must be able to (1) obtain information by leveraging historical data, or (2) in lieu of available data, generate its own $C_p$ distributions and perform the necessary operations to obtain chosen quantities of interest (QoIs).

As a first step in the development of the LAM, it is necessary to design a means to predict aerodynamic properties of airfoils, a requirement ubiquitous across fixed wings, rotorcraft, and turbomachinery. Traditionally, airfoil properties have been obtained by wind tunnel experiments or computational fluid dynamics (CFD) simulations. With recent developments in computational power and data-driven modeling, there have been efforts to map airfoil geometric information to its $C_p$ directly using machine learning (ML). For example, Yilmaz and German~\cite{yilmaz2017} applied a classification framework to develop a convolutional neural network (CNN) trained on airfoil pressures obtained from a panel method code. The model successfully predicted the $C_p$ distribution as a series of discretized values. Hui et al.~\cite{Hui_2020} proposed a five-layer CNN model to predict airfoil $C_p$, trained from their in-house airfoil database of RANS simulations. Zhang~\cite{Zhang_2023} utilized a variational autoencoder on a RANS dataset, where an encoder was used to extract latent features in varying airfoil geometries, angles of attack ($\alpha$), and freestream Mach numbers ($M_\infty$). A decoder was then used to reconstruct the $C_p$ within this latent space. Intrinsically, these models seek a latent variable representation.

An alternate approach is to explicitly employ dimensionality reduction techniques; active subspaces~\cite{constatine2014} and related methods have had tremendous success in identifying subspaces for QoIs derived from $C_p$ distributions. Active spaces were successfully used to map airfoil geometries and operating conditions to low-dimensional embeddings that capture the parameters' relations to lift and drag~\cite{grey_constantine}. The orthogonal complement of the \emph{active subspace}, termed the \emph{inactive subspace} can prove useful for identifying insensitivities and robustness. Wong et al.~\cite{wong2022} showed that samples from the inactive subspace could be used to ascertain whether a compressor blade with manufacturing variations or degradation could provide near identical performance to nominal. This answer to how ``forgiving'' an airfoil is to imperfections is a crucial component in robust aerodynamic design. Wong et al.~\cite{wong2020} proposed employing embedded ridge approximations to find the dimension-reducing subspace based on an underlying pressure field around an airfoil. 

There have also been multiple efforts to develop models that predict sectional lift, drag, and moment coefficients ($c_l$, $c_d$, and $c_m$). For example, Zhang et al.~\cite{Zhang_2018} predicted XFOIL-based $c_l$ and $c_d$ using deep learning methodologies such as CNNs and multi-layered perceptrons (MLP). Liu et al.~\cite{Liu_2022} coupled a CNN with Bayesian optimization to predict the aerodynamic coefficients obtained from OpeanFOAM simulations. These models, while two-dimensional in nature, can be applied to three-dimensional analyses (e.g. lifting line theory, blade element theory, and actuator disk model) in the form of lookup tables. A good example of this approach is the work by Cornelius and Schmitz~\cite{cornelius2024}, who adopted a feed-forward neural network trained on an extensive database of OVERFLOW simulations named PALMO. The model was used to generate a C81 table for an actuator disk model and was successfully coupled with CFD. However, these models treat each QoI independently, rather than taking advantage of the fact that $C_p$ inherently is related to all of the coefficients. Building a model that is capable of leveraging the underlying physics would enjoy the added benefit of improved interpretability. For instance, one could identify leading/trailing edge stall and reattachment, which translates pitch link loads, a critical element in rotor design.

A drawback found in some of the aforementioned approaches is the lack of a framework to characterize uncertainty. While it is not yet common to identify and rigorously propagate uncertainties into integrated metrics during design processes, incorporating uncertainty quantification techniques will be a crucial step towards making aviation much safer. Sources of uncertainty in airfoil data include the difference between theoretical and actual airfoil geometry, unsteadiness in the freestream, and unsteadiness introduced by flow separation. Neural networks, deterministic in nature, have difficulty quantifying such uncertainties. One way to develop an uncertainty-aware model is to take a Bayesian approach. For example, Anhichem et al., used Gaussian Process (GP) regression to build surrogate models for pressure distributions over an OAT15A airfoil~\cite{Anhichem_2024} and a RBC12 half-wing-fuselage multi-fidelity model~\cite{Anhichem_2022} via data fusion. 

From the survey of previous works, it becomes evident that existing ML models are dependent on an extensive sweep of CFD simulations for training. This is due to the fact that experimental data, the \emph{ground truth}, are much more limited in accessibility. The architecture of certain neural networks, such as CNN, also require $C_p$ values along a preset grid, which makes them unsuitable for training on data sourced from different experiments. Furthermore, the sole reliance on CFD may introduce numerical biases due to the researcher's modeling decisions, such as the choice of turbulence models. The limitations in existing literature of ML airfoil prediction tools can be summarized as the following:
\begin{itemize}
\item reliance on a particular CFD simulation to build a database, which may be prone to numerical biases, 
\item minimal leveraging of physical relationships between $C_p$ and aerodynamic coefficients, and
\item limited framework to characterize and propagate uncertainties within the training data and the model.
\end{itemize}

In addition to the AI analysis tool, the other requirement for a robust LAM is a rich historical database. Several open-source databases exist to train or adjust LLMs~\cite{commoncrawl, thepile, refinedweb}. There have also been datasets created to fine-tune existing LLMs for aerospace applications, such as aeroBERT-NER~\cite{aeroBERT-NER} which was used to adapt the BERT language model~\cite{BERT} to identify named entities in aerospace requirements. However, the vast, nearly century-old historical repository of experimental airfoil pressure data is largely left untapped by the wider aerospace community. This is because, while many reports and papers are available in NASA Technical Reports Server (NTRS) or Aerospace Research Central (ARC), the majority of the data is locked away in images or in poorly legible bulk scans. Digitizing the publicly available data into an accessible, open-source database is thus a paramount and time-critical task, as in a few decades, the next generation may be unaware of the existence of this wealth of experimental data. 

Building this database would also be pivotal in creating a ML model that can overcome existing limitations. Ideally, a probabilistic ML model would be able to capture and propagate experimental uncertainty when predicting $C_p$ and the related $c_l$, $c_d$, and $c_m$ for an airfoil under specific $M_\infty$ and $\alpha$. Such a model will be free of numerical biases from CFD simulations and can be used to supplement existing databases with $C_p$ predictions for untested operating conditions.

\begin{figure}[!ht]
\centering
\includegraphics[width=.85\textwidth]{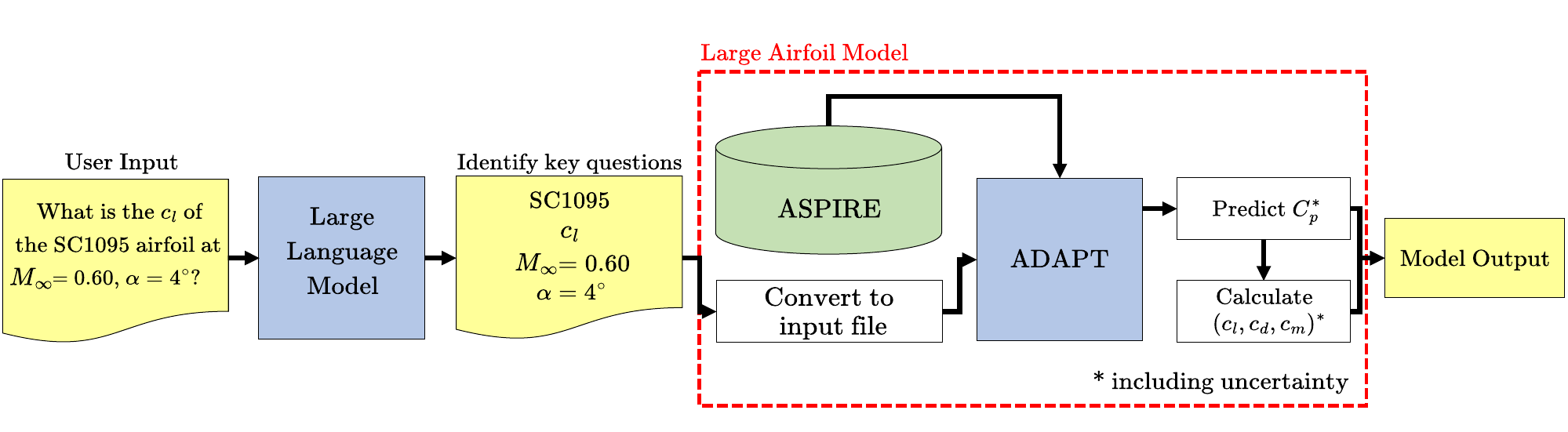}
\caption{Flowchart depicting the proposed forward problem workflow. The LAM is wrapped in red.}
\label{fig:forward_flowchart}
\end{figure}

In this paper, the primary components of the Large Airfoil Model\footnote{Accessible online on \url{https://large-airfoil-model.azurewebsites.net}} are introduced: a deep airfoil prediction tool (ADAPT) and the airfoil surface pressure information repository of experiments (ASPIRE). ADAPT is a novel machine learning model that employs a deep kernel learning architecture, combining deep learning for latent space identification and a probabilistic method for uncertainty characterization. Its input features are designed to incorporate the unstructured (relative to CFD) experimental data from ASPIRE, a digitized database of historical $C_p$ measurements. Figure~\ref{fig:forward_flowchart} depicts the workflow of the model. While only a forward problem is presented, the components can be applied to other tasks, including inverse problems. The full model, coupled with a LLM, allows users to perform a variety of design tasks based on the design questions that an aerospace engineer would ask, rather than requiring explicit technical inputs. Aspirationally, this represents a transformative shift in how aerodynamicists engage with data and models, paving the way for a more intuitive and innovative approach to aerodynamic design. 

The paper is organized as follows. In Sec.~\ref{2_LAM}, ADAPT, a novel deep probabilistic model to predict airfoil $C_p$ based on experimental training data, is proposed. In Sec.~\ref{3_database}, ASPIRE, providing the training data for ADAPT, is introduced. The methodology and challenges of the data digitization process are detailed, followed by a description of the range of available data.  The accuracy and computational efficiency of the model are assessed in Sec.~\ref{4_results}. 
 
\section{A Deep Airfoil Prediction Tool (ADAPT)}\label{2_LAM}
As discussed in Sec.~\ref{1_intro}, training on experiments instead of simulations allows the model to benefit from the reported measurement uncertainties. Additionally, the approach helps avoid potential biases associated with choosing a specific simulation tool—such as Reynolds-Averaged Navier--Stokes (RANS) solver with a particular turbulence model, which may not be suitable for all flow regimes—or relying on a set of tools with varying levels of fidelity.

\subsection{Machine Learning Model}
\begin{figure}[hbt!]
\centering
\includegraphics[width=.8\textwidth]{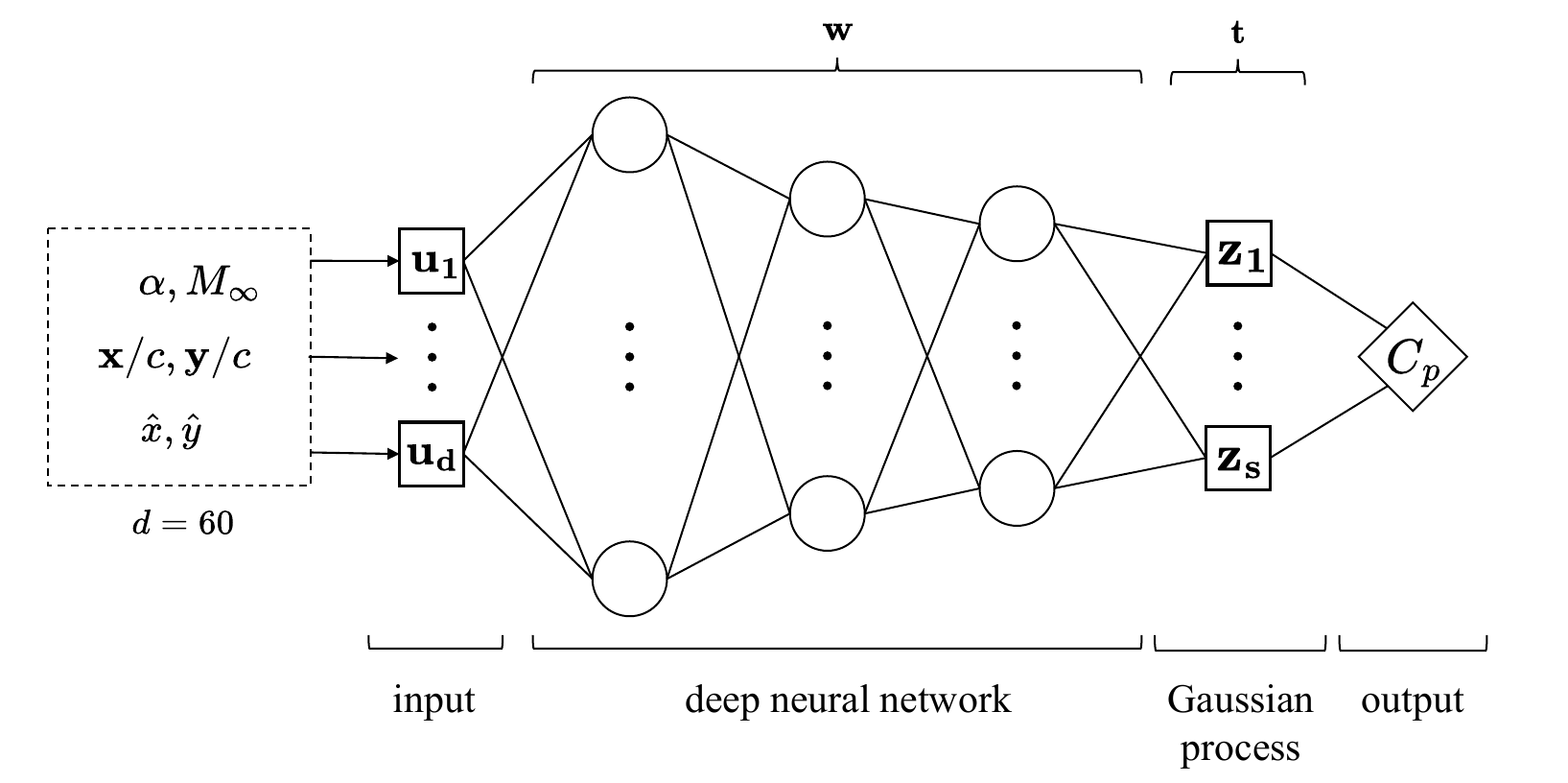}
\caption{Overview of the deep kernel learning model that maps a 60-dimensional training input to a 10-dimensional active space which is used to predict $C_p$ distributions.}
\label{fig:ModelOverview}
\end{figure}

ADAPT is deep kernel learning model, specifically designed to accommodate the training data structure made available by ASPIRE, as described in Sec.~\ref{3_database}. Its architecture is inspired by the modeling framework introduced by Wilson et al., 2013~\cite{Wilson_2016}. In ADAPT, a neural network is used to learn a nonlinear feature mapping from the input space to a latent space. Gaussian process model is applied on these transformed features, conditioned on the output observations of $C_p$. This approach allows the construction of kernels that capture the expressive power of deep neural network architectures. At its core, ADAPT is a Bayesian method, resulting in a multivariate Gaussian distribution over the space of $C_p$ values. This is written as
\begin{equation}
\begin{split}
    C_p \left(\underbrace{\alpha, M_\infty,}_{\textrm{operating conditions}} \underbrace{\mathbf{x}/c, \mathbf{y}/c,}_{\textrm{geometry}} \underbrace{\hat{x}, \hat{y}}_{\textrm{coordinates}} \right) & \equiv C_p \left( \underbrace{\mathbf{u}}_{\textrm{all inputs}}\right) \sim \mathcal{N} \left( \mu \left( \mathbf{z}, \mathbf{t} \right), \Sigma\left( \mathbf{z} , \mathbf{t} \right) \right),  \\
\textrm{where} \; \; \underbrace{\mathbf{z}}_{\textrm{latent variable}} & = \underbrace{f_{\mathbf{w}} \left(\mathbf{u} \right)}_{\textrm{deep neural network with weights\; $\mathbf{w}$}} ,
\end{split}
\end{equation}
where the notation $f_{\mathbf{w}}: \mathbf{u} \rightarrow \mathbf{z}$ denotes the deep neural network that is parameterized with weights $\mathbf{w}$, where $\mathbf{u} \in \mathbb{R}^{d}$ comprises the operating conditions, geometry and specific \emph{conformal} airfoil coordinates, i.e., $\mathbf{u} = \left\{ \alpha, M_\infty, \mathbf{x}/c, \mathbf{y}/c, \hat{x}, \hat{y}\right\}$. The outputs of the deep neural network are latent variables $\mathbf{z} \in \mathbb{R}^{s}$. In this latent space, a Gaussian process model is built, fully specified by a mean and covariance function. The covariance function is based on a two-point kernel function $k_{\mathbf{t}} \left(\mathbf{z}, \mathbf{z}'\right)$, where the subscript $\mathbf{t}$ indicates certain hyperparameters to parameterize the kernel function. The overall structure of the model is presented in Fig.~\ref{fig:ModelOverview}.

In ADAPT, a fully connected neural network (FCNN) with a [60-1000-1000-500-50-10] architecture maps the space of inputs $\mathbf{u}$ to the space of latent variables $\mathbf{z}$. The number of input variables is $d=60$ and the number of latent variables, which is driven by the network architecture, is $s=10$. For the Gaussian process layer, the Mat\a'ern 5/2 kernel function is used. The choice allows the model to more accurately capture aerodynamic phenomena on airfoils, such as the rapid pressure changes at the suction peak, transition from laminar to turbulent boundary layer, and compressibility/shock effects---all of which lead to large derivatives in $C_p$. Further details on the Gaussian process model and model training are provided in Appendices A--C. 

Since the model architecture only requires that the pressure measurement locations of the training data be defined by $\hat{x}$ and $\hat{y}$, the sources of training data are not necessarily restricted to experiments. As ADAPT is a Bayesian framework, it can take into account the inherent uncertainty in the training data during inference by defining the model likelihood (i.e. the noise model). In this context, experimental data enjoy the benefit of having easily identifiable uncertainty values, as quantified by measurement noise. For computational data, uncertainty quantification is far more nuanced and involved, requiring sophisticated techniques such as field inversion to infer modeling discrepancies in turbulence closures~\cite{cfd_uncertainty_quant}. As these uncertainties are deeply tied to the chosen computational model structure and spatial discretization, they are less directly applicable when constructing a noise model. For these reasons, this work focuses on the utilization of experimental training data rather than computational data, although by nature, the model can use either or both.

To this end, fixed noise values for each pressure measurements are first provided based on the reported accuracy of the original experiments. If the source material did not quantify the uncertainty of its measurements, the standard deviation was assumed to be 0.01, which was approximately the order of magnitude of errors observed in many sources. Some additional noise was also inferred to account for the human error introduced during the digitization process outlined in Sec.~\ref{sec:datamining}. It was found that the additional noise resulted in improved numerical stability of the predictions.

Another advantage of the model architecture lies in its grid flexibility, allowing users to tailor the number of discrete locations at which $C_p$ may be evaluated after training. This functional form offers significant benefits over deep neural networks that rely on a predefined discretization for the output, which also necessitates that the training data be fixed on a grid. This limitation of conventional neural networks contributes to the predominance of computational training data over experimental data, as users have direct control over the mesh. In contrast, experimental data feature varying pressure sensor locations for each experiment.

\subsection{Model Training}
The data were split into a training set and a test set, used to train the model and assess model accuracy, respectively. The sets were partitioned with respect to airfoils as opposed to withholding a certain percentage of available $C_p$. This ensures that the predictions are performed on airfoils that the model was not trained on. Experimental measurements of three airfoils were selected as the test set: SC1095~\cite{flemming1984}, NASA Supercritical Airfoil 9a~\cite{harris1979}, and NACA 63-415~\cite{bak2000}. The selections covered all applications within ASPIRE (general aviation, rotor, wind turbine) and supercriticality. In total, 40{,}416 $C_p$ data points were utilized to train the model and 879 data points were reserved for testing.

The training was performed on an NVIDIA A100 GPU. The Adam optimizer~\cite{adam} maximized the model's marginal likelihood, which served as the loss function, given the training data. The optimizer had an initial learning rate of $1.0 \times 10^{-3}$, with a step decay of 0.5 every 1000 epochs, resulting in a learning rate of $1.0 \times 10^{-4}$ after 3000 epochs of training. Beyond that point, the learning rate was set to decay linearly between $1.0 \times 10^{-3}$ and $1.0 \times 10^{-4}$ over 200-epoch cycles, for a total of 2000 additional epochs. Stochastic Weight Averaging (SWA), which equally averages the weights traversed by the optimizer, was performed by restoring the best weights for each cycle. As also reported by Refs.~\cite{izmailov2018} and~\cite{athiwaratkun2018}, the SWA procedure improved generalization of the LAM. 

The mean absolute error in the area enclosed by the predicted $C_p$ curves ($\text{MAE}_{\text{enclosed}}$) was found to be 0.033. The model accuracy is reported in terms of $\text{MAE}_{\text{enclosed}}$ due to the fact that mean absolute error (MAE) or the mean absolute percentage error (MAPE) of $C_p$ can be misleading. When using MAE, even slight misalignment between predicted and actual $C_p$ values across the entire airfoil can result in consistently large absolute errors, exaggerating the overall error metric. In the case of MAPE, small absolute errors can result in extremely high percentage errors when the true value is near zero, which occurs commonly in $C_p$. $\text{MAE}_{\text{enclosed}}$ avoids the limitations while remaining intuitive, since the error in the area enclosed by the pressure curves is analogous to an error in $c_l$. 

The model was trained via an exact GP inference, where the covariance matrix of the training set is directly inverted. The operation was bottle-necked by the high memory requirement and high computational complexity. To accommodate the continuous expansion of ASPIRE, it is essential to enhance the model's scalability. Approximate methods, such as stochastic variational inference \cite{hensman2014} and inducing point methods \cite{titsias09}, offer potential solutions. However, findings here indicate that these approaches result in considerably reduced accuracy, with only marginal improvements observed from increasing the number of inducing points. With 1000 inducing points, the $\text{MAE}_{\text{enclosed}}$ was 0.148, an order of magnitude greater than that of the exact method. Increasing the number to 10,000 yielded in approximately the same error. As the model is fully open-source, contributions from the community are encouraged to address this issue.

\subsection{Model Limitations}\label{sec:lim}
The main limitation of the LAM is the same as that of most other ML models; for inputs beyond the range of the training data, the predictions become significantly less reliable. Therefore, it is necessary to understand how the training data are distributed. For example, predictions for a wind turbine airfoil would be less reliable at higher freestream Mach numbers compared to results in the incompressible regime due to the relative lack of available training data. Sec.~\ref{sec:avail_data} discusses in detail the data distribution of the training database. An in-depth analysis of model predictions beyond this range is provided in Sec.~\ref{sec:beyond}.

Secondly, the airfoil thickness locations ($\mathbf{x}/c$) are fixed in the user input. With only 60 discrete chordwise locations, it can be challenging to accurately model small geometric features. Furthermore, existing training data are mostly from smooth airfoil geometries. This is because the experimentally tested profiles closely follow their respective design coordinates, which are typically streamlined for optimal performance. As a consequence of a relatively coarse input and a lack of training data, the model struggles to capture the effects of subtle variations in geometry, such as small cavities in the trailing edge, surface roughness and irregularities despite the fact that these features can influence aerodynamic performance of an airfoil. The model is thus best suited for predictions on smooth airfoils. Further analysis on the sensitivity of the current model to geometric perturbations is presented in Sec.~\ref{sec:beyond}.

A potential solution to this limitation is the incorporation of experimental data that capture the aerodynamic effects of surface irregularities, such as icing data or surface roughness data. Ice accretion, for example, introduces geometric perturbations that can significantly impact flow behavior, providing valuable training examples for the model. This would enhance the model’s applicability to non-smooth airfoils.

Another limitation is that the model is more memory-intensive than traditional neural networks. The GP element of ADAPT requires storing the full $40{,}416 \times 40{,}416$ covariance matrix of the training data. The model thus requires more overhead memory compared to neural networks which only require the storage of the weights and biases of each layer.  

The reduction of required memory could be achieved via more scalable methods such as Stochastic Variational Deep Kernel Learning (SV-DKL)~\cite{svdkl}. This variational framework significantly improves memory management by enabling mini-batch training, which allows for loading only a subset of data at a time. This reduces peak memory usage, making it feasible to train larger models. In addition, SV-DKL utilizes inducing points to approximate the full $N \times N$ covariance matrix ($K_{NN}$, where $N$ is the number of training data points). A lower-rank approximation is performed from $M$ inducing points ($M \ll N$) where $K_{NN}~\approx~K_{NM} K_{MM}^{-1} K_{MN}$. The approach reduces memory complexity from $\mathcal{O}(N^2)$ to $\mathcal{O}(NM)$, further facilitating scalable training on large data sets.

Lastly, the LAM is an airfoil aerodynamics model and its predictions are two-dimensional in nature. Capturing the aerodynamics of a three-dimensional lifting surface, such as a wing or a rotor blade, introduces new challenges arising from three-dimensional effects such as tip vortex effects and suffers from a significantly increased parameter space required to fully define the wing. Extending this model for three-dimensional applications thus requires a different model be trained from a comparatively sparse data set of experimental wing measurements available in the literature. However, it has been found that the LAM's two-dimensional predictions can be used to augment such a model during training. In our previous work~\cite{lee_2025}, it was concluded that a three-dimensional aerodynamics prediction model with strong generalization capability can be trained by leveraging the two-dimensional predictions of the LAM as a physics-driven prior. This approach effectively constrained the learning process, enabling the model to focus on capturing the three-dimensional correction to the base airfoil pressure distribution rather than learning the full mapping from scratch.

\subsection{Natural Language Interface for Model-Driven Predictions}\label{sec:lam_da}
As discussed in Sec.~\ref{1_intro}, this research seeks to pave the way in developing a ``question-driven'' airfoil aerodynamics analysis tool. Such a model must allow users to interact with the model through natural language, rather than structured input files. This is best achieved by integrating the LAM with a large language model interface. At the same time, it is essential to ensure that the LLM responses are grounded in real experimental data or LAM predictions. In this work, a Retrieval-Augmented Generation (RAG) framework~\cite{rag} is used, which augments response generation by incorporating an information retrieval process. In a RAG framework, the most appropriate data from a specific database is obtained as contextual information for the LLM, ensuring that its responses are accurate, up-to-date, and relevant.

\begin{figure}[hbt!]
\centering
\includegraphics[width=.9\textwidth]{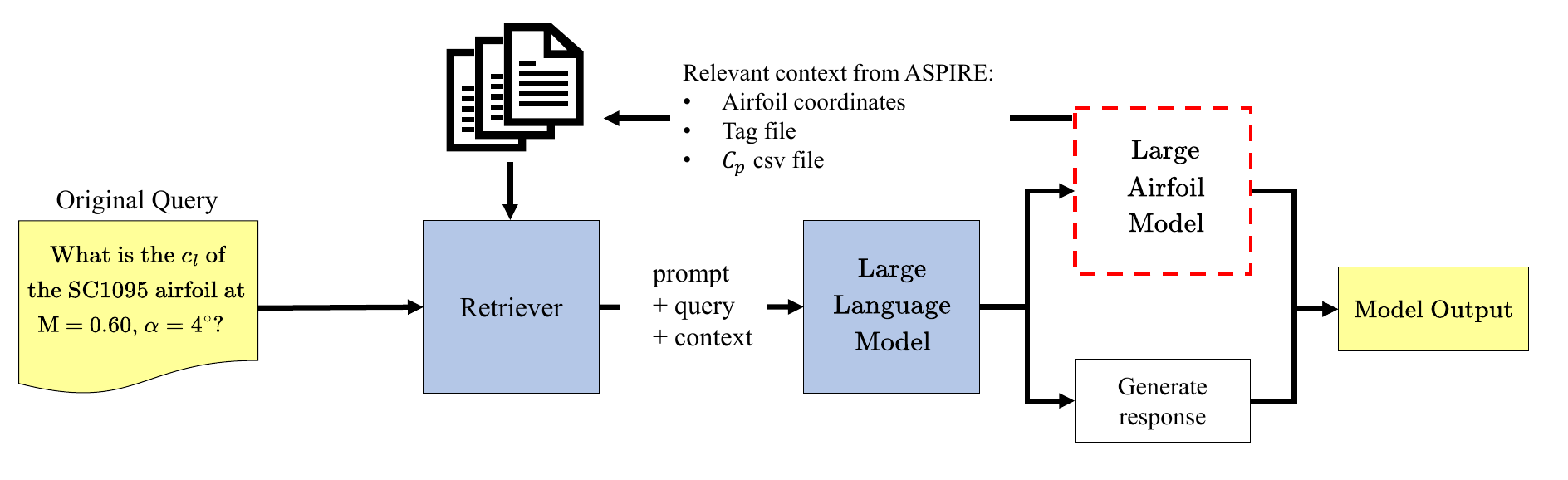}
\caption{Overview of the RAG framework powering the natural language interface for model-driven predictions. BERT~\cite{BERT} was used as the retriever and LLaMA 3.2 \cite{llama3} was used as the LLM.}
\label{fig:lam_da_overview}
\end{figure}

Fig.~\ref{fig:lam_da_overview} provides an overview of the RAG-enabled LLM. In this framework, the user query is initially passed through a retriever, which is designed to obtain the relevant airfoil and $C_p$ distribution information from the ASPIRE database. The retrieval process was facilitated by a BERT model~\cite{BERT}, selected for its advanced semantic understanding and effective handling of synonyms and paraphrasing. This capability is particularly critical in identifying aerodynamic parameters, which are often referenced using multiple terminologies. For example, ``angle of attack'' may be denoted as ``angle,'' ``AoA,'' or ``alpha.''

The retriever is used to extract aerodynamic parameters from the user query, starting with the airfoil. The extracted airfoil is compared with the list of available airfoils in ASPIRE. If there is a matching airfoil, BERT identifies additional parameters from the query: angle of attack, Mach number, and Reynolds number. The operating conditions of existing data are evaluated against the query parameters, applying tolerances of $0.1^\circ$, 0.01, and $1\times 10^5$ for $\alpha$, $M_\infty$, and $Re_c$, respectively. This enables the retriever to identify the index of the most relevant data file and the accompanying contextual sentences. For example, if the user asks for the $C_p$ distribution of a SC1095 airfoil at $\alpha = 6.2^\circ$ and $M_\infty = 0.6$, the retrieved index will reference the file \texttt{SC\_1095\_A6.2\_M0.60\_Re4.9e6\_A.csv}, which describes the $C_p$ distribution of the airfoil at the specified conditions. However, if a matching file is not found, the corresponding context entry would be: ``There is no matching airfoil available in the database.''

The context obtained from the retriever, along with the underlying prompt, is processed by the LLaMA 3.2 language model~\cite{llama3}, selected for its open-source nature and lightweight architecture. In addition to its standard capabilities for generating answers, the model is equipped with a predefined set of functions. If the retrieved context indicates the existence of a specific file (such as an airfoil's geometric profile or $C_p$ distribution) within ASPIRE, the model calls the relevant function to plot the relevant data from the corresponding CSV file. If the required $C_p$ distribution is not available within the database, the language model internally generates an input file based on the extracted key parameters. This is processed by ADAPT to generate the corresponding probabilistic predictions. The RAG-based approach enables a powerful AI-driven digital assistant to use the LAM, and offers the flexibility to incorporate additional functions in the future. Exemplary interactions with this interface are presented in \ref{app:chatbot}.

\section{Airfoil Surface Pressure Information Repository of Experiments (ASPIRE)}\label{3_database}
\subsection{Data Mining}\label{sec:datamining}
\begin{figure}[ht!]
\centering
\includegraphics[width=.75\textwidth]{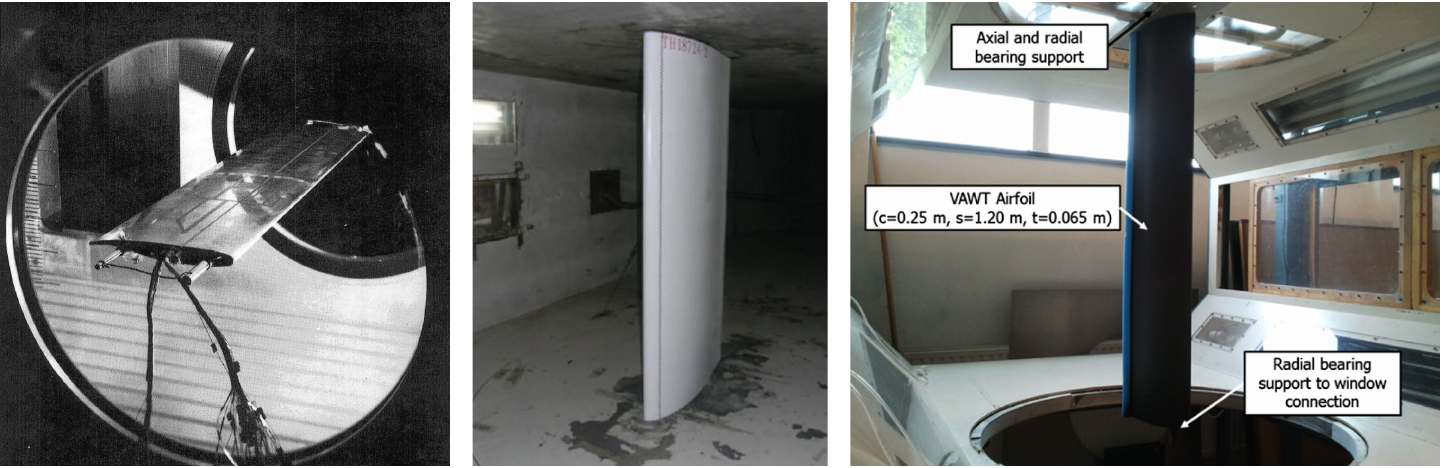}
\caption{``Infinite wing'' test articles for wind tunnel experiments. From left to right: Refs. [\citenum{spaid1983}, ~\citenum{chen2016},~\citenum{ragni2015}]}
\label{fig:infinitewing}
\end{figure}

A data mining campaign was undertaken to create a digital database of \emph{strictly experimental} pressure distributions for numerous airfoils. This large-scale, open-source repository, ASPIRE, consists of experimental $C_p$ measurements for various airfoil geometries at different angles of attack, chord-based Reynolds numbers, and freestream Mach numbers. The test articles of these experiments are of an \emph{infinite wing} configuration as seen in Fig.~\ref{fig:infinitewing}, where the wing spans the entire wind tunnel test section to minimize three-dimensional effects. ASPIRE's data sources range from nearly a century-old historical documents to more recent, accessible sources, including government reports, technical notes, and research articles.

\begin{figure}[h!]
\centering
\includegraphics[width=.85\textwidth]{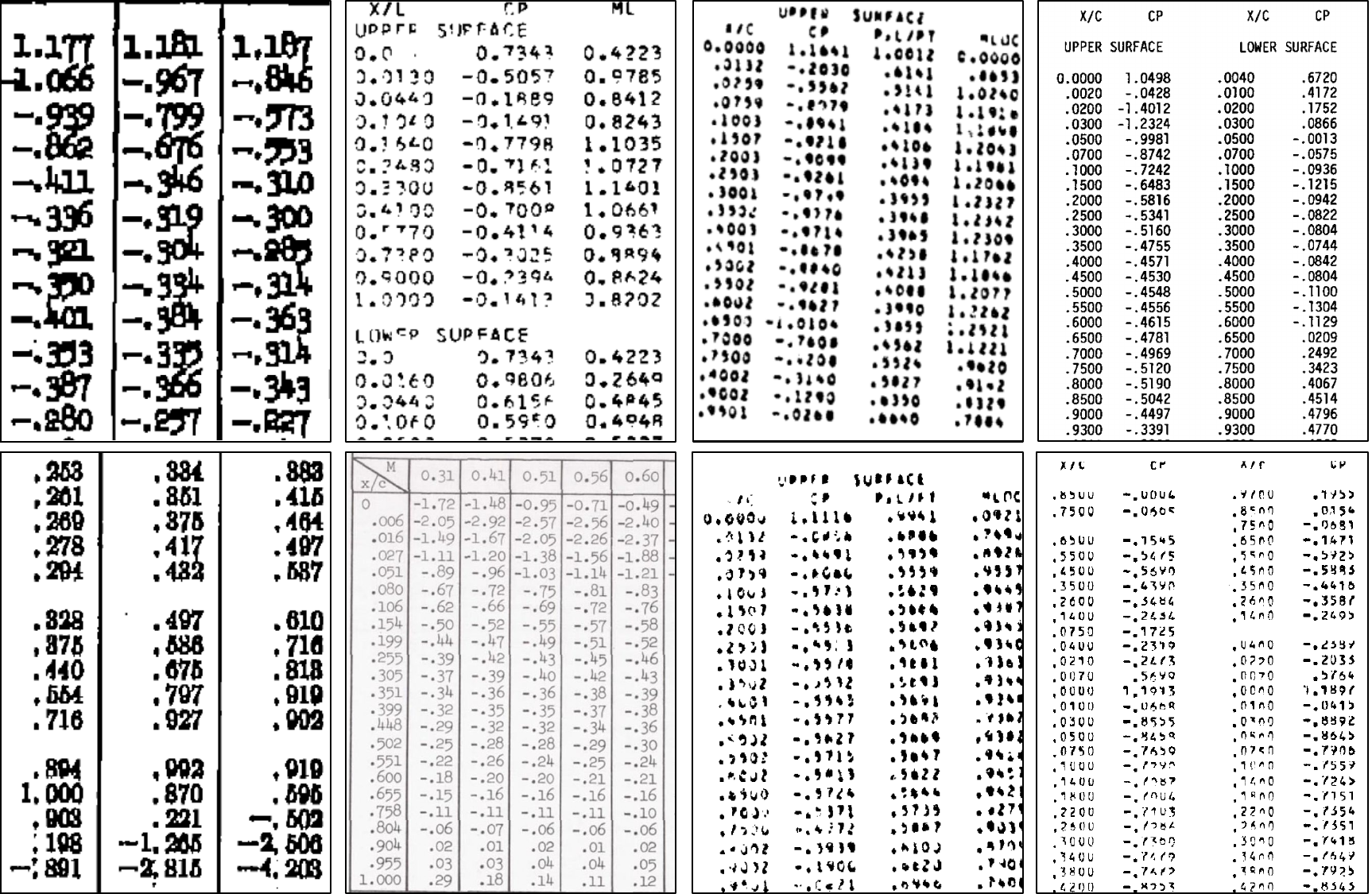}
\caption{Varying document legibility encountered during data mining. From left to right, top row: Refs.~[\citenum{johnson1950},~\citenum{stanewsky1979},~\citenum{ladson1987},~\citenum{harris1979}], bottom row: Refs.~[\citenum{pinkerton1938},~\citenum{stivers1954},~\citenum{johnson1985},~\citenum{cook1979}].}
\label{fig:table_limitation}
\end{figure}

In each source, the pressure distribution data was reported in either a tabulated or graphical format. These data were then digitized as a comma separated values (CSV) file. For tabulated results, the data were digitized using an online optical character recognition (OCR) tool called ExtractTables~\cite{extracttable}, followed by manual proofreading to correct any inaccuracies. The conversion accuracy was dependent on the legibility of the original document. Due to the early publication dates of many technical reports, the quality of PDF files were often poor. Examples of the varying legibility from different documents can be inferred from Fig.~\ref{fig:table_limitation}. In some cases, it was difficult to accurately identify experimental values. If an entry was considered unreliable, its value was determined indirectly by comparing it to the plotted results (if available), estimating based on the authors' best knowledge of airfoil $C_p$ trends, or omitting the data point.

\begin{figure}[h!]
\centering
\includegraphics[width=.37\textwidth]{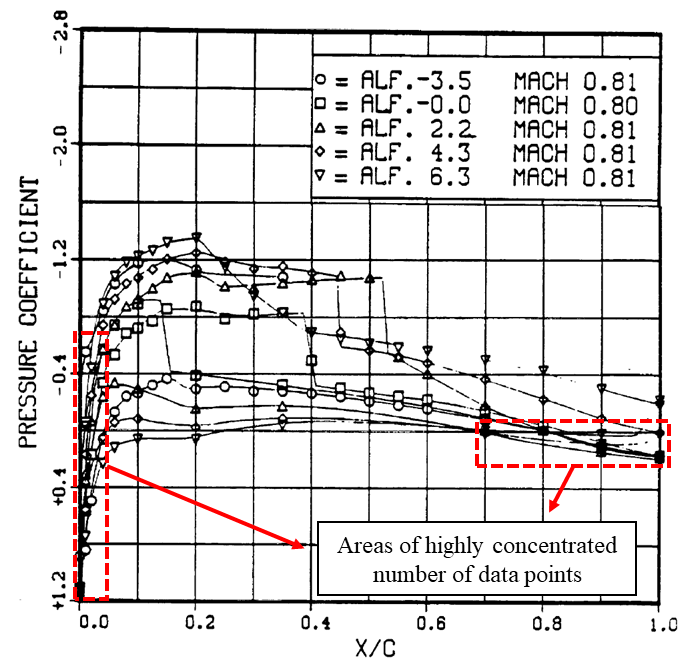}
\caption{Areas of highly concentrated data points typically found in airfoil pressure distribution plots. Original graph from Ref.~\cite{flemming1984}.}
\label{fig:graph_limitation}
\end{figure}

If presented in a graphical format with no accompanying tabulated results, the data were digitized using WebPlotDigitizer~\cite{webplotdigitizer}. This online tool enables manual extraction of individual data points. However, extracting certain points, particularly those at near the leading and trailing edges, proved challenging as the $C_p$ values converge. This issue was especially pronounced in graphs where multiple pressure distributions were plotted on the same axes. Figure~\ref{fig:graph_limitation} illustrates an example of such difficult cases. In these instances, the data points were carefully obtained by zooming in or by the authors' informed estimates based on airfoil physics. If neither method was viable, the point was omitted.  

\begin{figure}[hbt!]
\centering
\includegraphics[width=0.85\textwidth]{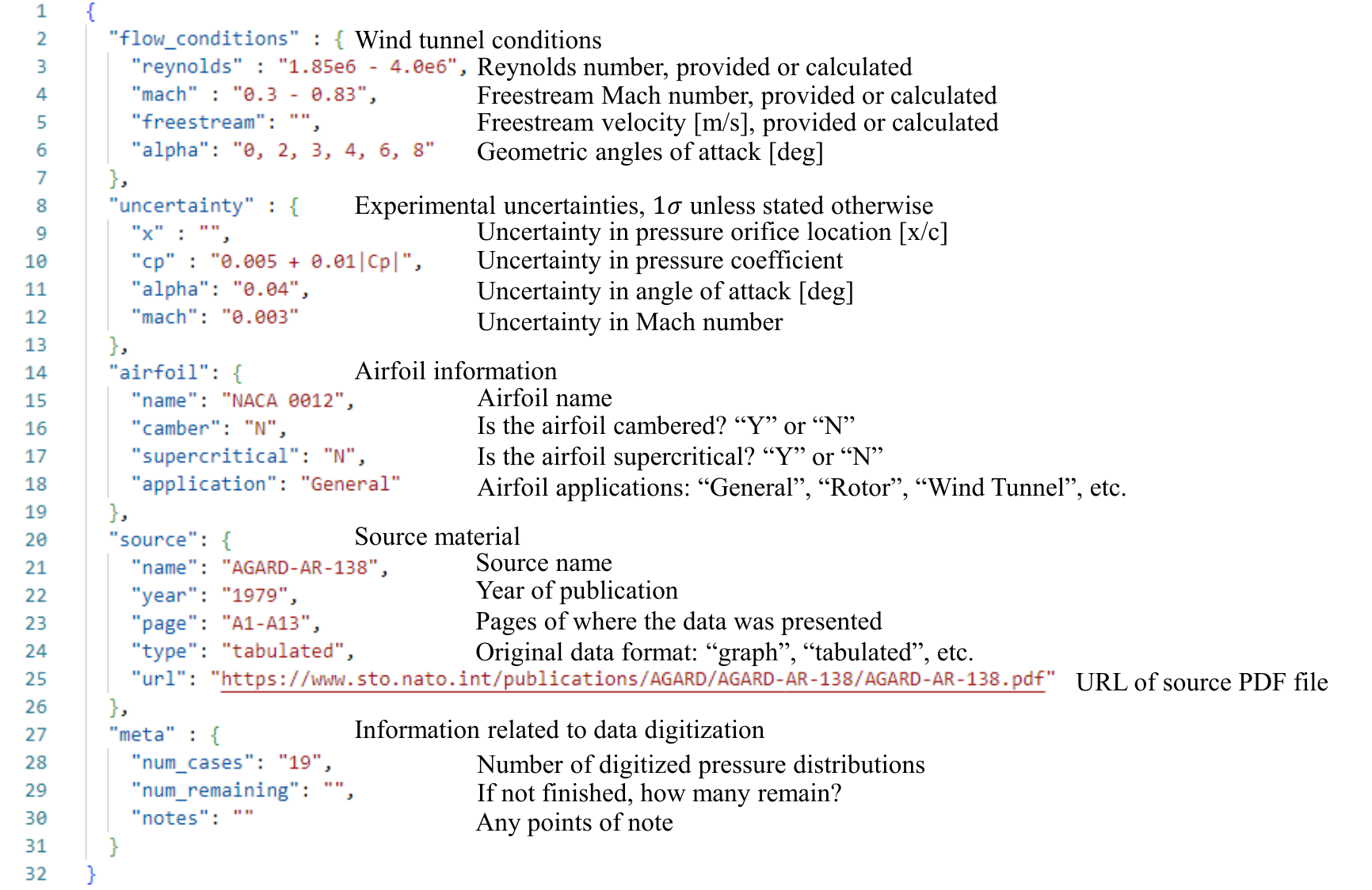}
\caption{An example ASPIRE tag file for a NACA 0012; experimental data from Ref.~\cite{thibert1979}.}
\label{fig:tag_file}
\end{figure}

In many sources, experimental accuracy was noted by the authors. These were often reported as a single scalar value in terms of the maximum magnitude of the error, or the maximum percentage error. In ASPIRE, these values were recorded in an accompanying tag file. Additionally, if the uncertainties in the independent variables such as those of the freestream Mach number, angles of attack, or pressure sensor locations were found in source documents, they were also recorded. An example of a tag file included in the database can be seen in Fig.~\ref{fig:tag_file}.

\subsection{Available Data}\label{sec:avail_data}
\begin{figure}[hbt!]
\centering
\includegraphics[width=.45\textwidth]{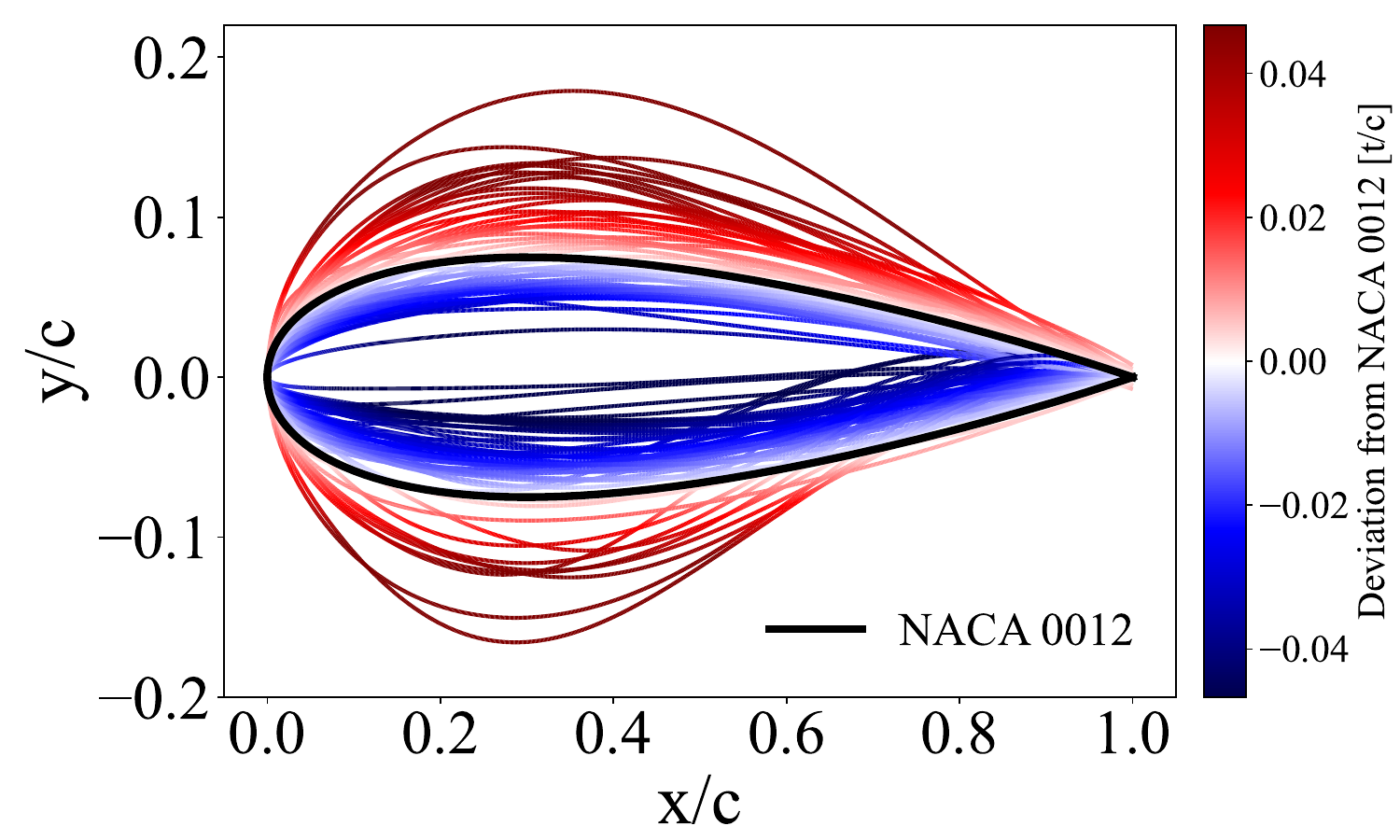}
\caption{Geometric profiles of airfoils in the database, colored by thickness compared to the baseline NACA 0012}
\label{fig:all_af}
\end{figure}

Currently, ASPIRE consists of 2917 unique pressure distributions from 69 airfoils taken across various airfoil families and applications. This makes the database a useful resource of experimental pressure measurements, all presented in an easily accessible format. All profiles of the airfoils included in the database, compared against the commonly studied NACA 0012, are illustrated in Fig.~\ref{fig:all_af}. Airfoil profiles thicker than a NACA 0012 at the given chordwise location are colored in shades of red. Profiles thinner than the NACA 0012 are colored in shades of blue.

\begin{figure}[!ht]
\centering
\begin{subfigure}{0.9\textwidth}
  \centering
  \includegraphics[width=0.8\linewidth]{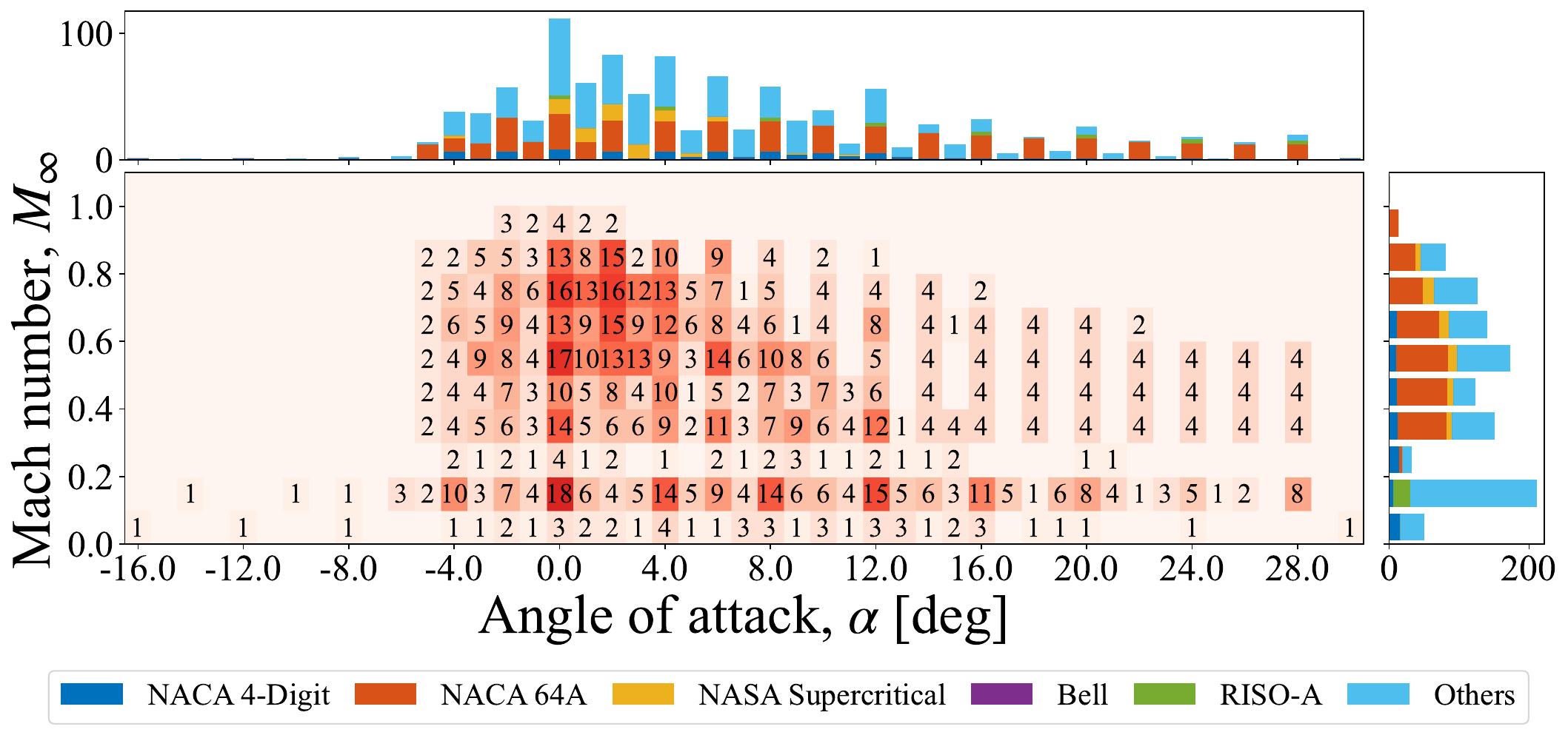}
  \caption{Distribution of available airfoil pressure data and the marginal histograms categorized by airfoil family}
  \label{fig:avail_data_family}
\end{subfigure} \\%
\vspace{3mm}
\begin{subfigure}{0.9\textwidth}
  \centering
  \includegraphics[width=0.8\linewidth]{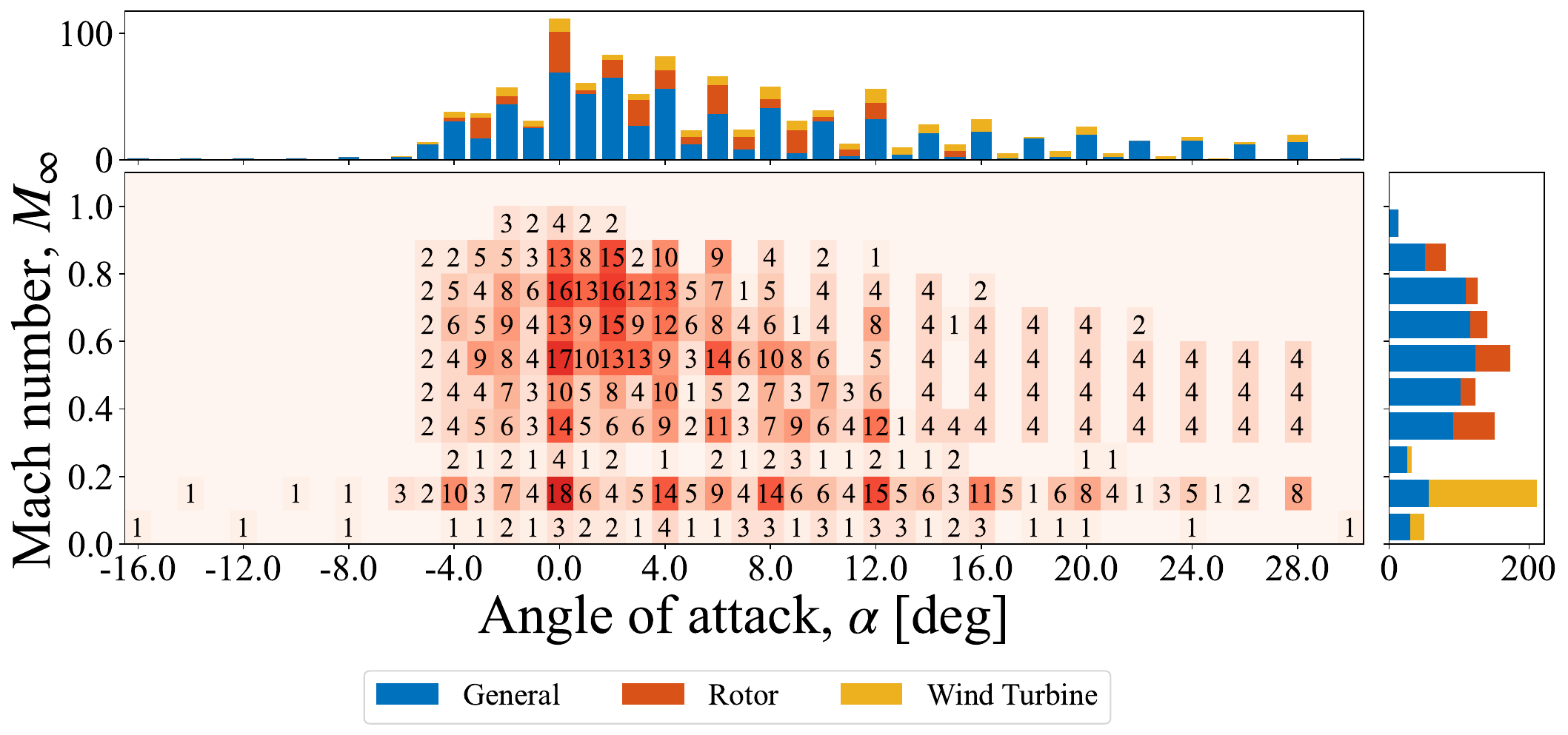}
  \caption{Distribution of available airfoil pressure data and the marginal histograms categorized by airfoil usage}
  \label{fig:avail_data_usage}
\end{subfigure} \\
\vspace{3mm}
\begin{subfigure}{0.9\textwidth}
  \centering
  \includegraphics[width=0.8\linewidth]{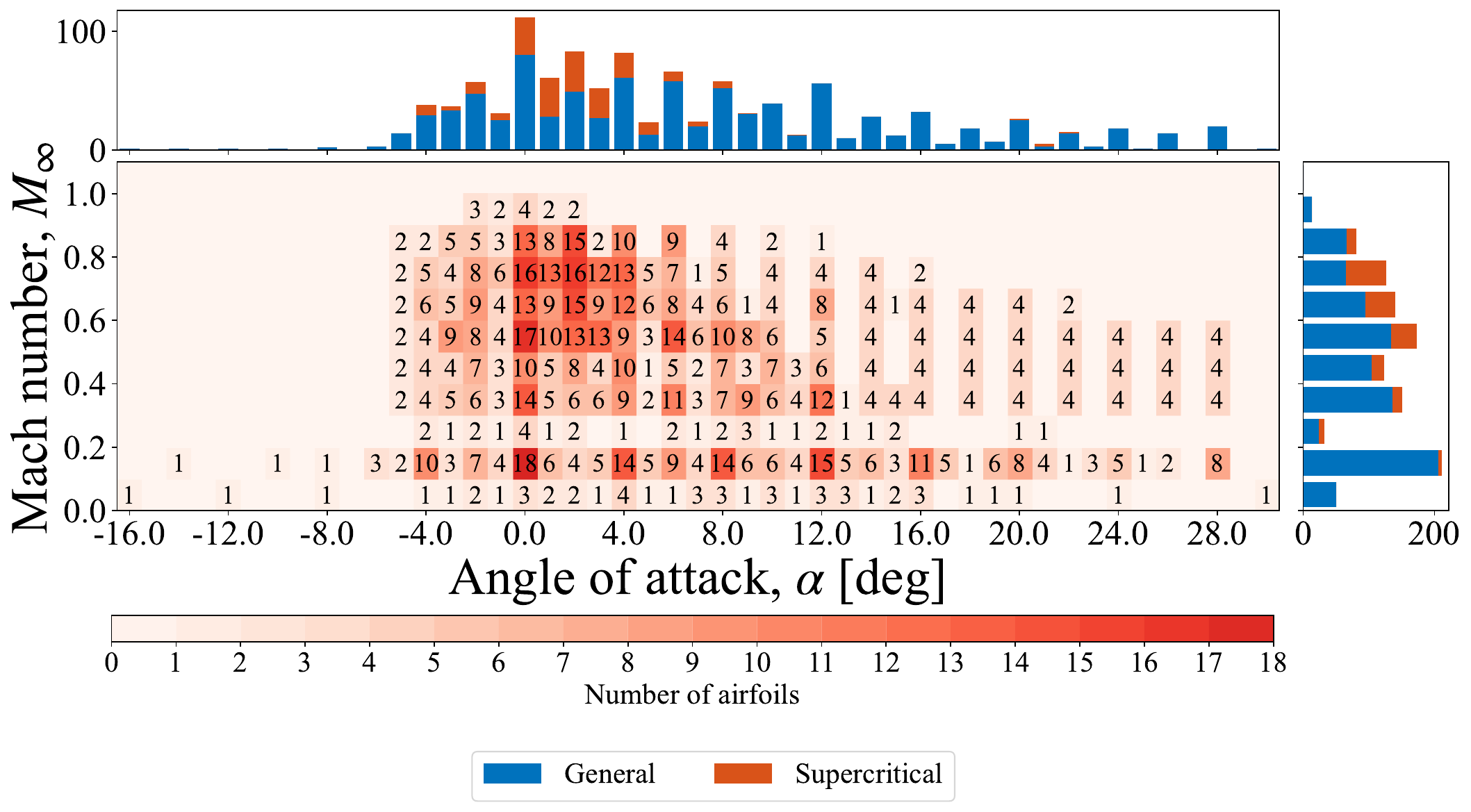}
  \caption{Distribution of available airfoil pressure data and the marginal histograms categorized by airfoil supercriticality}
  \label{fig:avail_data_supercrit}
\end{subfigure}
\caption{The distribution of available $C_p$ data in terms of the number of unique airfoils at each freestream Mach number and angles of attack. The marginal histograms are organized by different categories.}
\label{fig:avail_data}
\end{figure}

The available data ranges from $-30^{\circ}$ to $30^{\circ}$ in angles of attack ($\alpha$) and $0.0$ to $1.0$ for freestream Mach number ($M_\infty$) that span subsonic (including incompressible), transonic, and sonic regimes. The distribution of available $C_p$ data at a given operating condition is shown in Fig.~\ref{fig:avail_data}. The color and the number in the center plot denotes the number of unique airfoils available for a given $M_\infty$-$\alpha$ combination. Marginal histograms are plotted to provide a clear visualization of the available data at specific $\alpha$ or $M_\infty$. The data distribution is presented in terms of airfoil families, design usage, and supercriticality.

It is evident from the data distribution plots that there is a wide variety of airfoil families (Fig.~\ref{fig:avail_data_family}). It can also be observed that the vast majority of the available data (approximately 97\% of the database) are concentrated around $\alpha = -4.0^\circ - 12.0^\circ$ and $M_\infty = 0.3 - 0.9$. The distribution of $M_\infty$ follows a bimodal pattern, with one peak occurring in the compressible regime ($M_\infty = 0.60 - 0.70$) and another in the incompressible regime ($M_\infty = 0.10 - 0.20$), as shown in Fig.~\ref{fig:avail_data_usage}. This distribution is shaped by the types of airfoils that have been digitized. General aviation airfoils (e.g., NACA 4-series) and rotor airfoils (e.g., Sikorsky’s SC series) account for 82\% of the database. The design flight speeds of conventional aircraft and rotorcraft influence the experimental Mach numbers, leading to a higher concentration of data at elevated freestream Mach numbers. Similarly, wind turbine airfoils, which are typically operated at lower Mach numbers to ensure structural integrity and safety~\cite{cao}, contribute to the second peak in Fig.~\ref{fig:avail_data_usage}. Furthermore, the sizable number (24\% of the database) of supercritical airfoils, airfoils designed to operate in high $M_\infty$, further reinforces the bimodal nature of the distribution.

In comparison, the distribution of the angles of attack is unimodal. The number of available data drastically decreases at extreme angles of attack, both positive and negative. The relative dearth of data (3\% of the database) arises from the fact that these angles are well into the stall regime, which are typically not explored thoroughly in experimental literature unless the goal is to directly study the behavior of an airfoil under detached flow conditions. 

\begin{figure}[ht!]
\centering
\includegraphics[width=.9\textwidth]{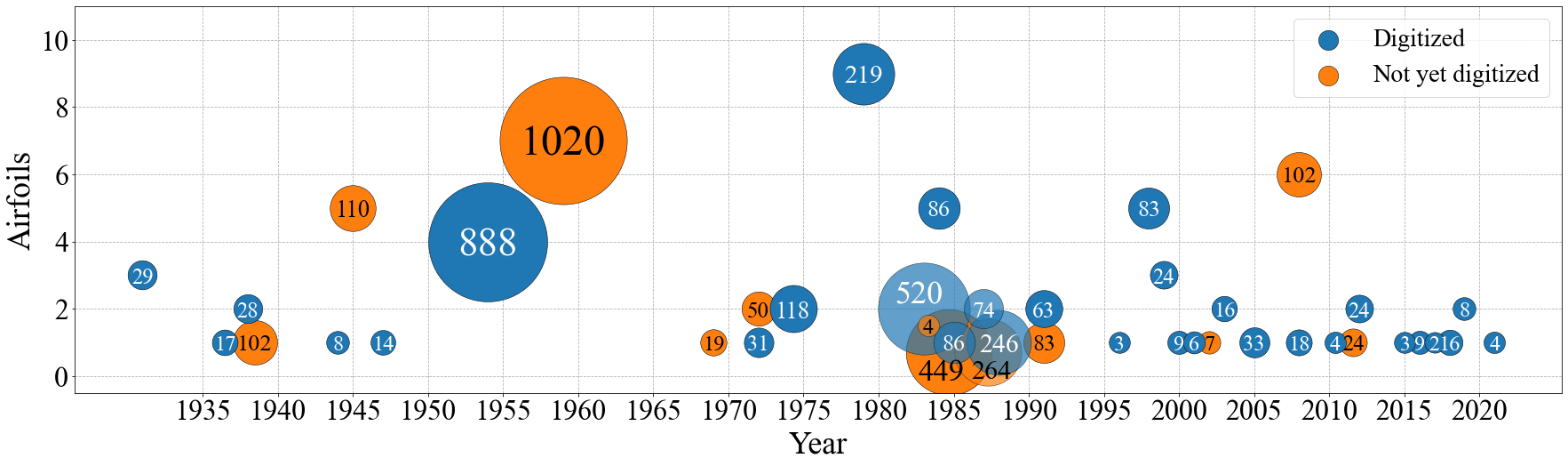}
\caption{Bubble plot of data that have and have not been digitized. The number on each circle represents the number of unique pressure distributions available within the source material.}
\label{fig:bubbleplot}
\end{figure}

In the future, it will be necessary to cover the underrepresented regions of the parameter space. This will ensure a more comprehensive training and validation for future ML models. The improved coverage can be achieved through multiple potential approaches. The first approach is targeted data mining, where the future digitization effort are focused on finding and digitizing experimental data from studies aimed at studying post-stall conditions such as Refs.~\cite{branch2023, Cebeci1995}. If no publicly available data can be found, targeted experiments can also be performed to directly obtain the necessary measurements. It is also possible to bolster the database via data augmentation, which would leverage existing data for symmetric airfoils. Specifically, the pressure coefficient ($C_p$) distribution at negative angles of attack can be assumed equivalent to that at positive angles of attack, with the upper and lower surfaces switched. This approach could effectively double the contributions from symmetric airfoil data. Lastly, releasing the database as a publicly available repository, the database can be expanded in terms of airfoil families and years of publication through contributions from the aerospace community. The distribution of digitized and not yet digitized data among the sources identified during the data mining operation is presented in Fig.~\ref{fig:bubbleplot}.

\subsection{Data Pre-processing}\label{sec:pre-process_conformal}
In many experimental reports, $C_p$ is plotted with respect to the normalized coordinates in the chordwise direction, $x/c$. The coordinate system spans from 0 to 1, where 0 corresponds to the leading edge and 1 to the trailing edge. Such a coordinate system does not permit a ML model to distinguish between the airfoil's upper and lower surfaces. Since a model trained on the dataset should adhere to the true \emph{physical} behavior of airfoils, it is necessary that the pressures at the leading and trailing edges match in lieu of severe trailing edge separation. As the training data does not currently include any cases with trailing edge separation, the periodicity at the boundaries must be maintained. 

\begin{figure}[ht!]
\centering
\includegraphics[width=.8\textwidth]{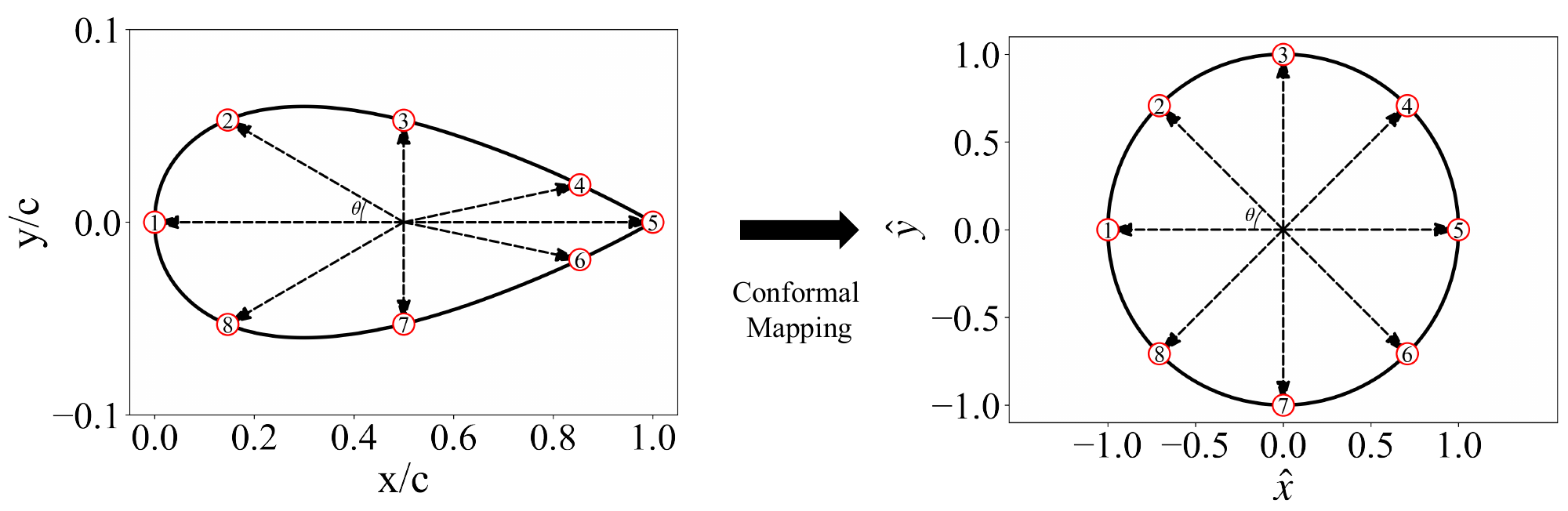}
\caption{Transformation of airfoil coordinates ($x/c$, $y/c$) to polar coordinates ($\hat{x}$, $\hat{y}$) via conformal mapping}
\label{fig:conf_mapping}
\end{figure}

To address this, a conformal mapping-based approach was adopted; the normalized airfoil chordwise location is transformed to $\hat{x}$ and $\hat{y}$ in a polar coordinate system. In this procedure, $x/c = \{x \in \mathbb{R} | 0 \le x \le 1 \}$ is transformed to $\hat{x} = \{x \in \mathbb{R} | -1 \le x \le 1 \}$, corresponding to the $x$-value of the equivalent unit circle. The equivalent angle is obtained via $\theta = \cos^{-1}(x/c)$, implying $\hat{y} = \sin{\theta}$, where $\hat{y} = \{y \in \mathbb{R} | -1 \le y \le 1 \}$ is the $y$-value of the unit circle. The transformation is visualized in Fig.~\ref{fig:conf_mapping}, where it is evident that the sign of $\hat{y}$ corresponds to the upper and lower surfaces. At the leading and trailing edges, $\hat{y} = 0$, which enforces periodicity.

The input data matrix for ADAPT was created by down-selecting the data in ASPIRE to constrain $-4.0^\circ \le \alpha \le 12.0^\circ$ and $0.0 \le M_\infty \le 0.75$. These ranges for $\alpha$ and $M_\infty$ were chosen due to the abundance of data available. Additionally, only data points with a chord-based Reynolds number ($Re_c$) on the order of $10^6$ were included. Fixing the $Re_c$ helped to eliminate Reynolds number effects and reduce the data dimensionality. A Reynolds number on the order of $10^6$ was chosen because, in this regime, the viscous effects of the freestream are considered minimal compared to the inertial effects~\cite{leishman}. In order to eliminate inconsistencies with the boundary layer transition behavior, data sets where the boundary layer was artificially tripped were excluded. The data were then pre-processed to extract the input variables necessary for the model to predict the pressure distributions. These input variables were: 28 $y$-values of the airfoil geometry at set chordwise locations for upper and lower surfaces respectively, $M_\infty$, $\alpha$, $\hat{x}$, and $\hat{y}$. This corresponded to a total of 60 input variables. This approach is ideal for datasets compiled from various experiments, as it enables explicit definition of pressure sensor locations, which often differ between sources.

\section{Results and Discussion}\label{4_results}
\subsection{Pressure Distributions}
\begin{figure}[p]
\centering
\includegraphics[width=.9\textwidth]{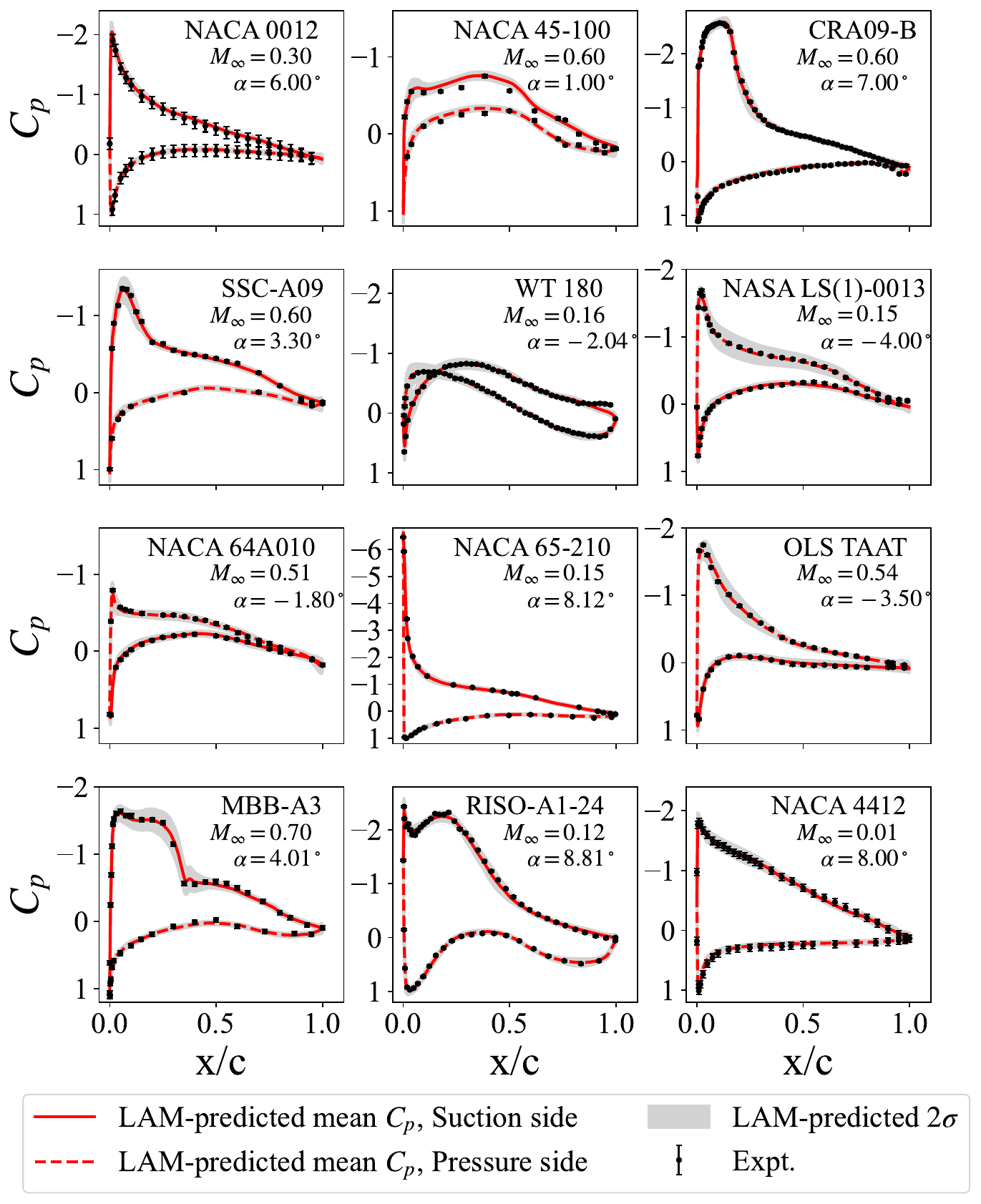}
\caption{Predicted $C_p$ for the training set. The error bars represent two standard deviations.}
\label{fig:training}
\end{figure}

To initially assess the capability of the LAM, a random subset of the training data was selected and input into the model. The predicted mean $C_p$ and two standard deviations are displayed in Fig.~\ref{fig:training}. The plots showcase the LAM’s capability to fit the training data, which is important as the ADAPT evaluations at the training data point need not necessarily interpolate the training data. Moreover, the 95\% confidence intervals of predicted pressure distributions differ significantly for each airfoil. Regions of higher uncertainty exist for cases where the noise reported by the source material is inherently high (e.g. the NACA 0012 airfoil), or for cases where the amount of training data is sparse (e.g. negative angles of attack for NASA LS(1)-0013 and OLS/TAAT cases). A significant increase in the uncertainty also occurs on regions of pronounced shock effects, as seen on the suction side of the MBB-A3 at $M_\infty=0.70$.   

\begin{figure}[h!]
\centering
\includegraphics[width=.6\textwidth]{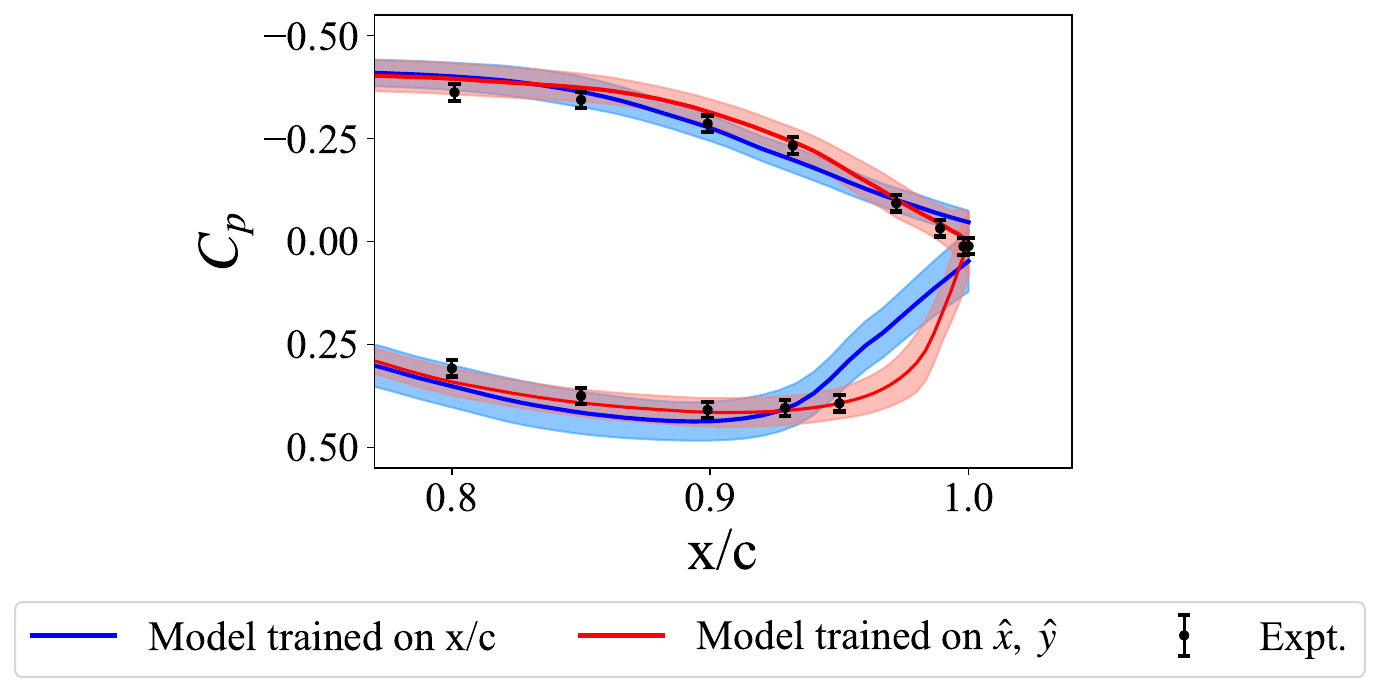}
\caption{Zoom view of a trailing edge $C_p$ and two standard deviations, predicted from models trained on normalized chordwise location (blue) and conformal coordinates (red).}
\label{fig:te_zoom}
\end{figure}

Figure~\ref{fig:te_zoom} confirms that, with the conformal transformation during data pre-processing, the $C_p$ at the upper and lower surfaces of the trailing edge match (red line). In contrast, the blue lines clearly show that a model trained on the one-dimensional input $x/c$ yields a mismatch in the trailing edge pressure, which is a non-physical behavior when there is no trailing edge separation.

\begin{figure}[p]
\centering
\includegraphics[width=.9\textwidth]{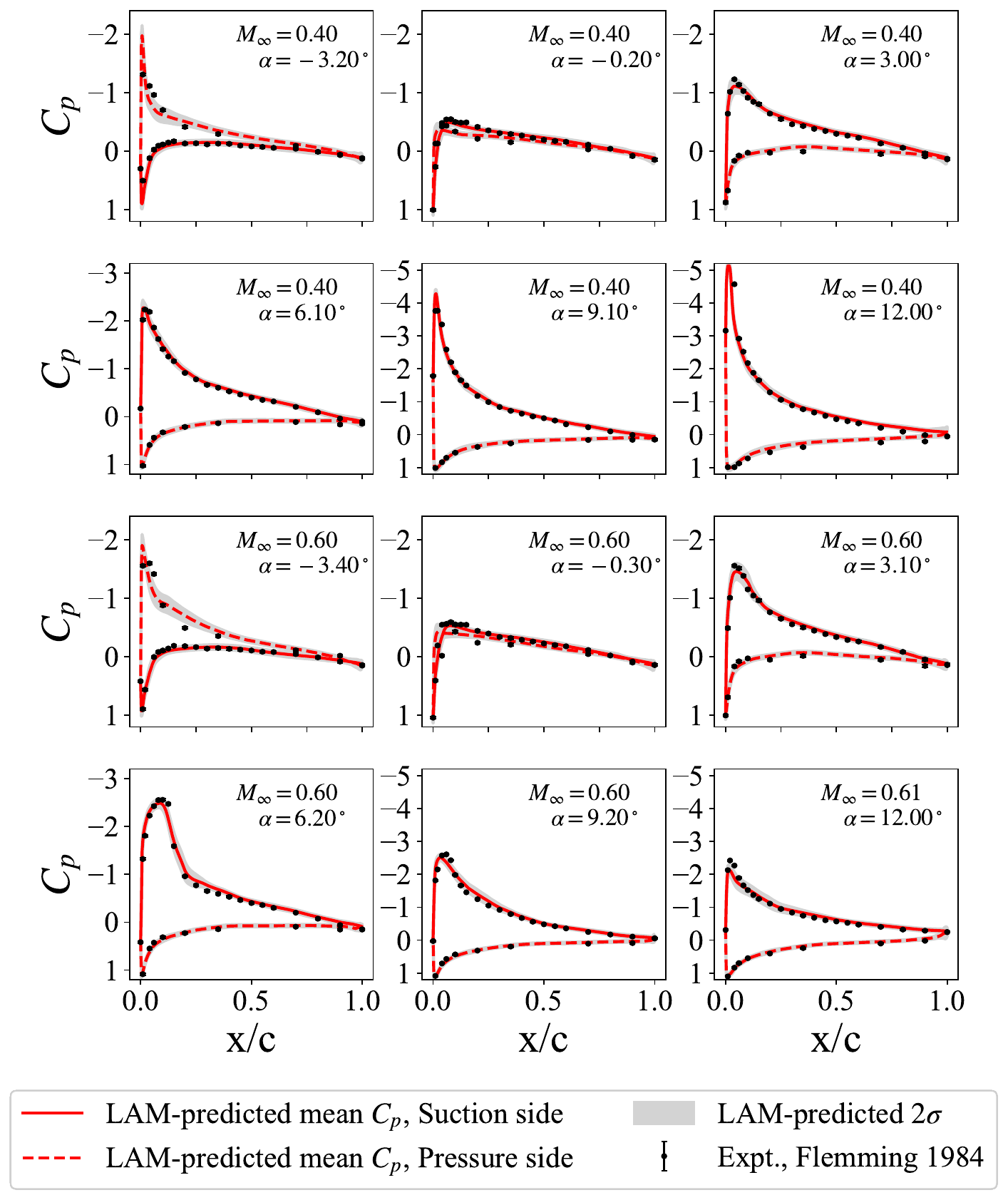}
\caption{SC1095: Predicted $C_p$ under various $M_\infty$ and $\alpha$. The error bars represent two standard deviations.}
\label{fig:sc1095}
\end{figure}

The remaining figures represent the predictions on the test set, airfoils previously untrained by the model. Figure~\ref{fig:sc1095} shows the predicted $C_p$ for the SC1095 airfoil, designed for rotorcraft applications. The comparison between predictions and the test data yielded a $\text{MAE}_{\text{enclosed}}$ of $0.027$ for the data subset. As the metric is directly correlated to the airfoil $c_l$, the error is equivalent to less than $2\%$ error relative to the airfoil $c_{l, \text{max}}$. It is notable that the model captures compressibility effects for $\alpha \geq 6.1^{\circ}$ when transitioning from $M_\infty = 0.40$ to 0.60. The model also predicts the onset of stall, as the change in the pressure distribution is minimal when increasing the angle of attack beyond $\alpha = 9.0^{\circ}$. While some inaccuracies in the mean $C_p$ can be observed for negative angles of attack, the experimental measurements fall within the 95\% confidence interval of the predictions.  

\begin{figure}[ht!]
\centering
\includegraphics[width=.78\textwidth]{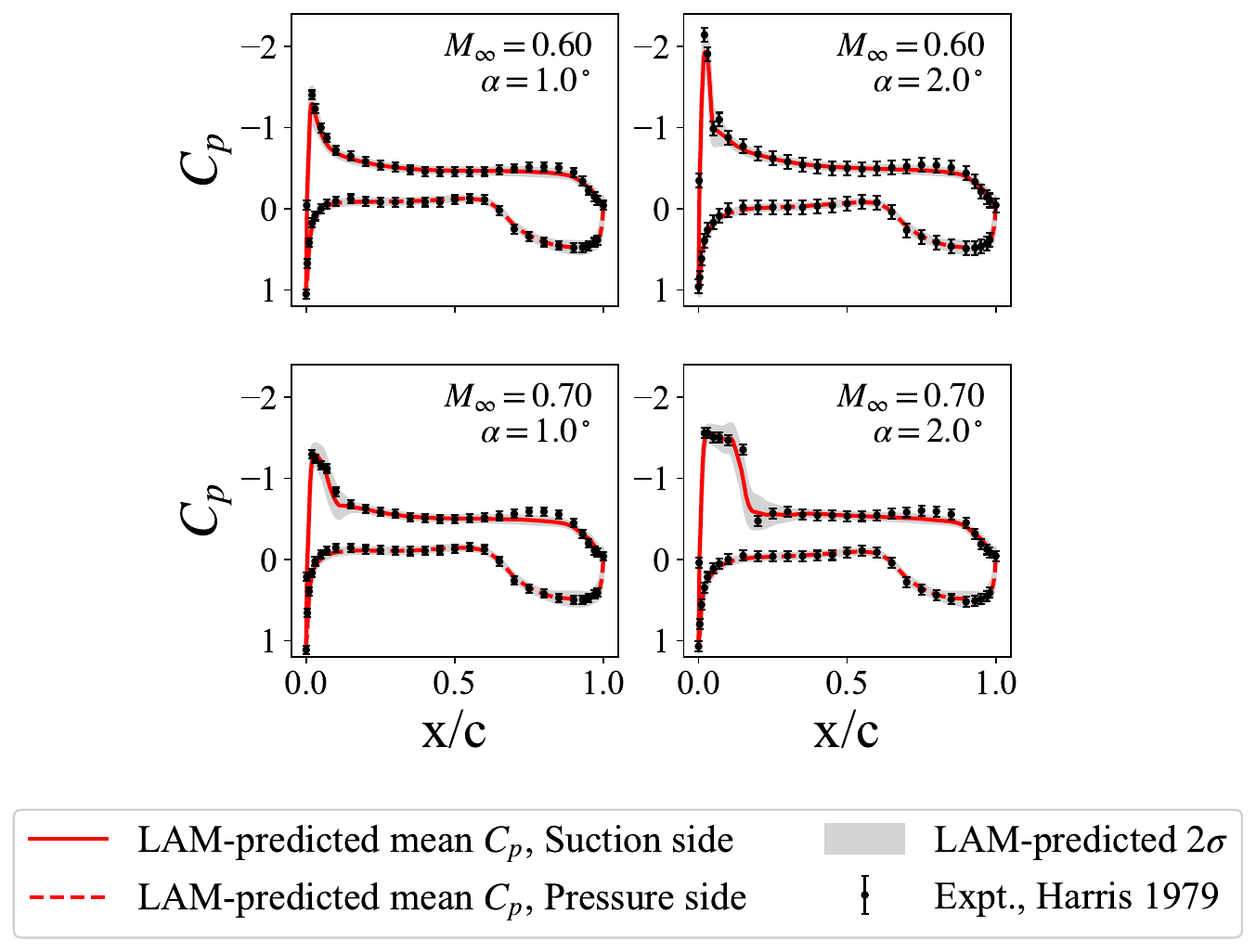}
\caption{NASA Supercritical Airfoil 9a: Predicted $C_p$ under various $M_\infty$ and $\alpha$. The error bars represent two standard deviations.}
\label{fig:sc9a}
\end{figure}

The predicted $C_p$ and the corresponding uncertainty for the Supercritical Airfoil 9a is presented in Fig.~\ref{fig:sc9a}. The $\text{MAE}_{\text{enclosed}}$ for the airfoil was found to be $0.018$. Surface pressures of this supercritical airfoil are significantly ``flatter''  on the suction side compared to those of conventional airfoils. The low error between the predictions and the experimental data (approximately $1.26\%$ of the airfoil $c_{l, \text{max}}$) suggests that the model accurately captures the aerodynamic characteristics associated with supercritical geometries. Furthermore, the model is able to accurately predict the onset of shock on the airfoil, as indicated by the rapid change in the pressure observed in the $M_\infty = 0.70$ cases (bottom row). The locations of shock formation are often very sensitive to various factors such as the surface roughness of the test article or chord-based Reynolds number. This inherent uncertainty is captured by the model through an increase in the confidence interval.

\begin{figure}[ht!]
\centering
\includegraphics[width=.78\textwidth]{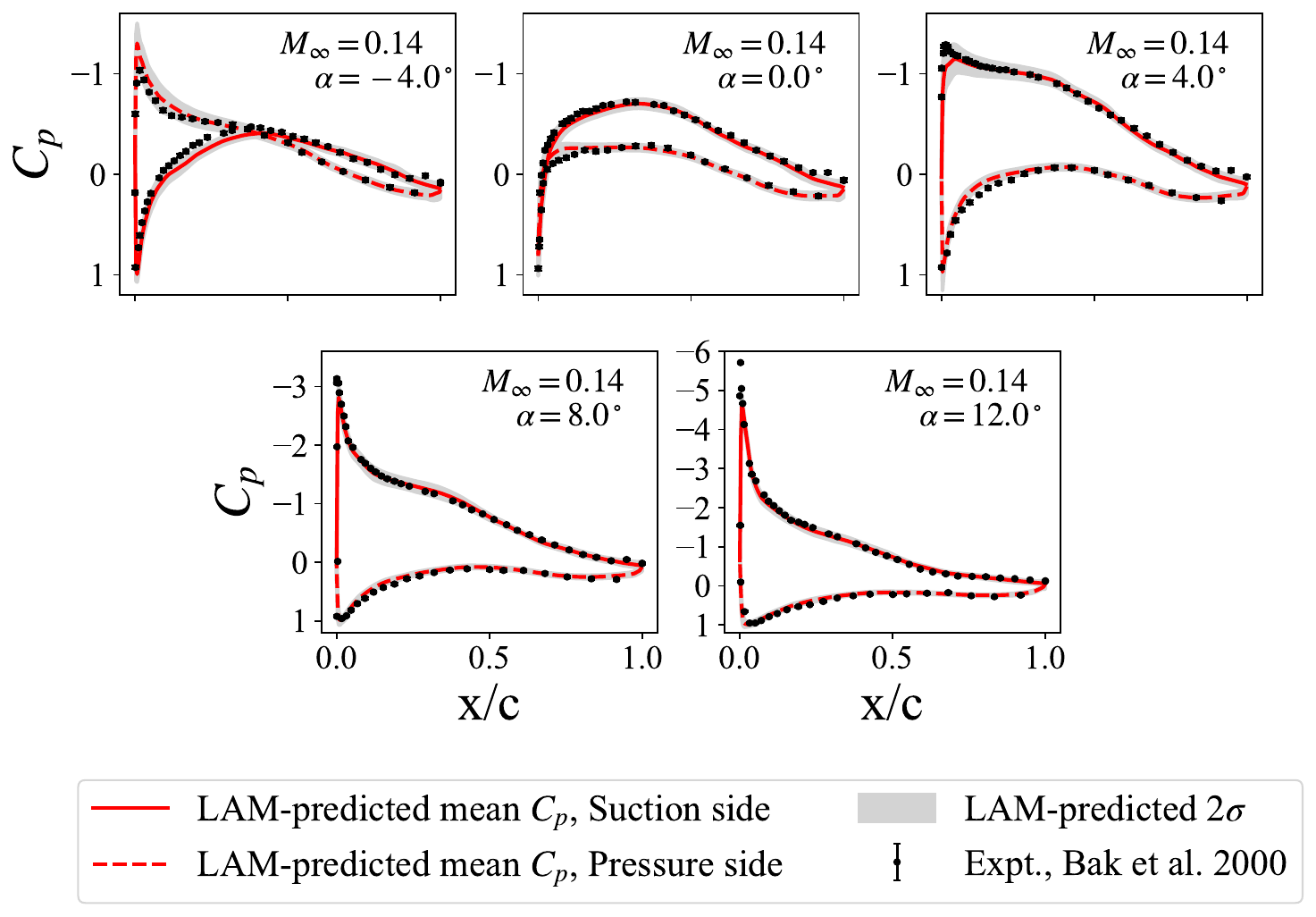}
\caption{NACA 63-415: Predicted $C_p$ under various $M_\infty$ and $\alpha$. The error bars represent two standard deviations.}
\label{fig:naca63-415}
\end{figure}

Lastly, Fig.~\ref{fig:naca63-415} demonstrates the model capability for the NACA 63-415, typically used for wind turbines. The $\text{MAE}_{\text{enclosed}}$ between the predictions and the experimental data from the test set was found to be $0.041$ (approximately $3\%$ of the airfoil's $c_{l, \text{max}}$), higher than that of the other two airfoils. The increased errors arise from regions of noticeable underprediction at the suction peak. This is most notable at $\alpha = 4.0^\circ$ and $12.0^\circ$. The reduced performance can be attributed to the relatively smaller size of wind turbine airfoil data, a total of 4055 $C_p$ data points. The number of data points for rotor airfoils, the next smallest subset of the training set, amounted to 11892. This number is approximately triple that of wind turbine airfoils. Expanding the database to match the quantity will likely make the model's predictions more robust.

In summary, the results clearly indicate that the LAM generates accurate predictions on both training and test sets. The model has the ability to generalize well to different airfoil geometries, while capturing their unique behaviors. The strong generalization capability arises from the fact that different airfoil families occupy separate regions within the model's active space. Appendix D provides a basic analysis of ADAPT's latent space for in-depth understanding of the model.

\subsection{Aerodynamic Force Coefficients}
The normal, axial, and moment coefficients about the quarter-chord for a two-dimensional body are: 
\begin{align}
    c_n &= \int_{0}^1{\left( C_{p, l}\left( \frac{x}{c} \right) - C_{p, u}\left( \frac{x}{c} \right) \right) \, d\frac{x}{c}}  + \int_{0}^1{\left( c_{f, u}\frac{d(y_u/c)}{d(x/c)}
    + c_{f, l}\frac{d(y_l/c)}{d(x/c)}  \right) \, d\frac{x}{c}}, \label{eq:cn_af} \\
    c_a &= \int_{0}^1{\left(C_{p, u}\left( \frac{x}{c} \right)\frac{d(y_u/c)}{d(x/c)}  
    - C_{p, l}\left( \frac{x}{c} \right)\frac{d(y_l/c)}{d(x/c)}\right) \, d\frac{x}{c}} + \int_{0}^1{\left(c_{f, u} + c_{f, l} \right) \, d\frac{x}{c}}, \label{eq:ca_af} \\
    c_{m} &= \int_{0}^1{\left[ \left( C_{p, l}\left( \frac{x}{c} \right) - C_{p, u}\left( \frac{x}{c} \right) \right) 
    \left(\frac{x}{c} - \frac{1}{4} \right) \right] \, d\frac{x}{c}} \nonumber \\
    &\quad - \int_{0}^1{\left( c_{f, u}\frac{d(y_u/c)}{d(x/c)}
    - c_{f, l}\frac{d(y_l/c)}{d(x/c)} \right) 
    \left(\frac{x}{c} - \frac{1}{4} \right)\, d\frac{x}{c}}  \nonumber \\
    &\quad + \int_{0}^1{ \left( C_{p, u} \left( \frac{x}{c} \right)
    \frac{d(y_u/c)}{d(x/c)} + c_{f,u} \right)\frac{y_u}{c} \, d\frac{x}{c}} \nonumber \\
    &\quad + \int_{0}^1{\left(-C_{p, l} \left( \frac{x}{c} \right)
    \frac{d(y_l/c)}{d(x/c)} + c_{f,l} \right) 
    \frac{y_l}{c} \, d\frac{x}{c}} \label{eq:cm_af}
\end{align} 

\noindent where $c_f$ is the skin friction coefficient, $\alpha$ is the angle of attack, and the subscripts $u$ and $l$ denote quantities of interest on upper and lower surfaces respectively. Here, $C_p$ is a function of the normalized chordwise location, $\frac{x}{c}$.

In ADAPT, $C_p$ is a Gaussian process; for a given set of inputs, the deep neural network projects $\mathbf{u}$ to the latent space $\mathbf{z}$, where $C_p$ has a mean and covariance defined over the conformal domain, as defined in Sec.~\ref{sec:pre-process_conformal}. If one introduces the assumption that contributions of skin friction are negligible---valid in most non-zero angles of attack and at higher $M_\infty$ \cite{Anderson2011Fundamentals}---then, $c_l$, $c_d$ and $c_m$ can be approximate as
\begin{align}  
c_n & \approx \frac{1}{2} \int_{-1}^1{\left( C_{p, l}\left(\hat{x} \right) - C_{p, u}\left( \hat{x} \right) \right) d\hat{x}},   \label{eq:cn} \\
c_a & \approx \frac{1}{2} \int_{-1}^1{\left(C_{p, u}\left( \hat{x} \right)\frac{d(y_u/c)}{d(x/c)} - C_{p, l}\left( \hat{x} \right)\frac{d(y_l/c)}{d(x/c)}\right) d\hat{x}}, \label{eq:ca} \\
c_{m} & \approx \frac{1}{2}\Biggl\{ \int_{-1}^1{\left( C_{p, l}\left( \hat{x} \right) - C_{p, u}\left( \hat{x} \right)  \right)  \left(\hat{x} + \frac{1}{2} \right) d\hat{x}} \nonumber \\ 
& \quad + \int_{-1}^1{ \left( C_{p, u} \left( \hat{x} \right) 
 \frac{d(y_u/c)}{d(x/c)} \frac{y_u}{c} \right) d\hat{x}} \nonumber \\
& \quad - \int_{-1}^1{\left(C_{p, l} \left( \hat{x} \right)  \frac{d(y_l/c)}{d(x/c)} \frac{y_l}{c} \right) d\hat{x} \Bigg\}}, \\   
c_l &= c_n \cos{\alpha} - c_a \sin{\alpha}, \label{eq:cl} \\
c_d &= c_n \sin{\alpha} + c_a \cos{\alpha}, \label{eq:cd} 
\end{align}

\noindent where $C_{p,u}\left( \hat{x} \right)$ and $C_{p,l}\left( \hat{x} \right)$ are model-predicted pressure coefficients with respect to the $x$-location in \emph{conformal coordinates}. The bounds of the integral differ from Eqs.~\ref{eq:cn_af}--\ref{eq:cm_af} due to the change of coordinate systems.

The above quantities are linear operators acting on a Gaussian process; therefore, they must also be Gaussian \cite{stats_book}. However, because the inputs are passed through a deep neural network, a series of non-linear operations, it is not possible to express the posterior distribution of the aerodynamic quantities analytically. They must be obtained by performing the integral numerically instead, using the Monte Carlo method.

\begin{figure}[h!]
\centering
\includegraphics[width=0.78\textwidth]{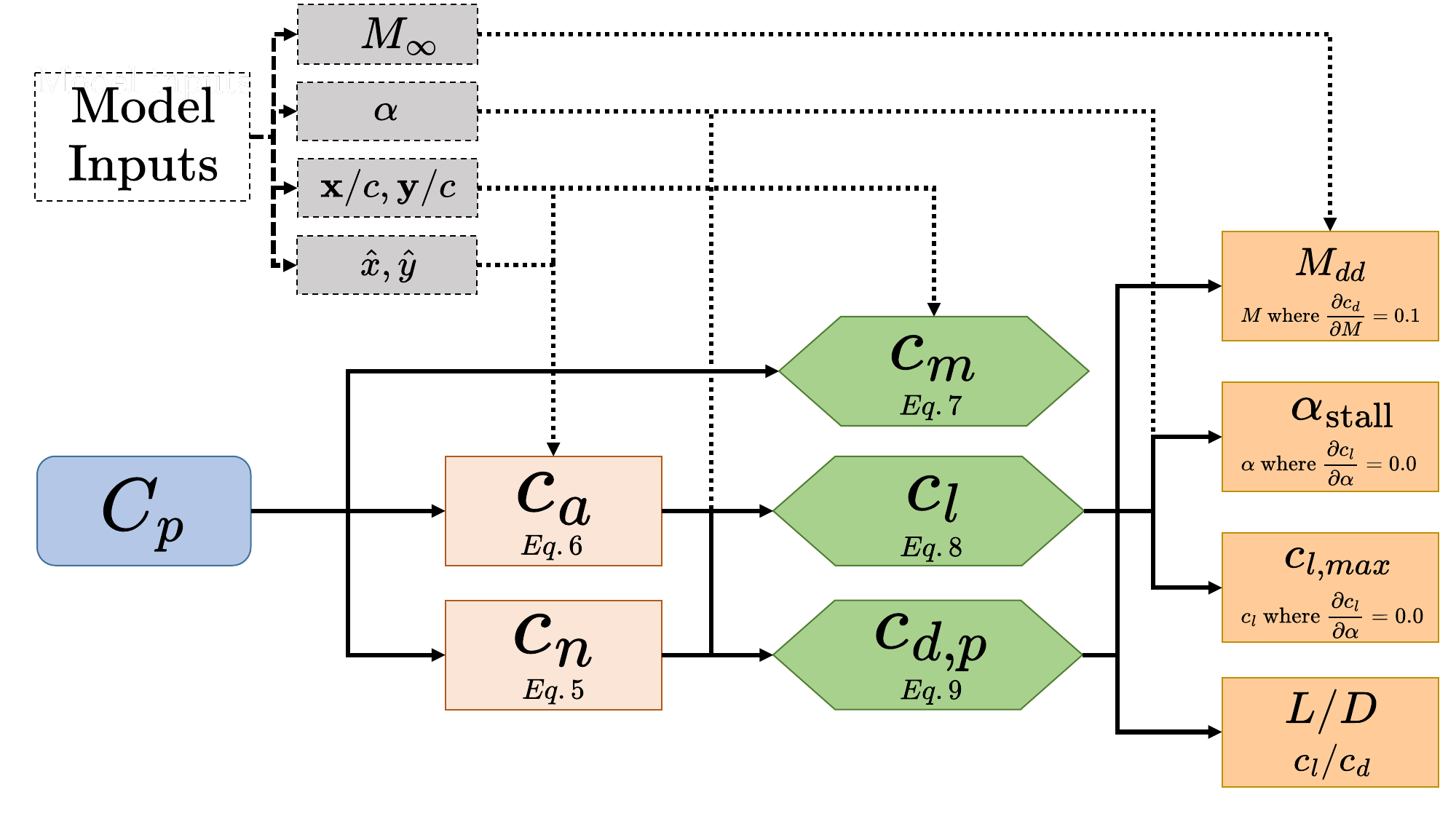}
\caption{Workflow of obtaining key aerodynamic quantities of interest based on model output $C_p$. While this investigation is limited to the green blocks in this paper, more intermediate design parameters in orange can be obtained similarly.}
\label{fig:aero_workflow}
\end{figure}

To begin, 10,000 samples are generated from the multivariate normal distribution associated with $C_p$, which was deemed sufficient for convergence. Each sample is pushed through Eqs.~\ref{eq:cn}--\ref{eq:cd} to obtain $c_l$, $c_d$, and $c_m$. From the integrated samples, the mean and the standard deviation of the aerodynamic coefficients are calculated. The overall workflow is captured in Fig.~\ref{fig:aero_workflow}. 

\begin{figure}[h!]%
    \centering
    \subfloat[\centering SC1095: Experimental $c_l$ of Ref.~\cite{flemming1984} derived from $C_p$ integration]{{\includegraphics[width=.33\textwidth]{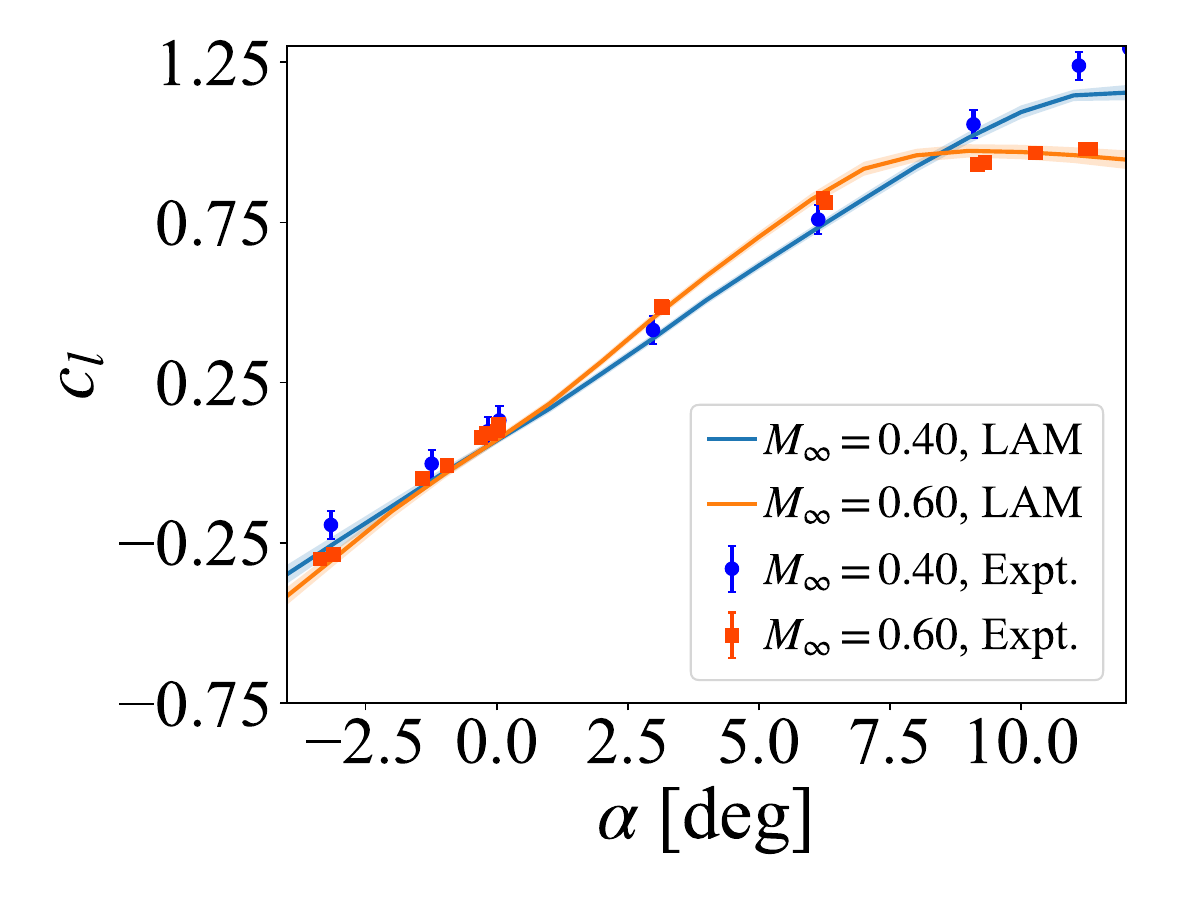} }\label{cl_vs_alpha_sc1095}}
    \subfloat[\centering NASA Supercritical Airfoil 9a: Experimental $c_l$ of Ref.~\cite{harris1979} from balance measurements]{{\includegraphics[width=.33\textwidth]{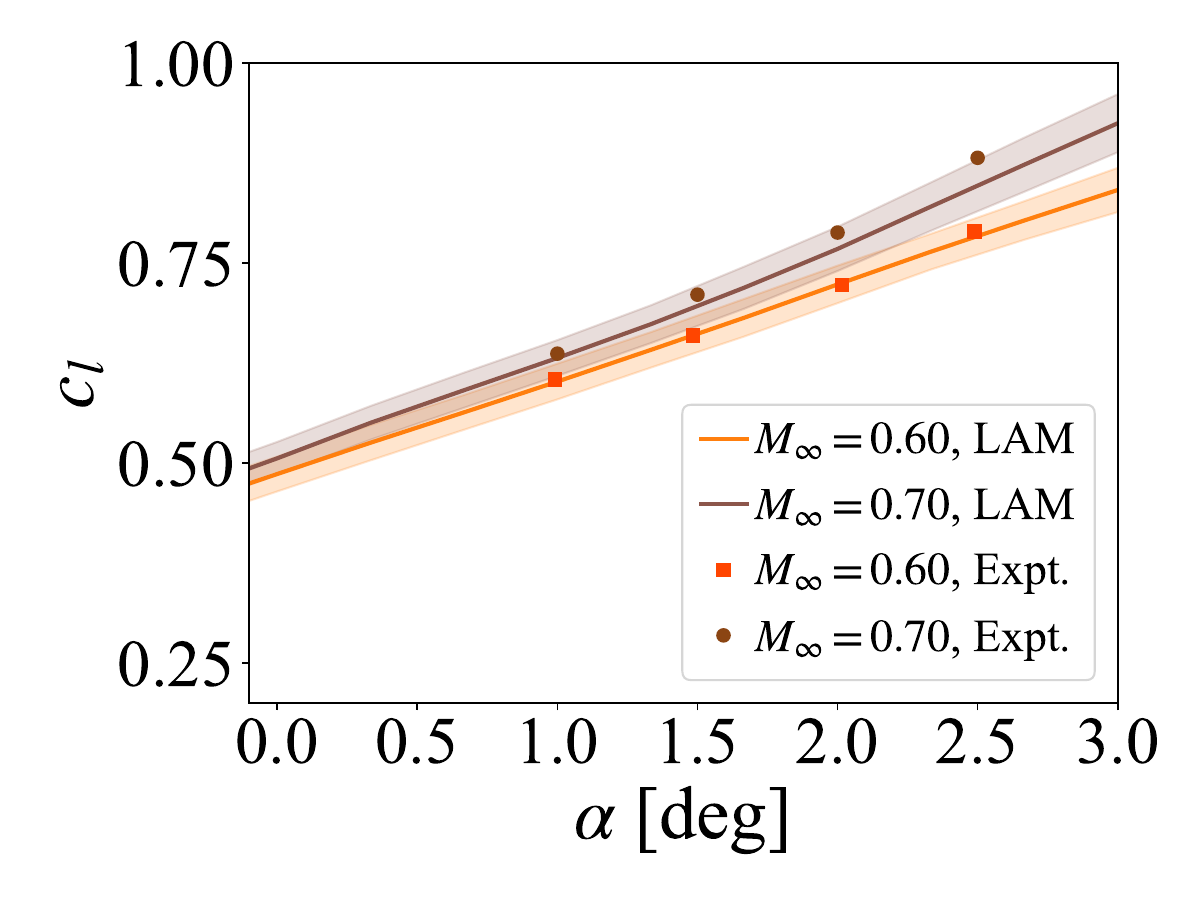} }\label{cl_vs_alpha_sc9a}}%
    \subfloat[\centering NACA 63-415: Experimental $c_l$ of Ref.~\cite{bak2000} from strain gauge balance]{{\includegraphics[width=.33\textwidth]{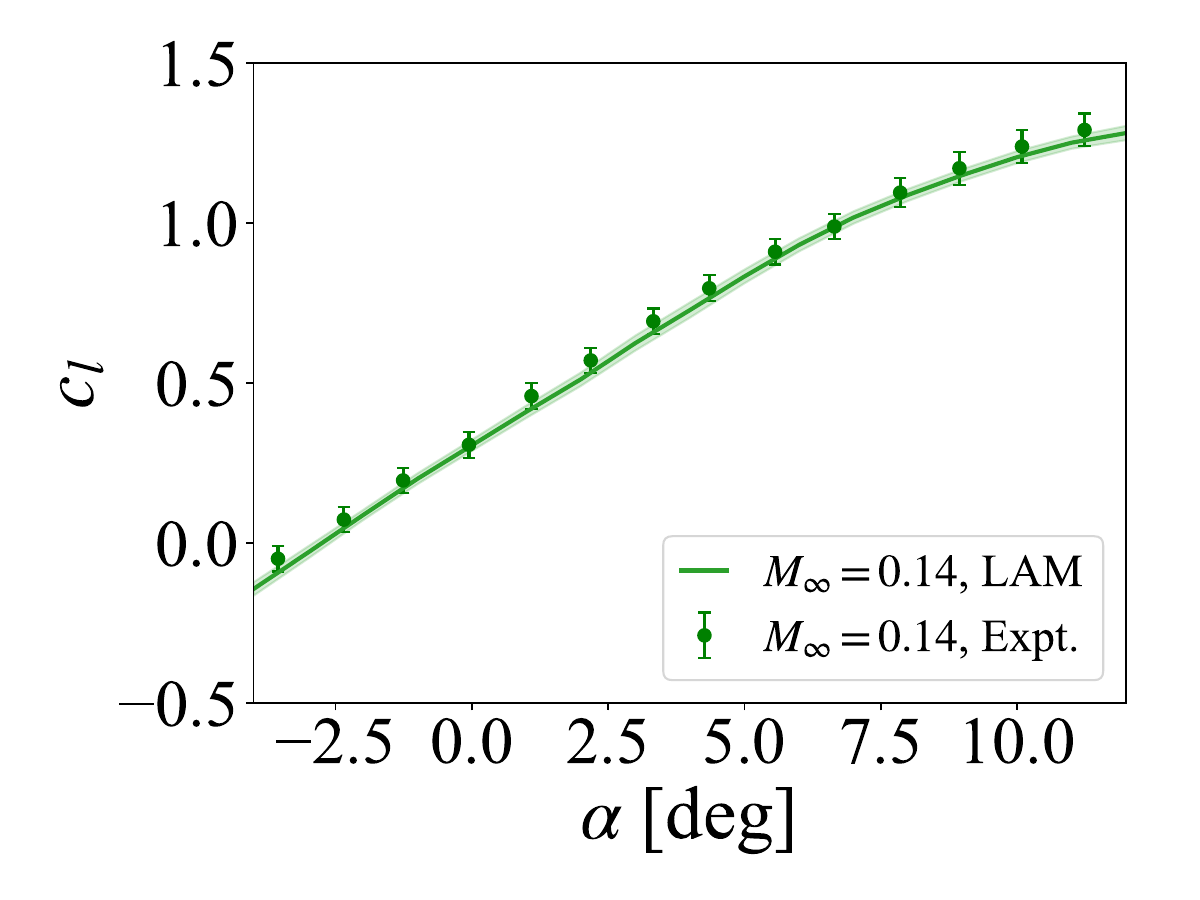} }\label{cl_vs_alpha_naca}}%
    \caption{Predicted $c_l$ at various angles of attack, freestream Mach numbers, and airfoils.  The error bars represent two standard deviations.}%
    \label{fig:cl_vs_alpha}%
\end{figure}

Figure~\ref{fig:cl_vs_alpha} shows the predicted sectional lift coefficients. Overall, good correlation between the predictions and the experimental measurements can be observed, where the deviation from each respective airfoil's $c_{l, \text{max}}$ never exceeds beyond $4\%$. Figures~\ref{cl_vs_alpha_sc1095} and \ref{cl_vs_alpha_sc9a} show that the lift curve slope increases and $c_{l, \text{max}}$ decreases with increasing freestream Mach number. These trends, in agreement with the  experimental data points, is the expected behavior of airfoils \cite{leishman}. Accurate predictions of stall are obtained, with the lift curve slope approaching zero at higher $\alpha$. However, it was found that there was some underprediction of $c_l$ in Fig.~\ref{cl_vs_alpha_sc9a} at $M_\infty = 0.70$ although the experimental data points are near the 95\% confidence interval. The error between the predicted mean and measurement was greatest at $\alpha = 2.5^\circ$ where $\Delta c_l = 0.05$. The discrepancy can be attributed to the fact that the model predicts a slightly earlier onset of shock as observed in Fig.~\ref{fig:sc9a}.

\begin{figure}[h!]%
    \centering
    \includegraphics[width=.33\textwidth]{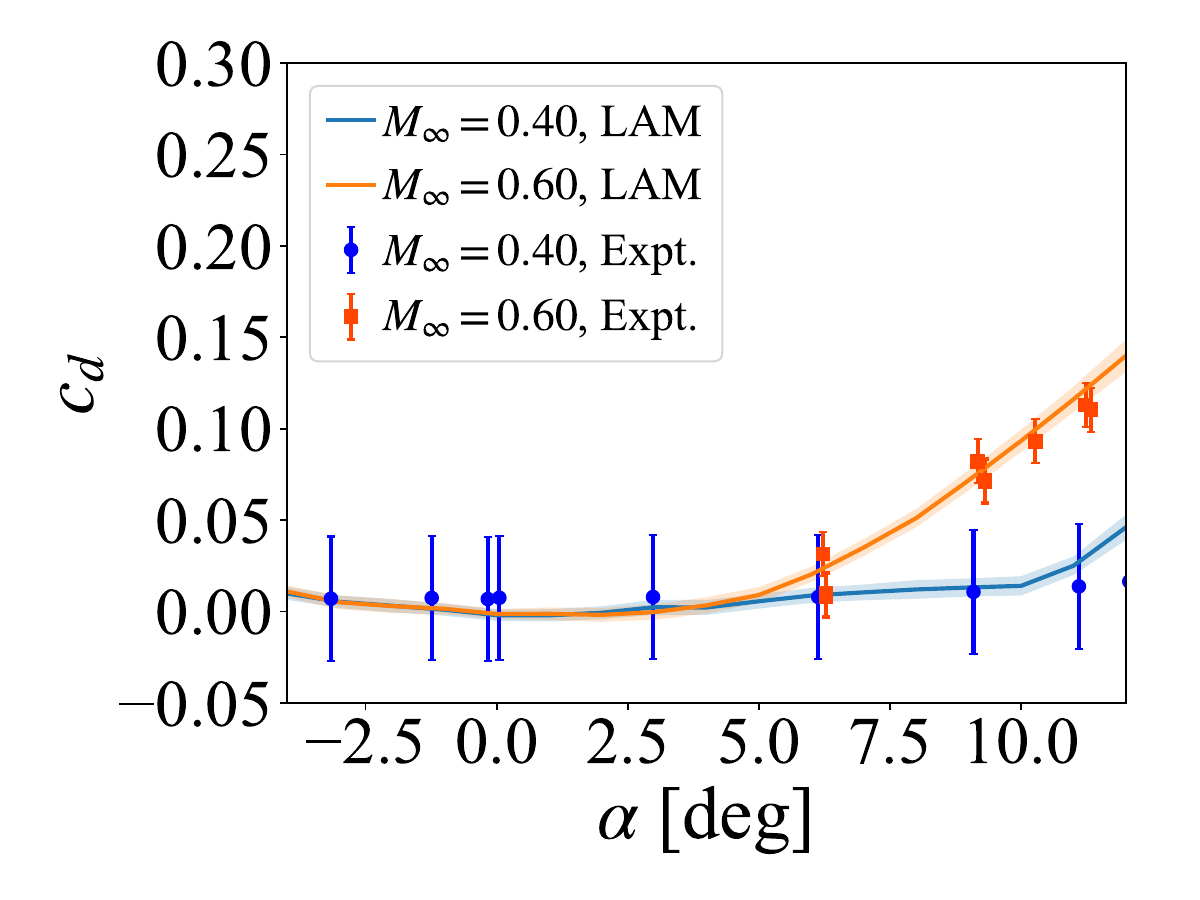}%
    \caption{Predicted $c_d$ at various angles of attack and freestream Mach numbers for SC1095. The error bars represent two standard deviations. Experimental $c_d$ are wake rake measurements from Ref.~\cite{harris1979}.}
    \label{fig:cd_vs_alpha}%
\end{figure}

Figure~\ref{fig:cd_vs_alpha} highlights the comparison between the predicted and experimentally measured $c_d$ for the SC1095 airfoil. The results from NASA Supercritical Airfoil 9a and NACA 63-415 are omitted as the assumption that the $c_{d, p} \approx c_d$ does not hold. As the only available measurements are from wake rakes that measure the total drag, it is difficult to validate the drag predictions from the model for these two cases. For the Supercritical Airfoil 9a, the operating conditions of the test data are at near-zero angles of attack. At the low values of $\alpha$, skin friction drag can comprise a greater portion of the total drag \cite{Anderson2011Fundamentals}. Supercritical airfoils are optimized for minimal pressure drag at high $M_\infty$ compared to conventional airfoils, which deviates from the assumption even further. As for the NACA 63-415, the experimental conditions are at a much lower freestream Mach number than the SC1095, which corresponds to a relatively higher skin friction drag. Thus, significant underprediction would be expected for these airfoils when using the method to calculate $c_d$. For the SC1095, where the assumption of a dominant pressure drag is valid, the maximum error ($\Delta c_d$) was found to be 0.018 at $\alpha = 6^\circ$ for one of the measurements. However, a second measurement at the same operating condition has a much lower error of 0.005. 

\begin{figure}[h!]%
    \centering
    \subfloat[\centering SC1095: Experimental $c_m$ of Ref.~\cite{flemming1984} derived from $C_p$ integration]{{\includegraphics[width=.33\textwidth]{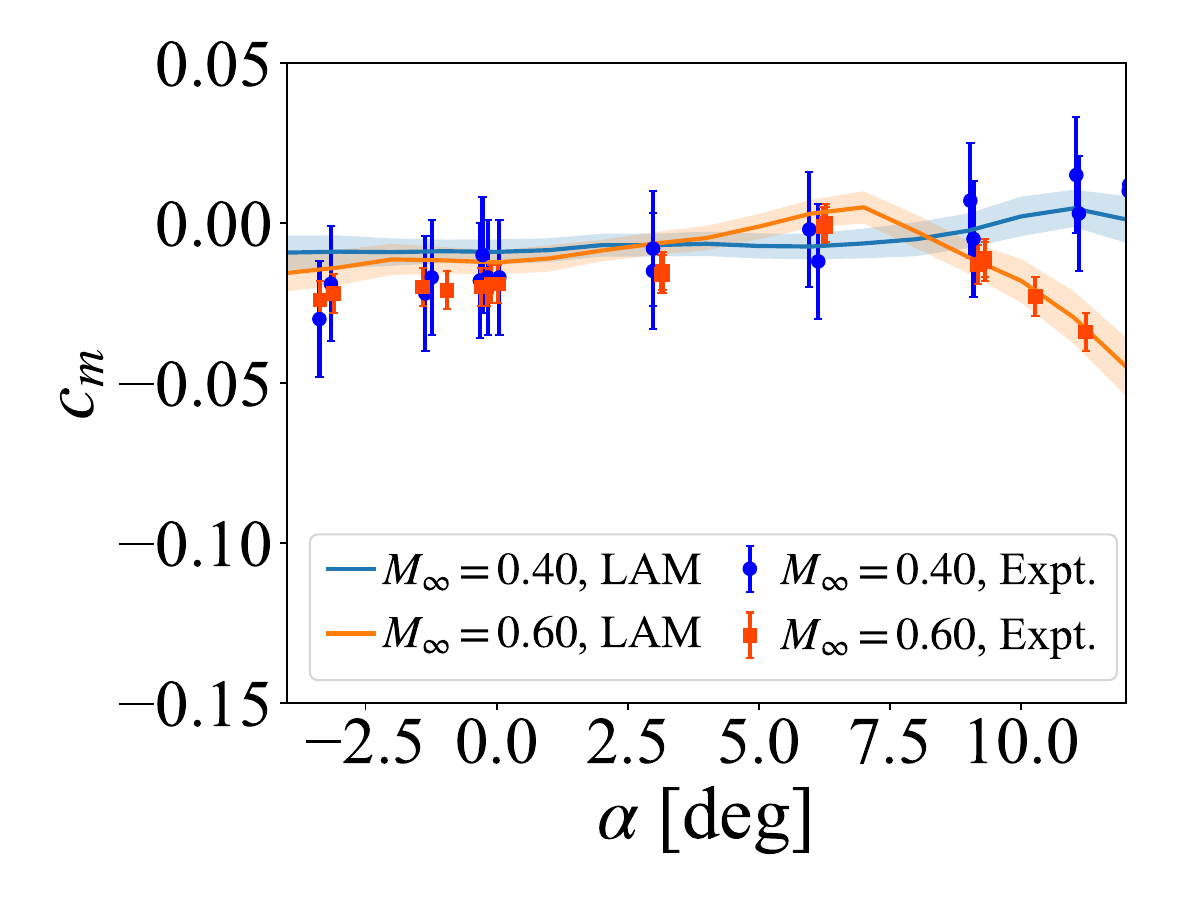}}\label{cm_vs_alpha_sc1095}}%
    \subfloat[\centering NASA Supercritical Airfoil 9a: Experimental $c_m$ of Ref.~\cite{harris1979} from balance measurements]{{\includegraphics[width=.33\textwidth]{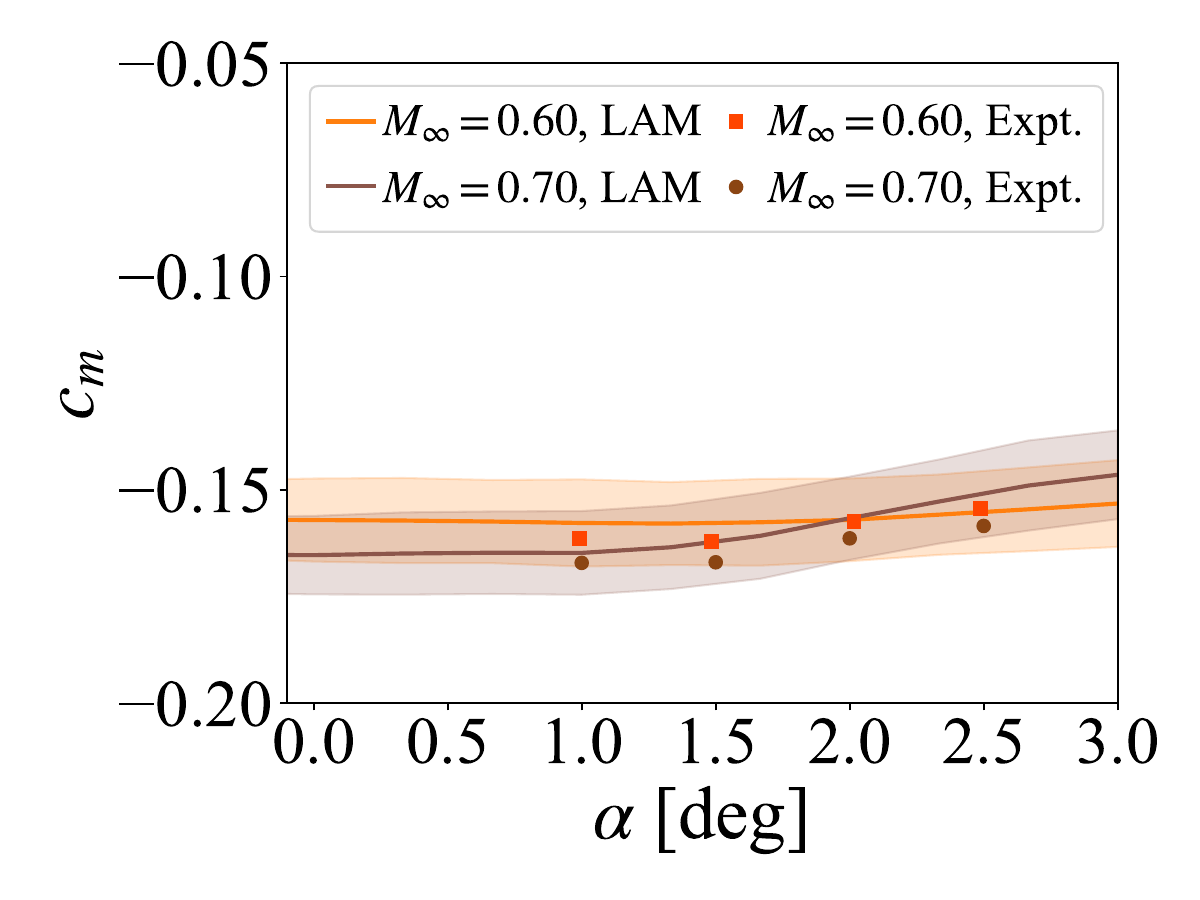}}\label{cm_vs_alpha_sc9a}}%
    \subfloat[\centering NACA 63-415: Experimental $c_m$ of Ref.~\cite{bak2000} from strain gauge measurements]{{\includegraphics[width=.33\textwidth]{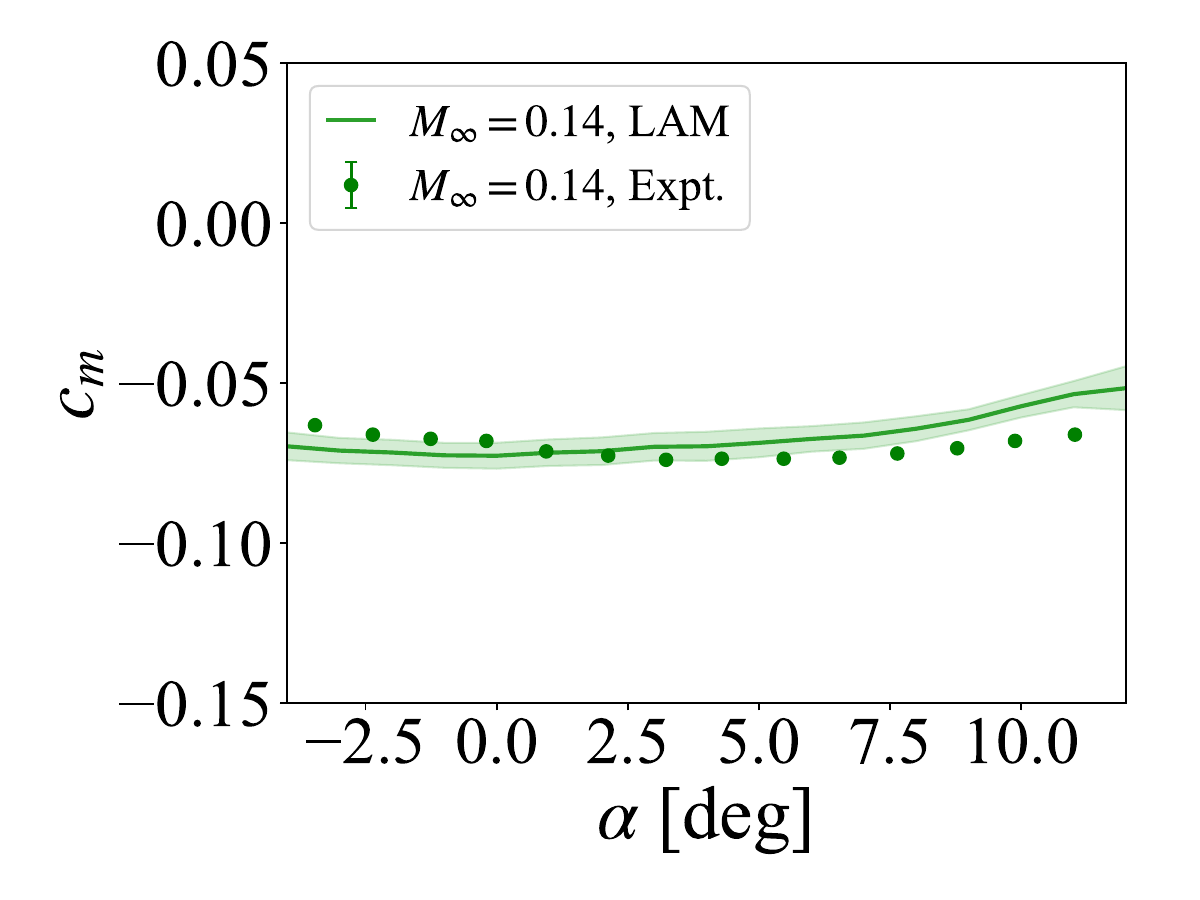}}\label{cm_vs_alpha_naca}}%
    \caption{Predicted $c_m$ at various angles of attack, freestream Mach numbers, and airfoils. The error bars represent two standard deviations.}%
    \label{fig:cm_vs_alpha}%
\end{figure}

In Fig.~\ref{fig:cm_vs_alpha}, the predicted $c_m$ is compared against the measurements. The error in $c_m$ was found to never exceeds 0.01 at all angles and airfoils, indicating the the overall trend is captured. Fig.~\ref{cm_vs_alpha_sc1095} shows that the the increase in the pitching moment magnitude of SC1095 with increasing $\alpha$ is captured especially well. Figures~\ref{cm_vs_alpha_sc9a} and \ref{cm_vs_alpha_naca} also show that the experimental measurements fall within or is close to the 95\% confidence interval. 

The presented figures demonstrate that the proposed method effectively predicts the aerodynamic force coefficients, with the predictions generally correlating well with experimental measurements. The trends in lift, drag, and moment coefficients match the expected behavior of the airfoils across different freestream Mach numbers and angles of attack. Overall, the model shows promise as a comprehensive aerodynamic analysis of an airfoil, when applied under conditions where the assumptions hold.

\subsection{Beyond the Training Data}\label{sec:beyond}
As discussed in Sec.~\ref{sec:lim}, the model demonstrates the highest accuracy when making predictions within the range of the training data. In the case of the LAM, this range corresponds to $-4.0^\circ \le \alpha \le 12.0^\circ$ and $0.0 \le M_\infty \le 0.75$. Recognizing the model's limitation and its reduction of performance beyond the training data is particularly valuable for users, as it facilitates informed decision-making about the appropriate use of the tool. 

\begin{figure}[!ht]
\centering
\includegraphics[width=.99\textwidth]{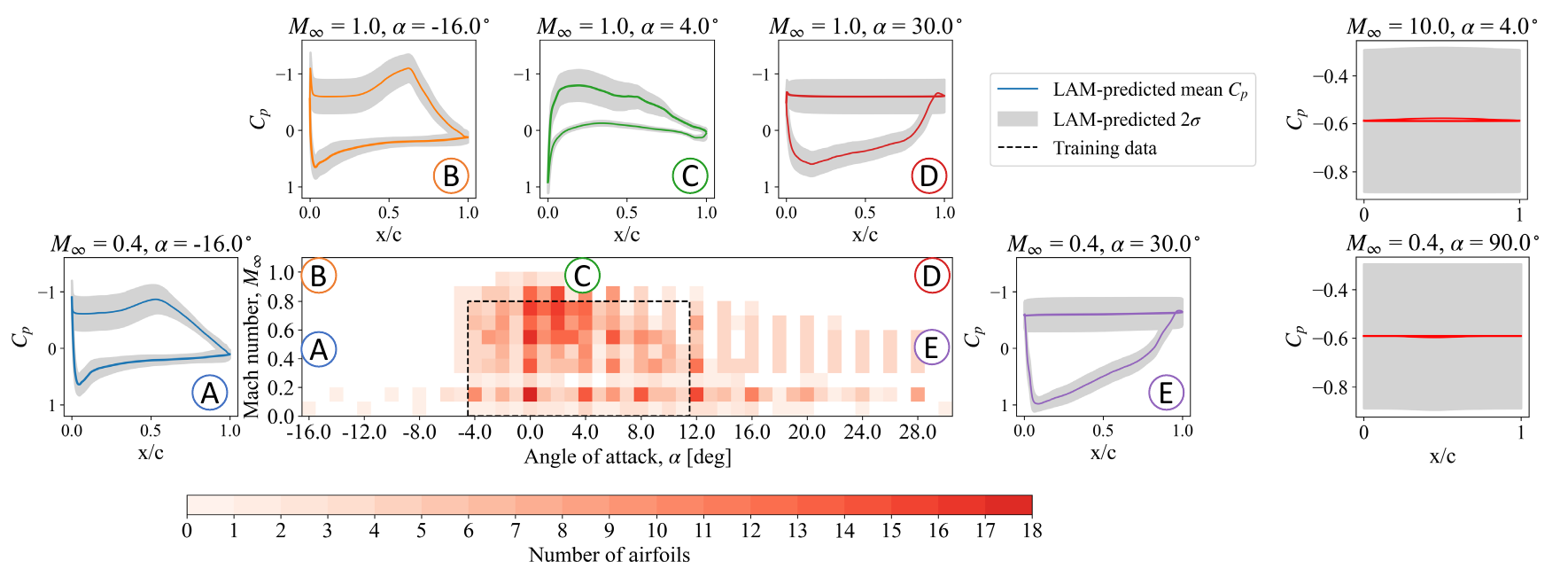}
\caption{Predictions of $C_p$ and two standard deviations for SC1095 under operating conditions beyond the training data. The extent of the training data in $\alpha$ and $M_\infty$ is highlighted by the black dotted box.}
\label{fig:beyond_training_data}
\end{figure}

Figure~\ref{fig:beyond_training_data} illustrates the model's performance beyond the training data with respect to angle of attack and freestream Mach number. For a SC1095 airfoil, the $C_p$ is predicted for high values of angles of attack and freestream Mach numbers not included in the training range. It is evident that, in this region, the unreliability is marked by a substantial increase in the posterior uncertainties. With further deviation from the training data---such as an extremely high freestream Mach number ($M_\infty=10.0$) or angle of attack ($\alpha=90^\circ$)---the bias-variance trade-off becomes apparent, leading to a physically implausible mean $C_p$, such as a flat line.

\begin{figure}[!ht]
\centering
\includegraphics[width=.45\textwidth]{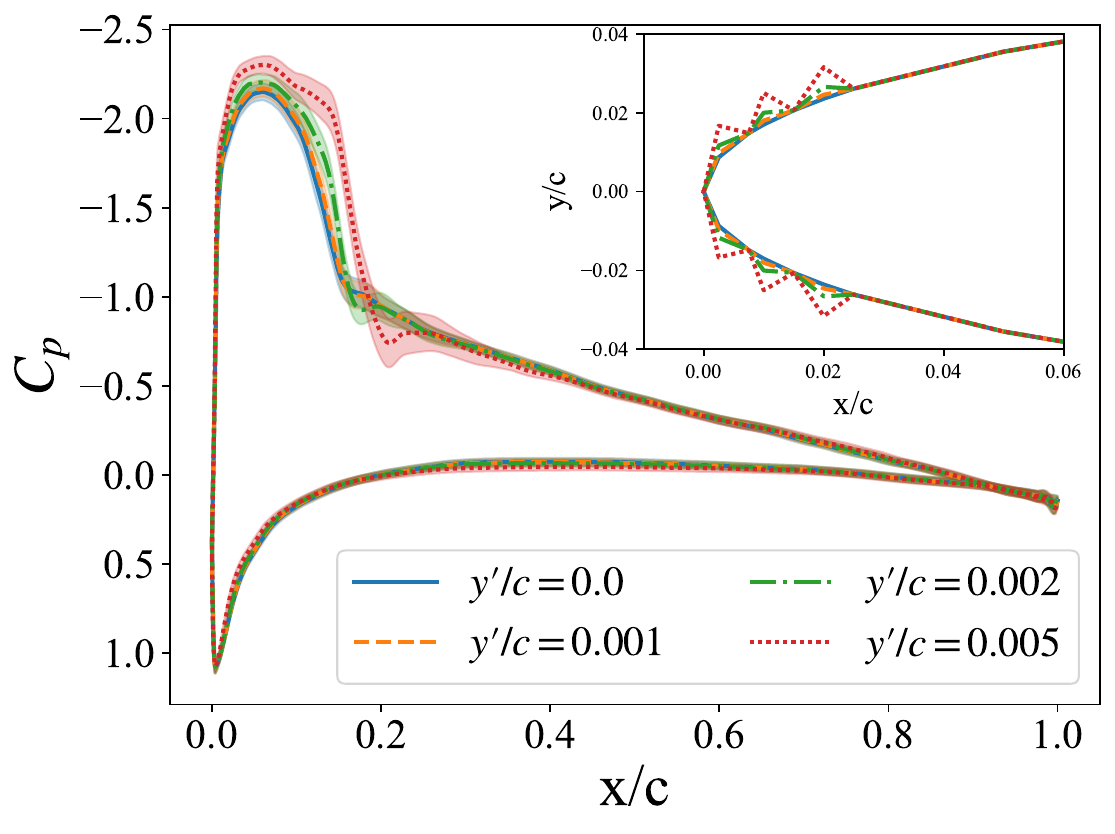}
\caption{The variation in predicted $C_p$ distribution based on the jaggedness of the NACA 0012 leading edge at $\alpha = 6^\circ$ and $M_\infty = 0.60$. $y^\prime/c$ indicates the magnitude of the leading-edge perturbation normalized by the chord length. The inset plot shows the geometry of the airfoil with the modeled perturbations.}
\label{fig:bump_study}
\end{figure}

Earlier discussions have noted that the model is limited by the fact that the training database is biased toward smooth airfoils. To assess its sensitivity to small geometric features, a numerical experiment is conducted to examine how predicted pressure distributions respond to leading-edge perturbations. A baseline airfoil (NACA 0012) is modified at three points chordwise locations: $x/c = 0.0025, 0.01, $ and $0.02$. The magnitude of these leading-edge perturbation normalized by the chord length, or $y^\prime/c$, are 0.001, 0.002, and 0.005. The results, shown in Fig.~\ref{fig:bump_study}, indicate that for small perturbations ($y^\prime/c = 0.001$), changes in $C_p$ are negligible. Increasing the magnitude to $y^\prime/c = 0.002$ begins to produce slight deviations in both the shape and magnitude of the $C_p$ distribution. When $y^\prime/c$ increases to $0.005$ and the airfoil becomes noticeably jagged, predicted pressure distribution deviates significantly from the baseline. The deviation is accompanied by a substantial increase in uncertainty bounds, which indicates reduced prediction reliability. These findings highlight the model's capability to capture the effects of geometric features down to $y^\prime/c = 0.001$, while also demonstrating that perturbations exceeding $y^\prime/c = 0.002$ result in increasingly uncertain predictions. Future work will look into leveraging surface roughness and ice-deposition datasets to better inform the impact of geometry perturbation on $C_p$.

\subsection{Grid Resolution}\label{subsec:grid_resolution}
The model allows the $C_p$ to be generated at an arbitrary set of locations. In contrast, results from CFD simulations depend on cell coordinates, and NN-based ML models typically produce outputs on pre-defined grids.

\begin{figure}[!ht]
\centering
\includegraphics[width=.6\textwidth]{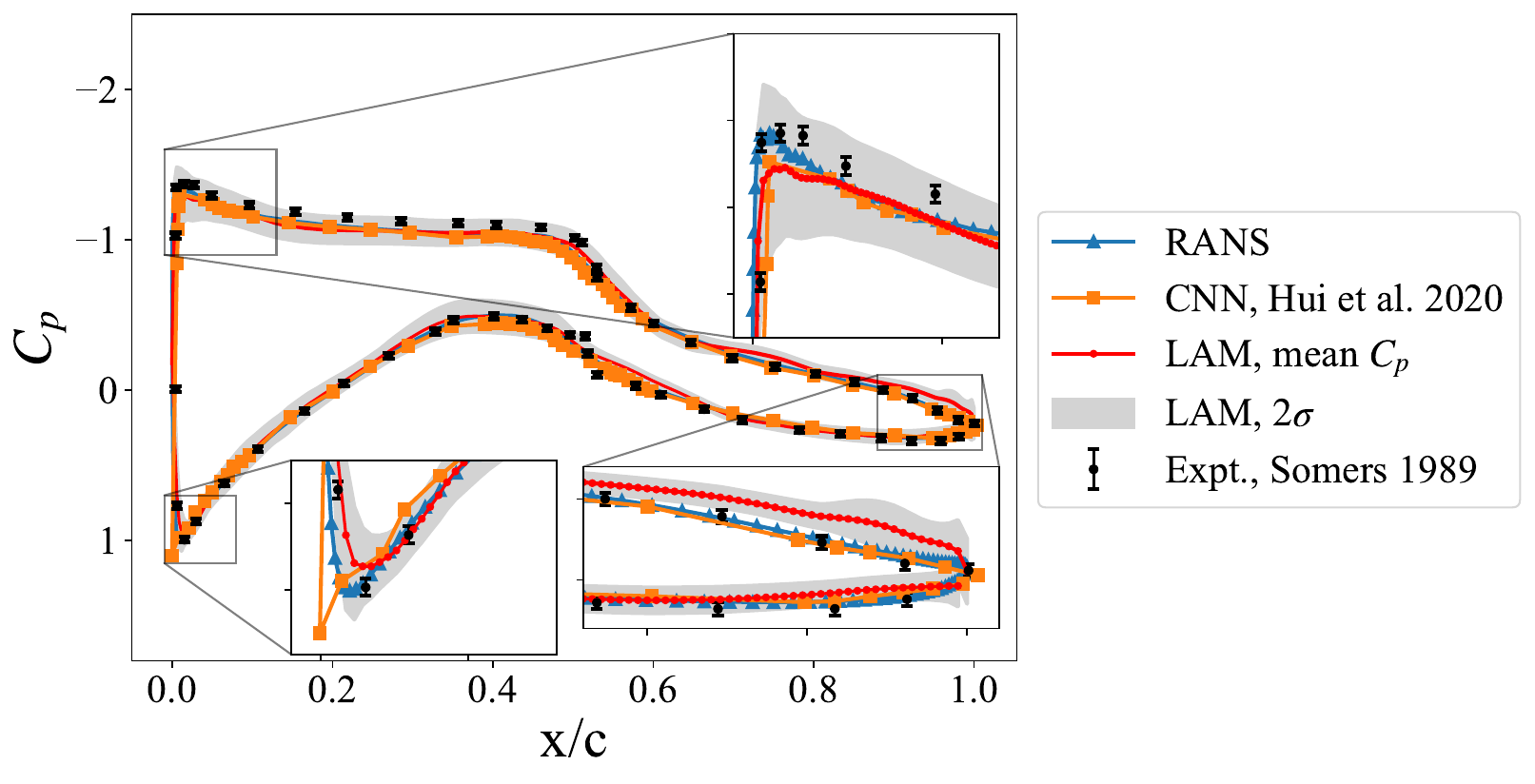}
\caption{Comparison of $C_p$ predictions between RANS, CNN-based model, and the LAM. Note the difference in grid resolution for each model. Experimental measurements are from Ref~\cite{s809_expt}.}
\label{fig:finer_grid}
\end{figure}

Figure~\ref{fig:finer_grid} illustrates the LAM's capability to generate results on a very fine grid. The $C_p$ distribution of a S809 airfoil at $\alpha = 5.13^\circ$ and $M_\infty = 0.15$ was predicted by the LAM and compared against results from a RANS simulation and CNN-based ML model presented by Hui et al~\cite{Hui_2020}. The RANS simulation was performed using SU2 \cite{SU2} on a single processor with a C-grid mesh that consists of 65,845 points. The run was considered converged when the variations in $c_l$ and $c_d$ fell below $1 \times 10^{-5}$. The figure confirms that the LAM has the finest resolution of points along the airfoil of 700 points, compared to 281 points of the SU2 simulation and 98 points of the CNN model. Moreover, increasing the resolution of the model is trivial, requiring only a change in the size of the input matrix. 

\subsection{Computational Performance and Capability}
Lastly, the computational cost and accuracy of the LAM are compared against two physics-based solvers: RANS (SU2) and XFOIL \cite{XFOIL}, per Table~\ref{table:runtime}. The comparison was performed for S809 at $\alpha = 5.13^\circ$ and $M_\infty = 0.15$ (Sec.~\ref{subsec:grid_resolution}) and NACA 0012 at $\alpha = 1.89^\circ$ and $M_\infty = 0.70$. The mesh for NACA 0012 was obtained from the SU2 tutorial repository and consists of 229,336 grid points. The numerical simulation was conducted at $\alpha = 1.50^\circ$, the angle that best matched the experimental validation data as determined in Ref.~\cite{naca0012_validation}. As with the S809 case, the convergence criteria were $\Delta c_l, \ \Delta c_d \le 1 \times 10^{-5}$. XFOIL results were obtained with 494 panels, the highest resolution possible in the program. While both airfoils were included in the training sets, the operating conditions were not. 
\begin{table}
\footnotesize
\centering
  \renewcommand{\arraystretch}{1.2}
\begin{tabular}{ |c||c|c|c|c|c|c|  }
 \hline
  Solver & Case & Wall Clock Time & $\text{MAE}_{\text{enclosed}}$ & MAE & Compressibility & Separation\\
 \hline
 \multirow{2}{*}{RANS (SU2)} & S809 & 1349.76s & 0.028 & 0.055 & \multirow{2}{*}{$\checkmark$} & \multirow{2}{*}{$\checkmark$}\\
  & NACA 0012 & 1021.20s & 0.016 & 0.034 & & \\
  \cline{1-7}
  \multirow{2}{*}{LAM} & S809 &  CPU: 4.13s / GPU: 0.881s & 0.030 & 0.056 & \multirow{2}{*}{$\checkmark$} & \multirow{2}{*}{$\checkmark$} \\
  & NACA 0012 & CPU: 4.30s / GPU: 0.887s  & 0.036 & 0.042 & & \\
  \cline{1-7}
  \multirow{2}{*}{XFOIL} & S809 & 1.98s  & 0.030 & 0.052 & \multirow{2}{*}{$\times$} & \multirow{2}{*}{$\times$} \\
  & NACA 0012 & 2.08s & 0.071 & 0.100 & & \\
 \hline
 \end{tabular}
 \caption{Computational efficiency and capabilities of different types of airfoil aerodynamic tools. All cases were run with a single processor CPU, except for the LAM GPU case, which was run on one NVIDIA A100 GPU.}
\label{table:runtime}
\end{table}

\begin{figure}[h!]%
    \centering
    \subfloat[\centering $C_p$ of S809 at $\alpha = 5.13^\circ$ and $M_\infty = 0.15$. Experimental measurements are from Ref.~\cite{s809_expt}.]{{\includegraphics[width=.35\textwidth]{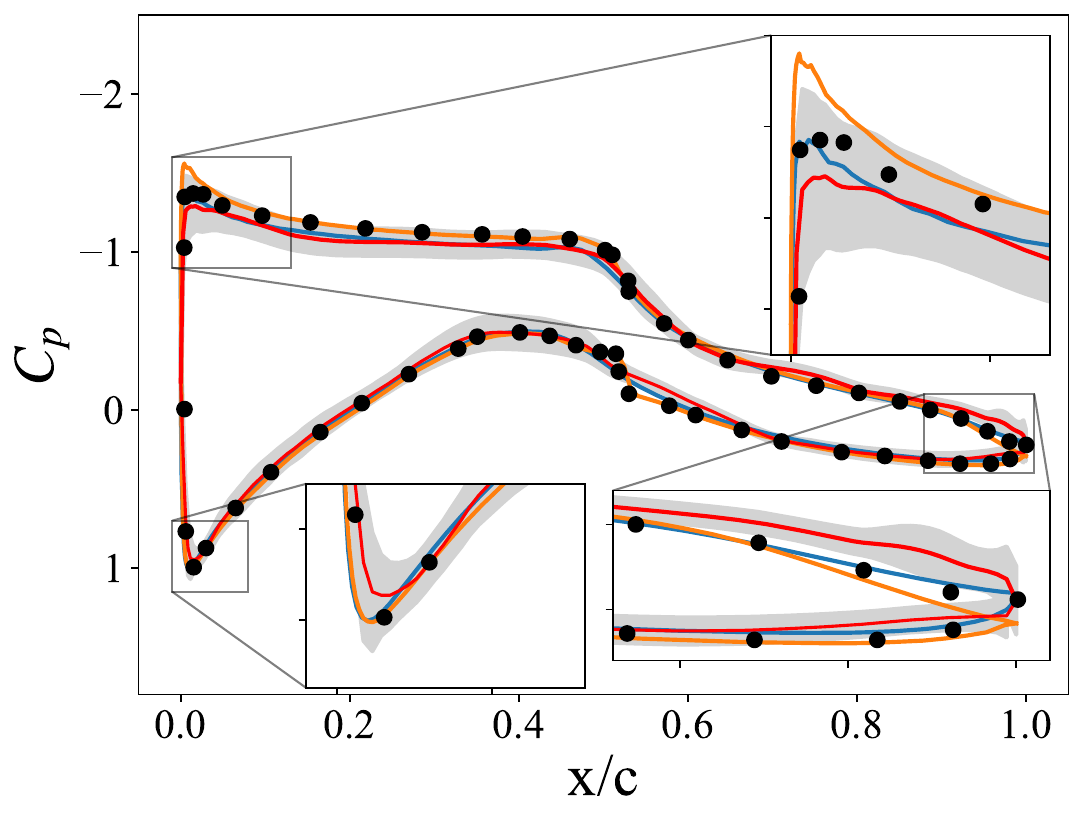} }\label{s809_comparison}}
    \hfill
    \subfloat[\centering $C_p$ of NACA0012 at $\alpha = 1.86^\circ$ and $M_\infty = 0.7$. Experimental measurements are from Ref.~\cite{naca0012_validation}.]{{\includegraphics[width=.35\textwidth]{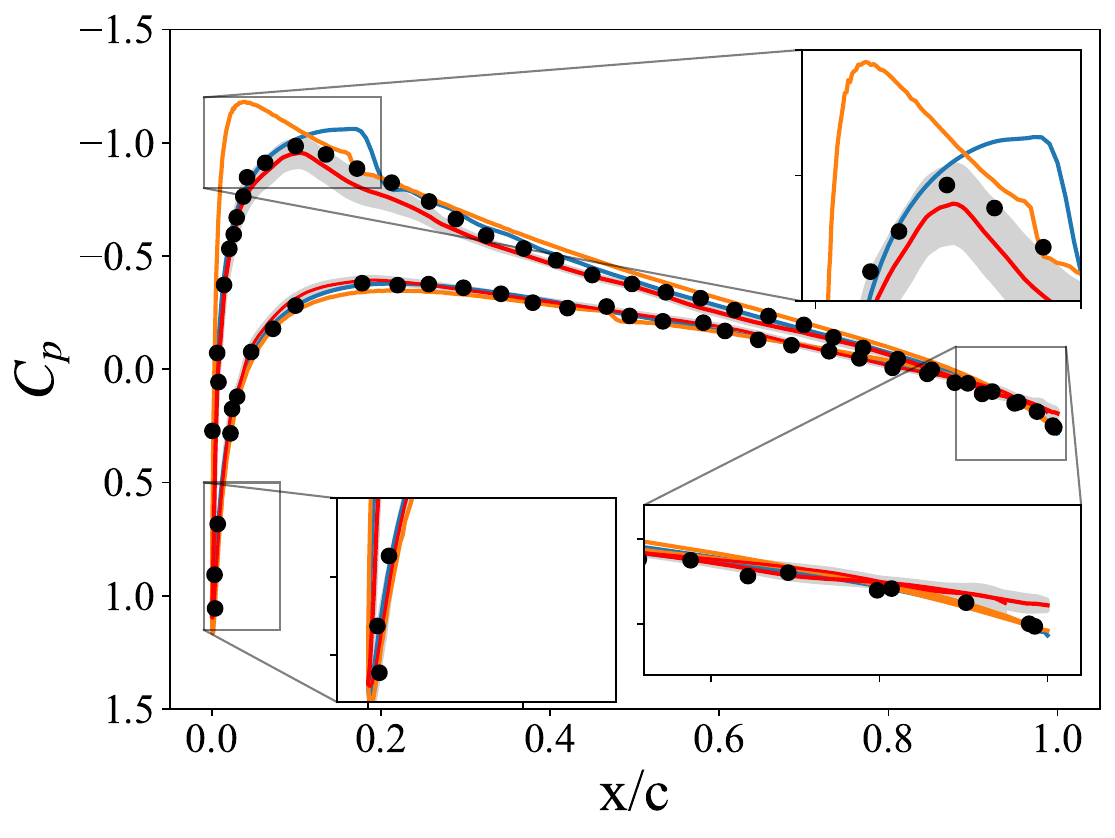} }\label{0012_comparison}}
    \captionsetup[subfloat]{labelformat=empty}
    \hfill\subfloat[]{{\vspace{12mm}\includegraphics[width=.2\textwidth]{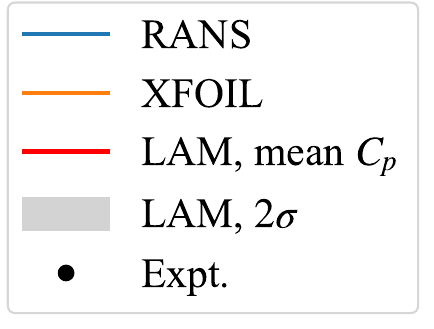}}}
    \caption{Comparison of predicted $C_p$ for different aerodynamic solvers: RANS simulations, the LAM, and XFOIL.}
    \label{fig:solver_comparison}
\end{figure}

Figure~\ref{fig:solver_comparison} illustrates the comparison between the predicted $C_p$ distributions. RANS simulations model the governing equations of fluid dynamics, which permits accurate capture of airfoil physics. This is reflected by the lowest error overall in both $\text{MAE}_\text{enclosed}$ and MAE. However, it consumes the longest wall clock time and requires the user's technical skills such as meshing techniques and turbulence modeling decisions. 

The S809 case is in the incompressible regime. One notable feature of the $C_p$ curve is the drop in pressure that occurs at mid-chord, which is caused by boundary layer transition. In this case, XFOIL was found to yield accurate results, on-par with CFD, with a low computational cost of 1.98 seconds. However, the program has certain limitations that affect its accuracy. Firstly, XFOIL requires controlling parameters such as the critical amplification factor ($N_\text{crit}$) and transition locations. Using these parameters to model precisely transition requires user expertise, and can still underperform compared to other methods such as the Langtry-Menter model in CFD. Secondly, as a potential flow solver, XFOIL cannot resolve complex aerodynamics such as compressibility effects and flow separation. This is observed in the NACA 0012 case at $M_\infty = 0.70$, where there is significant mismatch in $C_p$ at the suction peak and an increase in the errors. 

In contrast, the LAM has a significantly reduced runtime compared to that of RANS, and even outperforms XFOIL when utilizing a GPU. The model yields a similar accuracy to that of XFOIL at low freestream $M_\infty$, and successfully captures the boundary layer transition with no user input on transition modeling. Furthermore, the model also captures compressibility effects as demonstrated in the NACA 0012 case. 

\begin{figure}[!ht]
\centering
\includegraphics[width=.4\textwidth]{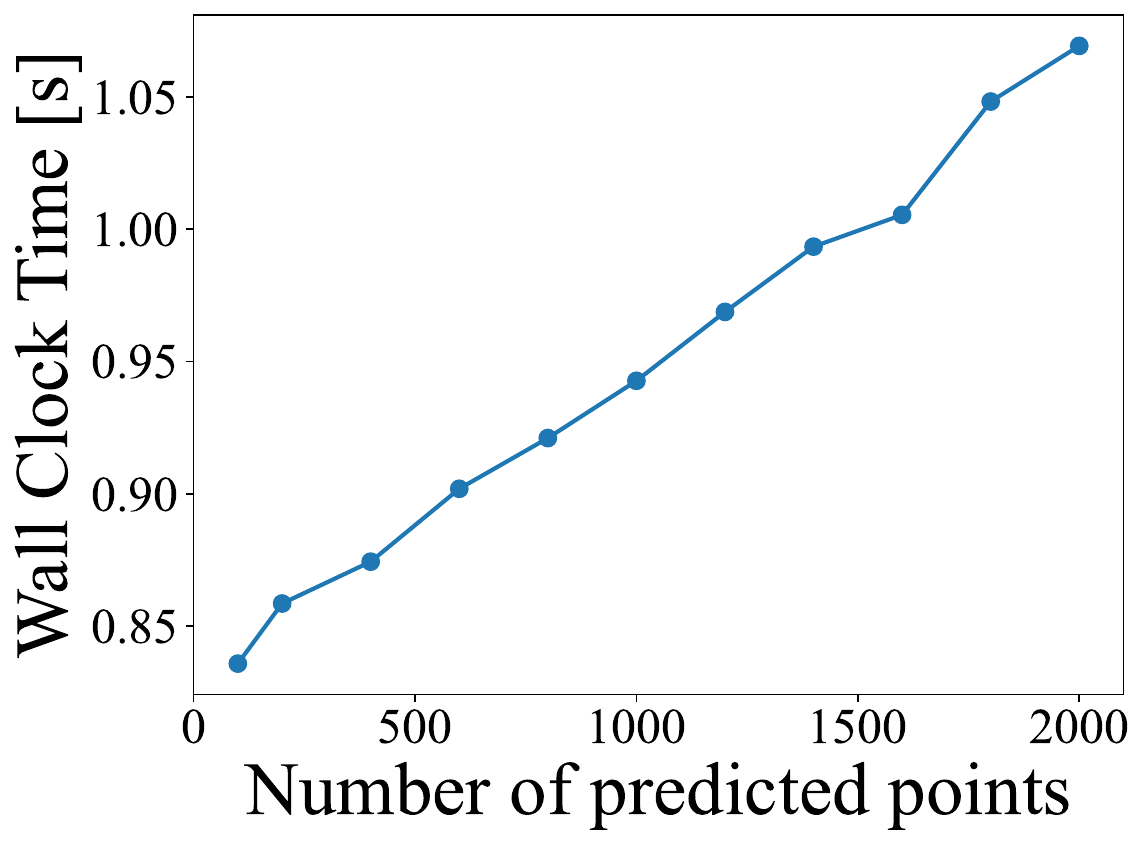}
\caption{Computational cost of the LAM as a function of the number of predicted $C_p$ points over an airfoil. The computations were performed on a single A100 GPU.}
\label{fig:runtime_scalability}
\end{figure}

It is also valuable to understand how the model scales with the size of the test data set, or the resolution of the predicted $C_p$. A Gaussian Process model is typically bottlenecked by the inversion of the $N \times N$ covariance matrix, where $N$ represents the number of training data points. This operation exhibits a computational complexity of $\mathcal{O}(N^3)$. For this model, since the inverted covariance matrix remains unchanged with respect to the test data, it can be pre-computed and stored to bypass the computational burden. With this approach, the primary computational cost for prediction arises from the multiplication of the covariance between the training and test points, which scales as $\mathcal{O}(NM)$, where $M$ is the number of test points. Consequently, the prediction runtime scales linearly with the size of the test data set. This trend is corroborated by Fig.~\ref{fig:runtime_scalability}, which illustrates the wall clock time required to generate $C_p$ predictions at varying resolutions.

\section{Conclusion}\label{5_conc}
In this study, the development of a Large Airfoil Model (LAM) was presented\footnote{also accessible online on \url{https://large-airfoil-model.azurewebsites.net}}. The novel machine learning model is able to answer technical questions on aerodynamic characteristics of airfoils, paving the way for new and intuitive aerodynamic design tools. Central to the effort is A Deep Airfoil Prediction Tool (ADAPT), a new machine learning model that adopts a probabilistic approach through deep kernel learning, enabling it to make uncertainty-aware predictions of airfoil pressure coefficient $C_p$ distributions. Alongside ADAPT, Airfoil Surface Pressure Information Repository of Experiments (ASPIRE), the first large-scale, open-source repository of airfoil experimental data, was presented. It was developed to be a vast dataset that the LAM can easily access and serve as the training data for ADAPT.

ADAPT was designed and developed to best accommodate the unstructured nature of a dataset built from various experimental sources, though the flexibility of the model architecture allows the ingestion of any type of airfoil $C_p$ distributions, including those from CFD data. The Gaussian process layer of the deep kernel learning architecture, which learns the underlying function rather than outputs at pre-defined locations discretely, enables the flexibility. Additionally, the model, trained on experimental data as opposed to numerical simulations, offers a Bayesian means of pressure prediction. This allows the proper characterization of the uncertainties associated with measurements and digitization. The model yielded a $\text{MAE}_{\text{enclosed}}$, the mean absolute error in the area enclosed by the pressure curves, of $0.029$ on the test data. This was equivalent to under 4\% error in each airfoil's maximum lift coefficient, indicating good correlation. While a single general model was created to leverage the breadth of data made available from this work, it would also be possible to create multiple specialized LAMs depending on the user's requirements. 

The database ASPIRE was created from a data mining campaign, extracting wind tunnel experimental data that were locked behind bulk scans of tables and images. It includes digitized pressure measurements for various airfoil families and applications such as general aviation, rotorcraft, and wind turbines. The database covers a wide variety of operating conditions, ranging from $-30^{\circ}$ to $30^{\circ}$ in $\alpha$ and $0.0$ to $1.0$ in freestream Mach number, $M_\infty$. It contains 2917 unique pressure distributions of 69 airfoils, obtained from sources that span the past century. The goal is to continue expanding the database and address any existing gaps. Contributions can be made on the aforementioned website. 

The LAM was interfaced with a large language model, where the user provides a natural language query rather than specific technical inputs. The capability was enabled by a Retrieval-Augmented Generation (RAG) framework, which augments LLM response generation by incorporating an information retrieval process. This ensured that the model responses are based on either the experimental data within ASPIRE or the probabilistic predictions from ADAPT.

As a natural consequence of the $C_p$ predictions, the LAM predictions permitted the calculation of aerodynamic force coefficients using a Monte Carlo approach. These one-dimensional metrics are crucial as they underpin all aerodynamic design processes. The predicted coefficients, compared to the experimental lift, drag, and moment coefficients ($c_l$, $c_d$, and $c_m$) yielded maximum errors of 0.05, 0.018, and 0.01, respectively.

Since the LAM is fundamentally a Gaussian process model, the predictions can be performed at an arbitrary set of locations. This allows the model to yield $C_p$ values at an infinite grid resolution, which provides greater flexibility than convolutional neural networks based airfoil aerodynamics models, which predict $C_p$ on a preset grid. 

Finally, the model was found to be computationally efficient, requiring an average of 0.884 seconds on a single CPU to generate predictions for a S809 airfoil at $M_\infty=0.15$, $\alpha=5.13^\circ$ and NACA 0012 at $M_\infty=0.70$, $\alpha=1.86^\circ$. The developed model required an average of 0.883 seconds when accelerated by a single A100 GPU. This outperforms XFOIL, a potential-flow based solver. Unlike the low-fidelity method, the LAM has the capability to capture complex aerodynamic phenomena, such as compressibility effects, transition, and flow separation, having been trained on such effects. This also means that the LAM does not require the user's expertise on a particular software to model specific aerodynamic phenomena accurately.

\section*{Acknowledgments}\label{ack}
The authors would like to thank Aaditya Hingoo, Ryan McKee, Tom Hoang, Bipin Koirala, and Grant Ewing for their assistance in the digitization of airfoil experimental data, and Chris Silva, Vengalattore Nagaraj, and Robert Scott for their guidance on key design questions. 

This research was partially funded by the U.S. Government under Cooperative Agreement No. W911W6-21-2-0001. The U.S. Government is authorized to reproduce and distribute reprints for Government purposes notwithstanding any copyright notation thereon.

The views and conclusions contained in this document are those of the authors and should not be interpreted as representing the official policies or position, either expressed or implied, of the U.S. Army Combat Capabilities Development Command (DEVCOM), Aviation \& Missile Center (AvMC), or the U.S. Government.

\bibliographystyle{elsarticle-num}  
\bibliography{references.bbl}

\newpage
\appendix \label{appendix}
\section{Mathematical Preliminaries of Gaussian Process}\label{app:GP}
First, it is assumed that a dataset $\mathcal{D}$ exists, which contains $N$ input vectors related to airfoil geometry and operating conditions (i.e. the latent variables obtained from ADAPT's FCNN) and the corresponding pressure coefficient, $C_p$, at a given chordwise location; $(\textbf{z}_n, y(\textbf{z}_n))$ for $n = 1,\ \hdots ,\ N$. It is assumed that each observation of $C_p$ is a noisy version of the underlying ``true" value, $y_n = f(\textbf{z}_n) + \epsilon_n$, where $\epsilon \sim \mathcal{N}(0,\,\sigma_n^{2})$. The values of $\sigma_n^2$ of the likelihood model is obtained from the experimental accuracy recorded in ASPIRE. It will also be useful to assume that the predictive pressure coefficient values at other locations are required. This training data and testing data is given by 
\begin{equation}
    \textbf{Z} = \begin{bmatrix} \textbf{z}_1\\ \vdots \\ \textbf{z}_N \end{bmatrix}, \
    \textbf{y} = \begin{bmatrix} y(\textbf{z}_1)\\ \vdots \\y(\textbf{z}_N) \end{bmatrix} \text{and} \
    \textbf{Z}^\ast = \begin{bmatrix} \textbf{z}_1^\ast\\ \vdots \\ \textbf{z}_M^\ast \end{bmatrix}, \
    \textbf{f}^\ast = \begin{bmatrix} f^\ast(\textbf{z}_1)\\ \vdots \\ f^\ast(\textbf{z}_M) \end{bmatrix},
\end{equation}

\noindent where the asterisk indicates the testing locations. In GP regression, the target variable is modeled as a Gaussian process, which is fully characterized by its mean function ($\mu(\mathbf{z})$) and covariance function ($k(\mathbf{z}, \mathbf{z'})$). The GP regression model can be defined as 
\begin{equation}
f(\mathbf{z}) \sim \mathcal{GP}(\mu(\mathbf{z}), k(\mathbf{z}, \mathbf{z'})).
\end{equation}

Then, the joint density of the observed data and the latent, noise-free function on the test points is given by 
\begin{equation}
    \begin{pmatrix} \textbf{y} \\ \textbf{f}_\ast \end{pmatrix} \sim \mathcal{N}\begin{pmatrix} \begin{pmatrix} \mu_Z \\ \mu_\ast \end{pmatrix}, \begin{pmatrix} \textbf{K}_{Z,Z} + \sigma_n^2\textbf{I} & \textbf{K}_{Z,\ast} \\ \textbf{K}_{Z,\ast}^T & \textbf{K}_{\ast, \ast} \end{pmatrix} \end{pmatrix}.
    \label{eq:gp}
\end{equation}

\noindent $\textbf{K}_{\cdot,\cdot}$ denotes the covariance matrix between the GPs evalulated at either $\textbf{Z}$ or $\textbf{Z}^\ast$. For example, $\textbf{K}_{Z,Z}$ represents the $N \times N$ covariance matrix between the GP at training data $\textbf{Z}$ while $\textbf{K}_{Z,\ast}$ represents the $N \times M$ covariance matrix between the GP at the training data $\textbf{Z}$ and a set of test points $\textbf{Z}^\ast$. 

Given the training data of input-output pairs ($\mathbf{Z}$, $\mathbf{y}$), the posterior Gaussian distribution at $\mathbf{Z_*}$ can be written as
\begin{align}
\mu_{*|Z} &= \mathbf{K}_{Z,*}^T (\mathbf{K}_{Z,Z} + \sigma_y^2\mathbf{I})^{-1}\mathbf{y}  \label{eq:posteriorMean} \text{ and} \\
\Sigma_{*|Z} &= \mathbf{K}_{*,*} - \mathbf{K}_{Z,*}^T (\mathbf{K}_{Z,Z} + \sigma_y^2\mathbf{I})^{-1} \mathbf{K}_{Z,*}. \label{eq:posteriorCov}
\end{align}
Due to numerical instability that may arise from directly calculating the inverse $(\mathbf{K}_{Z,Z} + \sigma_n^2\mathbf{I})^{-1}$, a Cholesky decomposition is performed where, $\mathbf{K}_\sigma = \mathbf{L}\mathbf{L}^T$. This allows the rewriting of Equations \ref{eq:posteriorMean} and \ref{eq:posteriorCov} as
\begin{equation}
\mu_{*|Z} = \mathbf{K}_{Z,*}^T \mathbf{\alpha}, \text{ and}
\label{eq:posteriorMeanChol}
\end{equation}
\begin{equation}
\Sigma_{*|Z} = \mathbf{K}_{*,*} - \mathbf{v}^T\mathbf{v},
\label{eq:posteriorCovChol}
\end{equation}
where $\mathbf{\alpha} = \mathbf{L}^T \backslash (\mathbf{L}\ \mathbf{y})$ and $\mathbf{v} = \mathbf{L} \backslash \mathbf{K_{Z,*}}$.

\section{Kernel Function}\label{app:kernel}
As observed in the previous equations, the covariance function (also known as the kernel function) is a crucial ingredient in a Gaussian process regression, as it encodes the assumptions about the function that is learned~\cite{RW2006}. One of the most widely used kernel function is the Squared Exponential (SE) kernel, also known as the Radial Basis Function (RBF) kernel. The function is of the form
\begin{equation}
k_{SE}(\mathbf{z}, \mathbf{z^\prime}) = \sigma_{f}^2 \exp \left(-\frac{1}{2l^2}\dist{\mathbf{z}-\mathbf{z^\prime}}^2 \right),
\end{equation}

\noindent where the hyperparameter $\sigma_{f}^2$ denotes the kernel amplitude (or variance) and $l$ denotes the length scale. One property of the SE kernel is that it is  infinitely differentiable, indicating that any GP employing this covariance function has mean square derivatives of all orders. Consequently, the model outputs have strong smoothness. This property of the SE kernel can prove detrimental in modeling some physical phenomena~\cite{Stein1999}.

A commonly used alternative to the SE kernel is the Mat\a'ern class of covariance functions. The class includes a smoothness parameter, often denoted by $\nu$, which allows the user to control the smoothness of the resulting GP sample paths. The parameter allows the function to perform better in modeling processes with discontinuities or sharp changes. The covariance is given by 
\begin{equation}\label{eq:matern}
k_{\text{Mat\a'ern}}(\mathbf{z}, \mathbf{z^\prime}) = \sigma_f^2 \frac{2^{1-\nu}}{\Gamma(\nu)} \left( \frac{\sqrt{2\nu}\dist{\mathbf{z} - \mathbf{z^\prime}}}{l} \right)^\nu K_\nu \left( \frac{\sqrt{2\nu}\dist{\mathbf{z} - \mathbf{z^\prime}}}{l} \right)
\end{equation}

\noindent where $\Gamma$ is the gamma function and $K_{\nu}$ is the modified Bessel function of the second kind. The most commonly used cases of the Mat\a'ern classes in machine learning are those where $\nu$ is a non-negative half-integers \cite{RW2006}, namely $\nu = 1/2, 3/2, \text{ and } 5/2$. In this work, $\nu = 5/2$ is used, which simplifies Eq.~\ref{eq:matern} to 
\begin{equation}\label{eq:matern52}
k_{\nu=5/2}(\mathbf{z}, \mathbf{z^\prime}) = \sigma_f^2 \left(1 + \frac{\sqrt{5}}{l}\dist{\mathbf{z} - \mathbf{z^\prime}} + \frac{5}{3l^2}\dist{\mathbf{z} - \mathbf{z^\prime}}^2\right) \exp\left(-\frac{\sqrt{5}}{l}\dist{\mathbf{z} - \mathbf{z^\prime}}^2\right).
\end{equation}

\noindent With the Mat\a'ern 5/2 kernel function, the Gaussian Process $f(\mathbf{z})$ is twice differentiable in the mean-square sense. This allows the model to capture the rapid changes in the pressure curves, often occurring at the suction peak or shocks, while maintaining some level of smoothness.

\section{Training Objectives}\label{app:objective}
Training the model involves finding optimal values of the weights and biases of the fully connected network and GP hyperparameters $\mathbf{t}$ ($\sigma_f^2$ and $l^2$), which are obtained by maximizing the marginal likelihood given the training targets $\mathbf{y}$: 
\begin{equation}\label{eq:mll}
\log p(\mathbf{y}|\mathbf{Z}, \mathbf{t}) = -\frac{1}{2}\mathbf{y}^T(\mathbf{K}_{Z,Z} + \sigma_n^2\mathbf{I})^{-1}\mathbf{y} - \frac{1}{2}\log|\mathbf{K}_{Z,Z} + \sigma_n^2\mathbf{I}| - \frac{N}{2}\log(2\pi),
\end{equation}

\noindent where $N$ is the number of training data, $\textbf{K}_{Z,Z}$ represents the $N \times N$ covariance matrix between the GP at training data $\textbf{Z}$, and $\sigma^2_n$ is the reported variance of the training data. 

The model \textit{jointly} learns all deep kernel hyperparameters, weights of the neural network ($\mathbf{w}$) and the parameters of the base kernel ($\mathbf{t}$), in an end-to-end fashion. The approach allows the incorporation of all components of the DKL process into a single model.

\section{Analysis of Model Active Space}\label{app:active_space}
The LAM predicts an airfoil's pressure distribution by first projecting the input data into a 10-dimensional latent space and then mapping these latent variables to an output multivariate normal distribution as a Gaussian process.  In this section of the Appendix, the model's behavior in this active space is analyzed. This is important for ensuring that the model operates with a robust and interpretable parameterization \cite{grey2023}. Dimensionality reduction, the process of simplifying the data while preserving its essential structure, plays a key role in this analysis.

In many studies, Principal Component Analysis (PCA) is the preferred technique for improving the interpretability of the active space. The method decomposes a data matrix $\mathbf{Z}$ into $\mathbf{Z} = \mathbf{U} \mathbf{\Sigma} \mathbf{V}^\top$ where $\mathbf{U}$ is the matrix of left singular vectors, $\mathbf{\Sigma}$ is the diagonal matrix of singular values, and $\mathbf{V}$ is the matrix of right singular vectors. The decomposition allows for the identification of the principal components, the directions in the data that exhibit the most variance.

However, real-world data is often represented through noisy measurements, including the digitized data in ASPIRE. The presence of noise can significantly alter the principal component directions. Therefore, in this work, Robust Principal Component Analysis (RPCA) is employed as opposed to the standard PCA. RPCA is an optimization strategy where a data matrix is decomposed into a low-rank component and a sparse component, effectively separating the underlying data structure from the outliers.

Mathematically, RPCA solves the following optimization problem:
\begin{align}
\min_{\mathbf{L}, \mathbf{S}} \|\mathbf{L}\|_* + \lambda \|\mathbf{S}\|_1 \quad \text{subject to} \quad \mathbf{Z} = \mathbf{L} + \mathbf{S},
\end{align}
where $\mathbf{L}$ is the low-rank component, $\mathbf{S}$ is the sparse component, $\|\mathbf{L}\|_*$ denotes the sum of singular values of $\mathbf{L}$, and $\|\mathbf{S}\|$ denotes the $L_1$ norm of $\mathbf{S}$. The parameter $\lambda$ balances the trade-off between the low-rank approximation and the sparsity of the outliers. When $\lambda$ is set to a high value, the influence of $\mathbf{S}$ is minimized, and $\mathbf{L}$ dominates the decomposition. In this case, the decomposition closely resembles that of standard PCA. Once the decomposition is achieved, $\mathbf{L}$ is subjected to standard PCA procedure to obtain the principal components. These components, denoted as $L_s$, serve as the primary directions that capture the most variance in the cleaned, low-rank data.

\begin{figure}[h!]
\centering
\includegraphics[width=.45\textwidth]{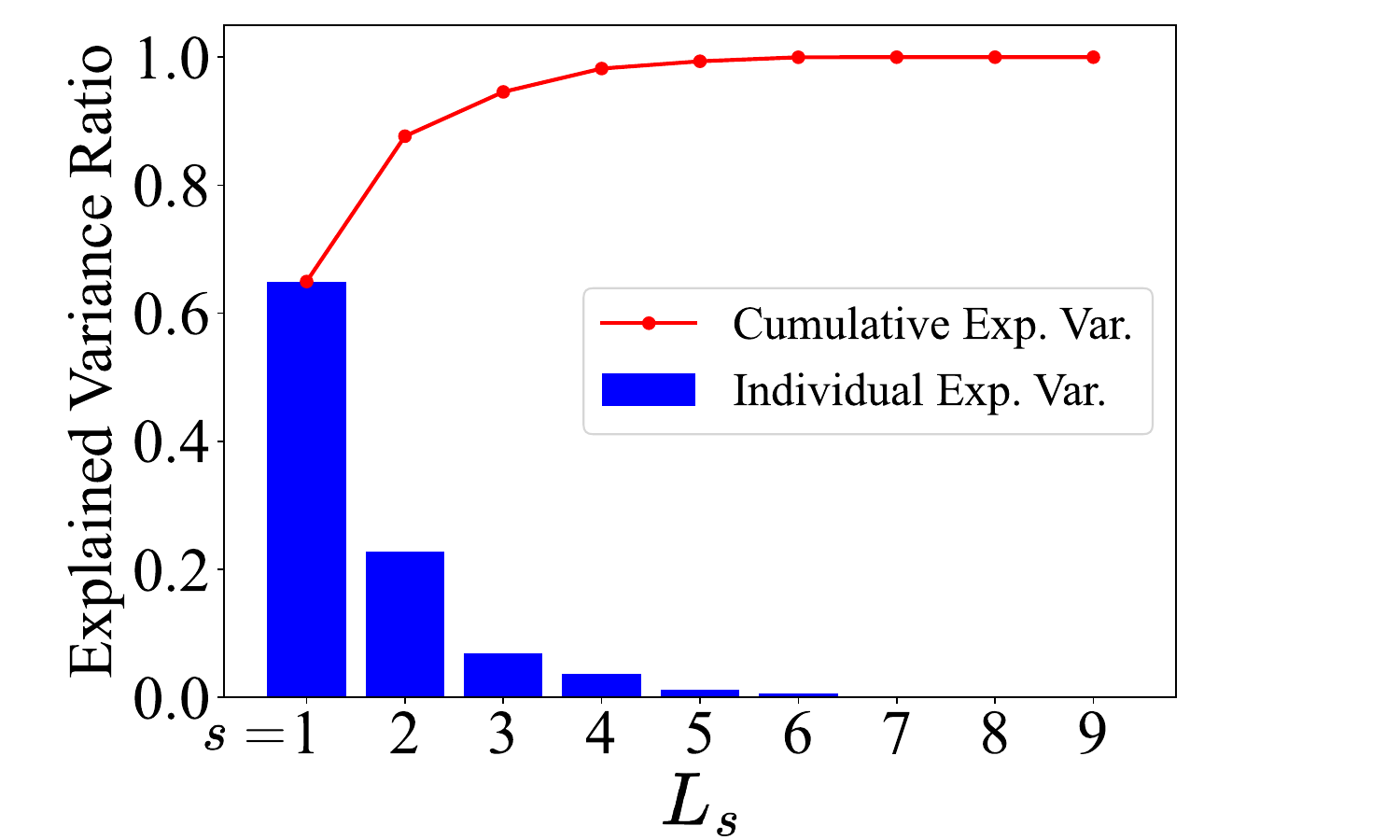}  
\caption{Individual and cumulative explained variances of the integrated latent variables}
\label{fig:cumsumvar}
\end{figure}

In this investigation, an artificial test dataset was generated, consisting of all airfoils from the training set. The operating conditions ranged from $-4.0^\circ \le \alpha \le 12.0^\circ$ and $M_\infty$ = 0.30, 0.60, 0.70. To better understand the overall behavior of the airfoils, $c_l$ was studied instead of individual $C_p$ values. The process required the generation of 601 points along the entire airfoil for each $\alpha$-$M_\infty$ pair. The artificial test set was fed into the deep neural network which produces the 10 latent variables, resulting in a matrix $\mathbf{Z}$ of size $1{,}060{,}164 \times 10$. The same integral operator used to calculate $c_l$ was then applied to this matrix. The final $1{,}764 \times 10$ array of integrated latent variables was processed through RPCA to obtain $\mathbf{L}$. From the cumulative explained variance by individual $L$ components (Fig.~\ref{fig:cumsumvar}), it was found that 3 components were sufficient to explain approximately 95\% of the variance. This allowed us to analyze contributions of all variables using the $1{,}764 \times 3$ matrix, which conveniently could be visualized in a 3D plot.

\begin{figure}[h!]
\centering
\begin{subfigure}{0.95\textwidth}
  \centering
  \includegraphics[width=0.9\linewidth]{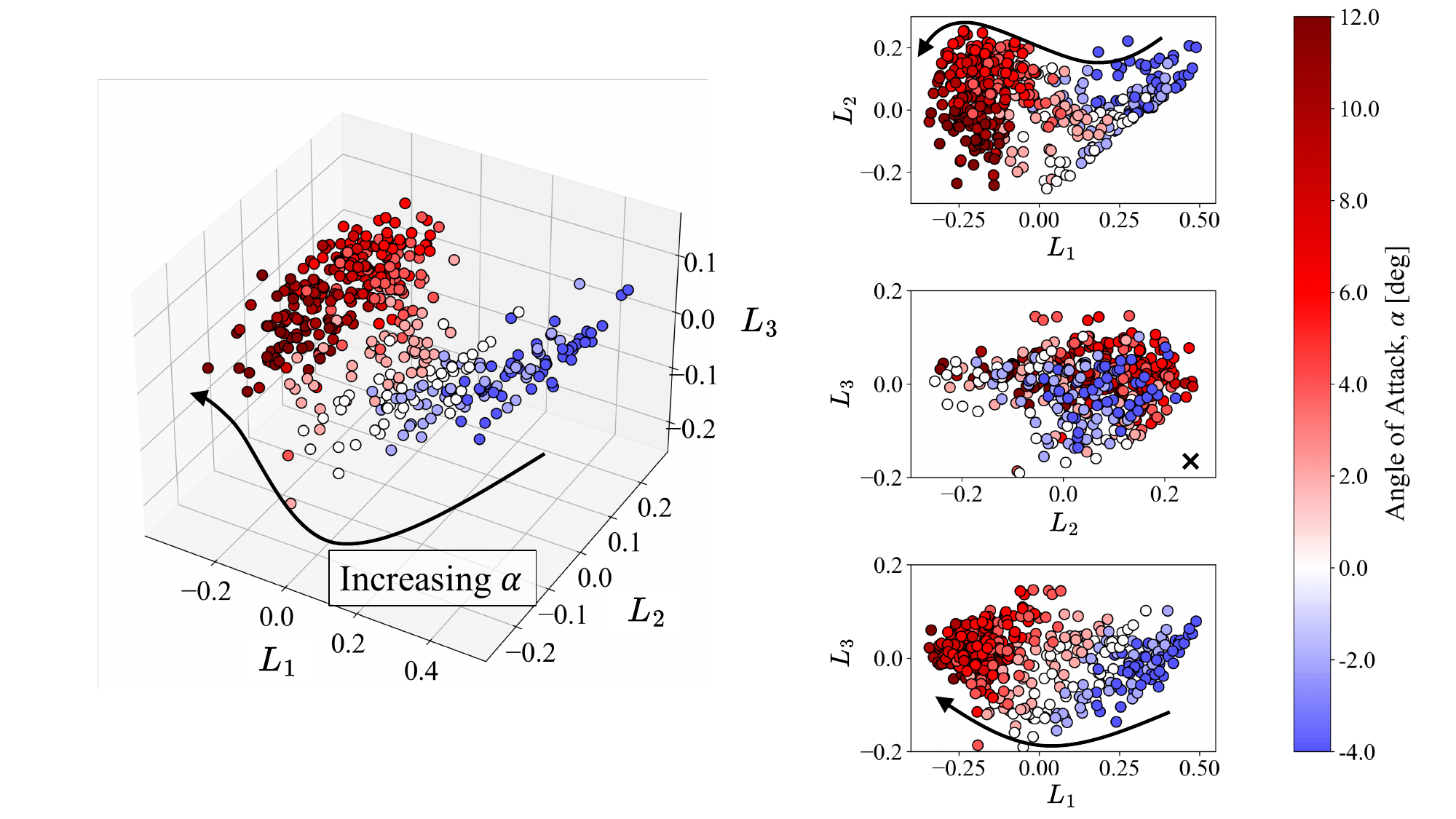}
  \caption{Distribution of the latent variables within the active space, colored by $\alpha$}
  \label{fig:pca_alpha_show_all}
\end{subfigure} \\%
\begin{subfigure}{0.95\textwidth}
  \centering
  \includegraphics[width=0.9\linewidth]{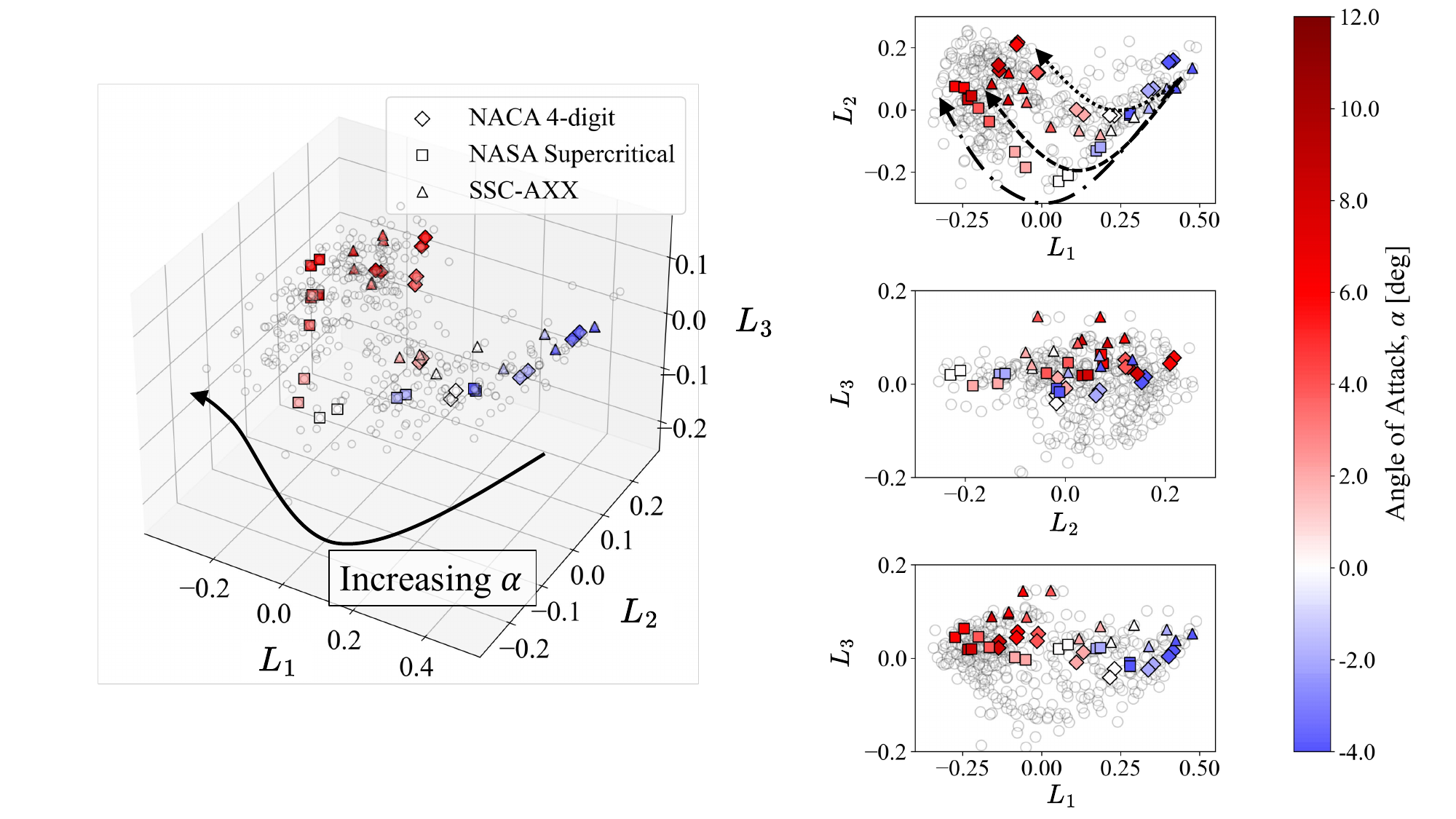}
  \caption{Trajectories followed by NACA 4-digit, NASA Supercritical and SSC-AXX families in response to changing $\alpha$}
  \label{fig:pca_alpha_by_family}
\end{subfigure} \\
\caption{Analysis of the model active space with varying $\alpha$ and $M_\infty = 0.70$}
\label{fig:pca_alpha}
\end{figure}

Figure~\ref{fig:pca_alpha} depicts the distribution of the latent variables within the active space, where each point is colored by its  angle of attack. A clear trajectory with respect to $\alpha$ can be observed. The trend is clearest in the $L_1$--$L_2$ plane, where increasing $\alpha$ in the pre-stall regime corresponds to a parabolic path in the plane. The latent variables deviate from this crescent curve as the angle of attack approaches stall. At $\alpha = 12^\circ$, it can be seen that the latent variables decrease along $L_2$ with relatively smaller changes in the $L_1$ direction.

Figure~\ref{fig:pca_alpha_by_family} presents 3 airfoil families in isolation in the same active space at $M_\infty = 0.70$. NACA 4-digit series, NASA Supercritical airfoils, and SSC-AXX airfoils were chosen as they were designed for different purposes and thus exhibit different behaviors with respective to $\alpha$ and $M_\infty$. For instance, the 3 airfoil families have different degrees of camber and airfoil thicknesses. Consequently, with changing $\alpha$ and a fixed $M_\infty = 0.70$, it is observed that the 3 airfoil families follow distinct trajectories. This is most noticeable in the $L_1$--$L_2$ plane, where NASA Supercritical airfoil follows the outermost, SSC-AXX the middle, and NACA 4-digit series the innermost trajectories respectively.

\begin{figure}[h!]
\centering
\begin{subfigure}{0.95\textwidth}
  \centering
  \includegraphics[width=0.9\linewidth]{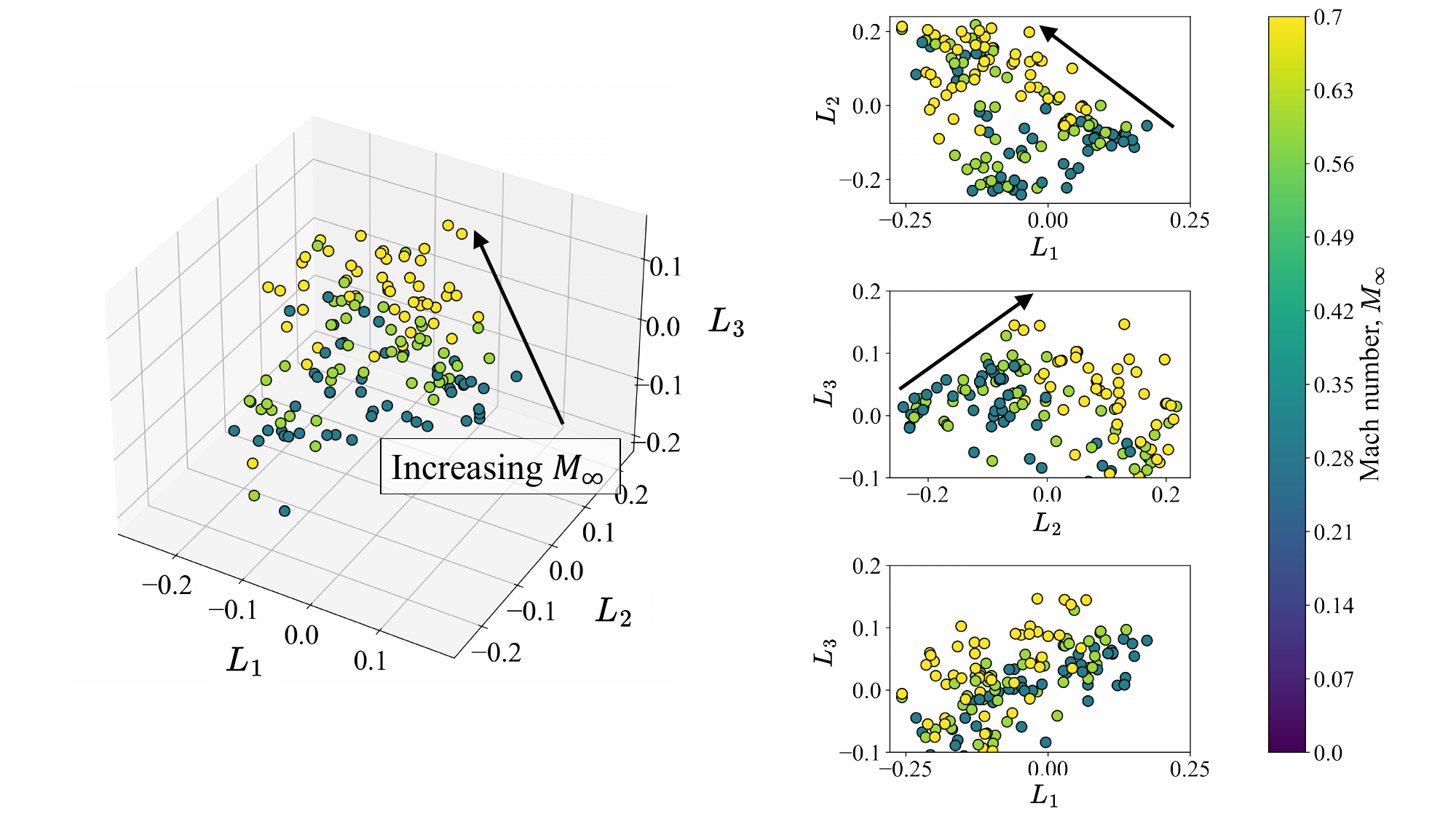}
  \caption{Distribution of the latent variables within the active space, colored by $M_\infty$}
  \label{fig:pca_mach_show_all}
\end{subfigure} \\%
\begin{subfigure}{0.95\textwidth}
  \centering
  \includegraphics[width=0.9\linewidth]{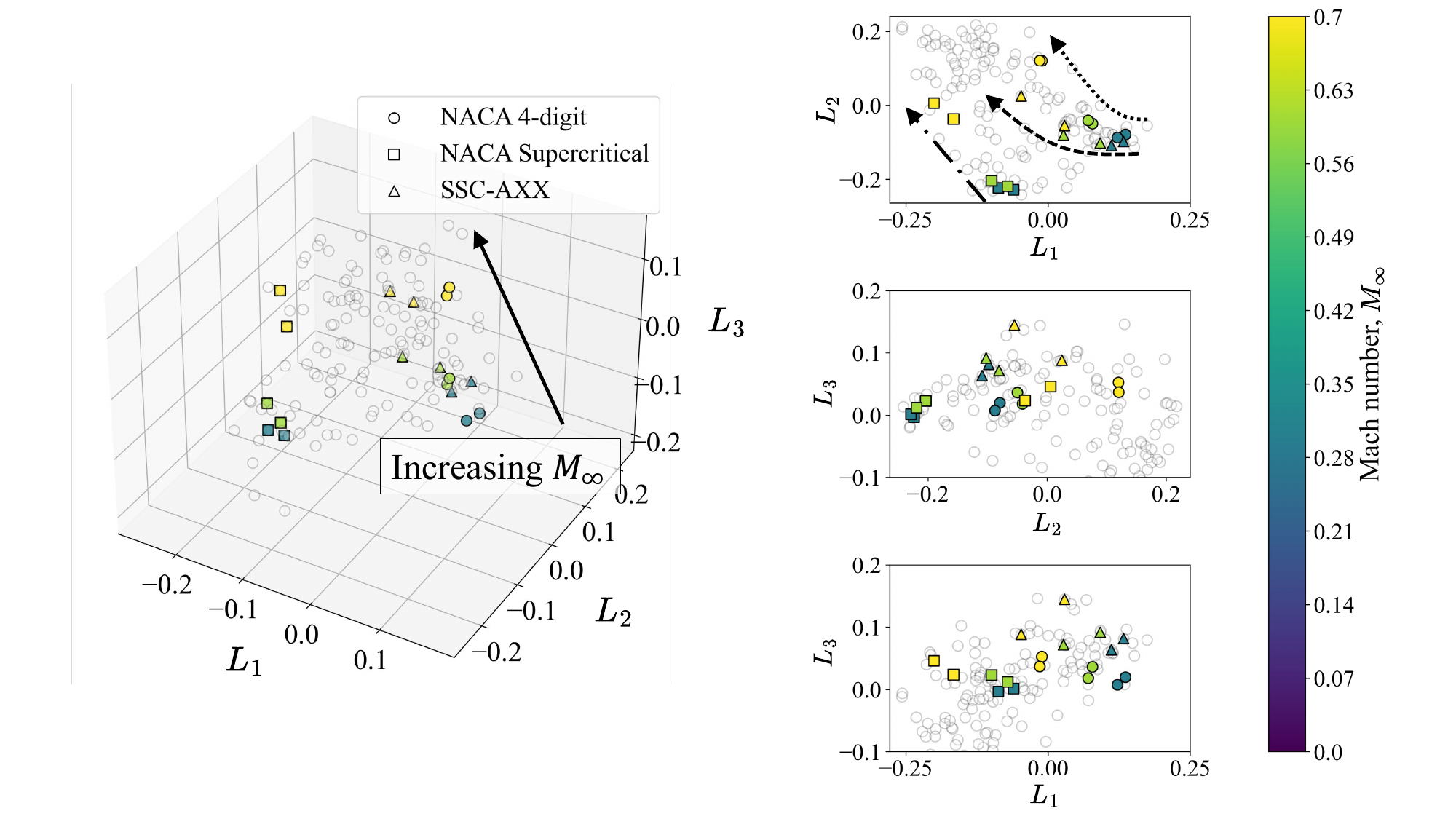}
  \caption{Trajectories followed by NACA 4-digit, NASA Supercritical and SSC-AXX families in response to changing $M_\infty$}
  \label{fig:pca_mach_by_family}
\end{subfigure} \\
\caption{Analysis of the model active space with varying $M_\infty$ and $\alpha = 4.0^\circ$}
\label{fig:pca_mach}
\end{figure}

The behavior of the latent variables with respect to varying freestream Mach numbers is illustrated in Fig.~\ref{fig:pca_mach}. With increasing $M_\infty$, a decrease in the $L_1$ direction, and an increase in $L_2$ and $L_3$ directions can be observed. The identified trends are less distinct than those in relation to $\alpha$. This is likely due to the fact that angle of attack, in general, has a greater impact on $c_l$ compared to the freestream Mach number. In Fig.~\ref{fig:pca_mach_by_family}, the 3 separate trajectories are followed by different airfoil families. NASA Supercritical airfoil follows the outermost, and NACA 4-digit series the innermost diagonal paths. The SSC-AXX family follows a path between these two airfoil families. As with Fig.~\ref{fig:pca_alpha_by_family}, the clearly discernible trajectories are due to the fact that the airfoils are optimized for different freestream Mach number regimes. NACA 4-digit series airfoils were designed for incompressible flows and NASA Supercritical airfoils for transonic regimes. SSC-AXX airfoils, on the other hand, is a rotorcraft airfoil which must perform well in a wide range of $M_\infty$. 

The analysis of the active space learned by ADAPT's deep neural network showed that the latent variables are grouped into distinct clusters in terms of $\alpha$ and $M_\infty$. While the overall behavior remains similar, the trajectories taken by an airfoil within the active space differs based on the airfoil family. The results serve as the evidence that the model has captured an interpretable and physical mapping between the input space and the latent space during training. 

\newpage
\section{Example Interactions with the Natural Language Interface}\label{app:chatbot}
\begin{figure}[!ht]
\centering
\begin{subfigure}{0.9\textwidth}
  \centering
  \includegraphics[width=0.7\linewidth]{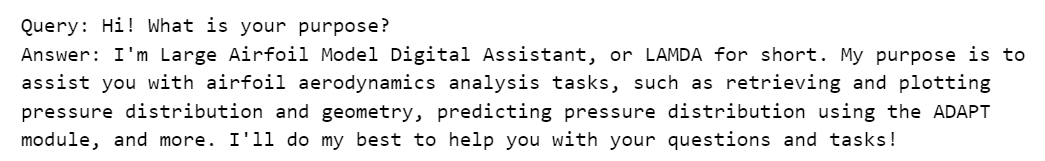}
  \caption{Response to a general user inquiry pertaining to the interface's capabilities}
  \label{fig:general_inquiry}
\end{subfigure} \\%
\begin{subfigure}{0.9\textwidth}
  \centering
  \includegraphics[width=0.7\linewidth]{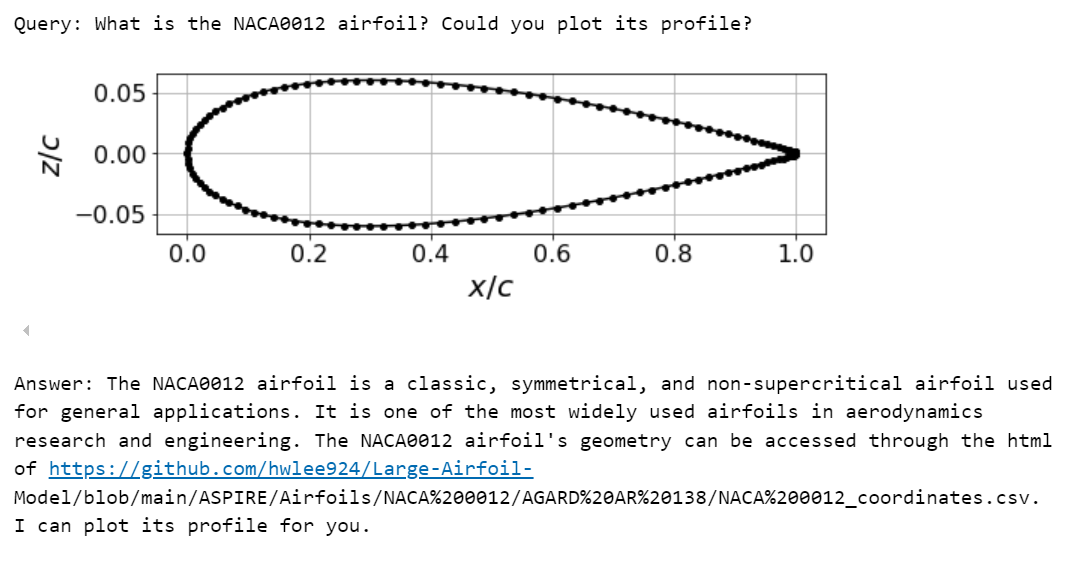}
  \caption{Retrieval of existing airfoil geometric profile and related information from the ASPIRE database.}
  \label{fig:airfoil_inquiry}
\end{subfigure} \\
\begin{subfigure}{0.9\textwidth}
  \centering
  \includegraphics[width=0.7\linewidth]{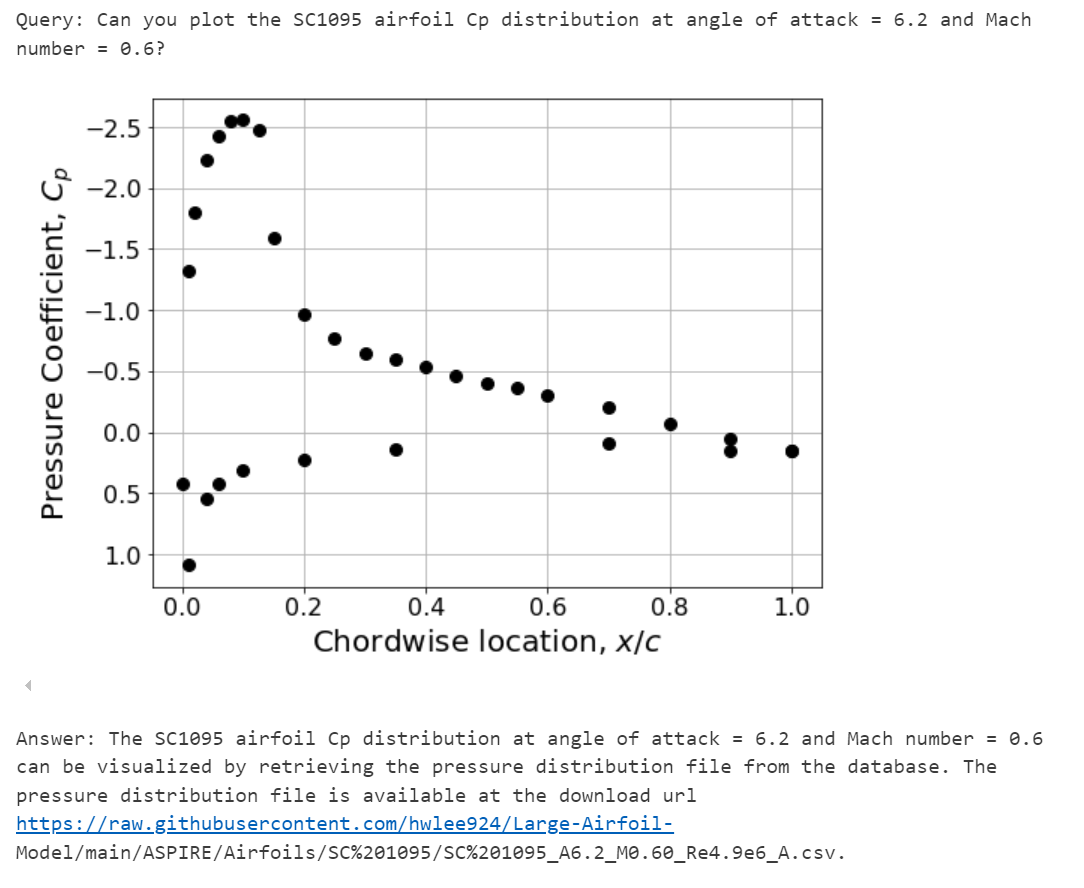}
  \caption{Retrieval of existing $C_p$ data from the ASPIRE database}
  \label{fig:cp_retrieval}
\end{subfigure}
\caption{Example interactions with the model's natural language interface, which can retrieve general information and digitized data from the ASPIRE database}
\label{fig:example_interaction_1}
\end{figure}

\begin{figure}[!ht]
\centering
\begin{subfigure}{0.9\textwidth}
  \centering
  \includegraphics[width=0.64\linewidth]{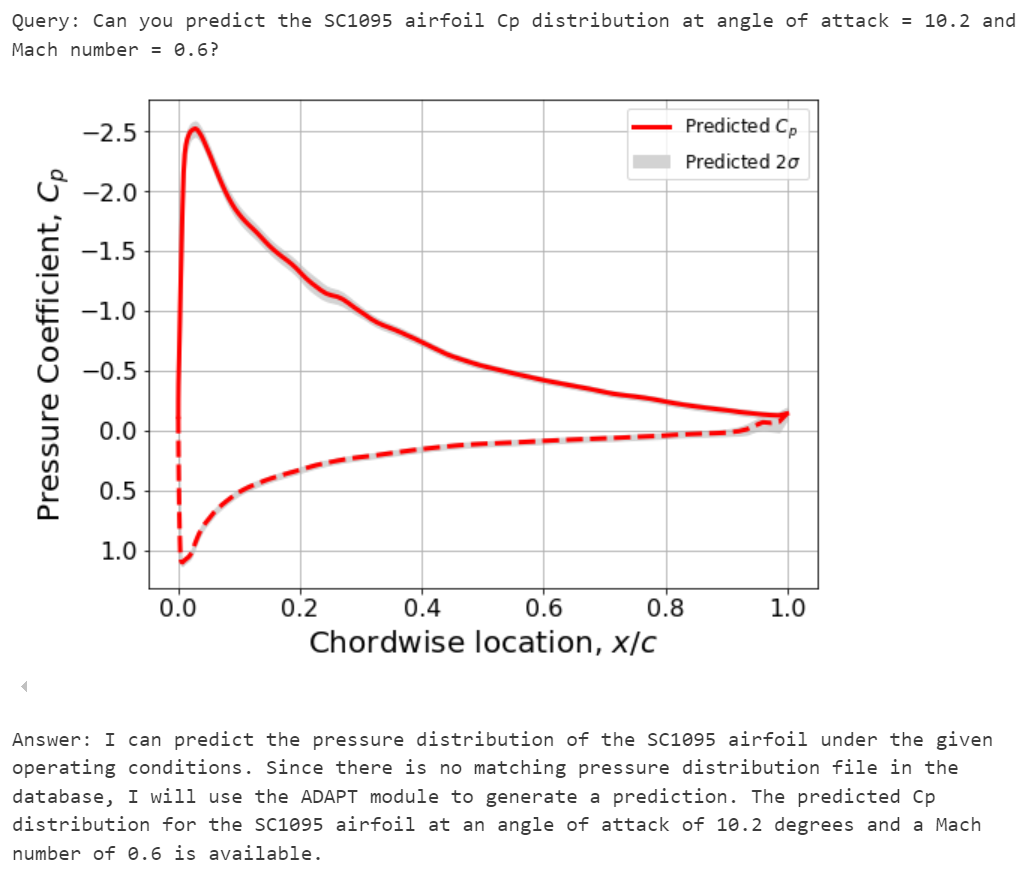}
  \caption{Prediction of $C_p$ distribution under new operating conditions for an airfoil that exists within the ASPIRE database}
  \label{fig:prediction_inquiry}
\end{subfigure} \\%
\begin{subfigure}{0.9\textwidth}
  \centering
  \includegraphics[width=0.64\linewidth]{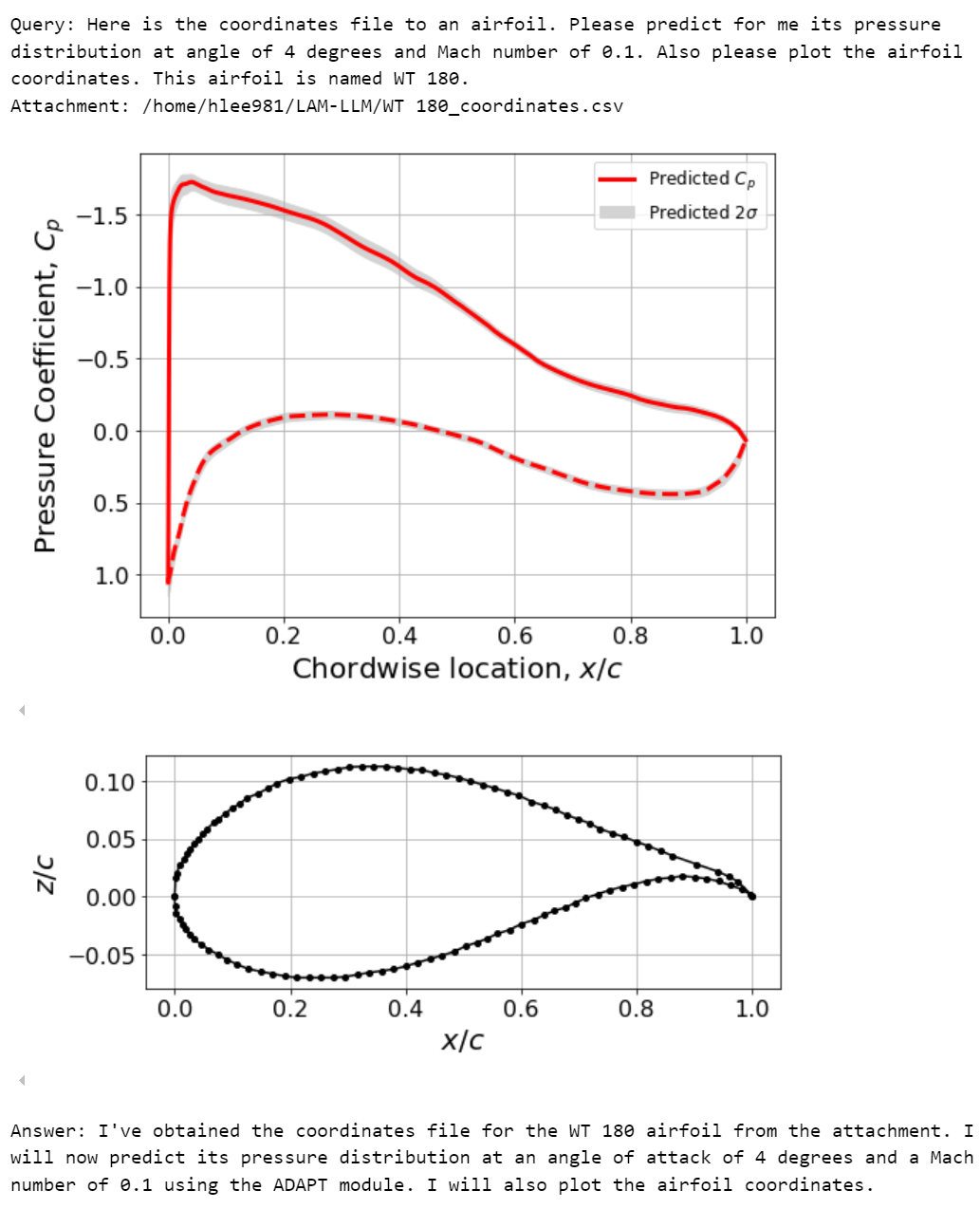}
  \caption{Prediction of $C_p$ distribution under new operating conditions for an airfoil provided manually by the user}
  \label{fig:attachment_inquiry}
\end{subfigure} \\
\caption{Example interactions with the model's natural language interface, which can utilize ADAPT to predict airfoil aerodynamics}
\label{fig:example_interaction_2}
\end{figure}

This section presents a number of exemplary interactions with the Large Airfoil Model's natural language interface introduced in Sec.~\ref{sec:lam_da}. The examples captured in Figs.~\ref{fig:example_interaction_1} and \ref{fig:example_interaction_2} illustrate the model's capacity to respond to queries on airfoil aerodynamics. The examples include retrieving and predicting pressure distributions, interpreting airfoil geometries, and other relevant tasks.

\end{document}